\documentclass[a4paper,11pt]{article}

\usepackage[T1]{fontenc}
\usepackage[utf8]{inputenc}
\usepackage[backend=biber, sorting=none]{biblatex}
\usepackage[linkcolor = black]{hyperref}
\usepackage{dsfont}
\usepackage{fullpage}
\usepackage{scalerel}
\usepackage[title]{appendix}

\usepackage{xcolor}
\usepackage{graphicx}
\usepackage{amsmath}
\usepackage{amssymb}
\usepackage{mathtools}
\usepackage{units}
\usepackage{slashed}
\usepackage{physics}
\usepackage{cancel}
\usepackage[normalem]{ulem}
\usepackage{marginnote}

\title{Relativistic and spectator effects in leptogenesis with heavy sterile neutrinos}
\author{
Björn Garbrecht$^1$, %
Philipp Klose$^2$, %
Carlos Tamarit$^1$
}
\hypersetup{colorlinks=true}

\DeclareMathOperator{\p}{P}

\DeclareMathOperator{\SU}{SU}

\let\i\undefined\DeclareMathOperator{\i}{i}
\DeclareMathOperator{\dz}{\frac{\rm d}{{\rm d} z}}

\newcommand{\smallpara}{ {\scaleto{\parallel}{5pt}} }

\begin{document} 

\thispagestyle{empty}
\begin{minipage}{.45\linewidth}
\begin{flushleft}                          
{TUM-HEP-1198-19}
\end{flushleft} 
\end{minipage}
\hfill
\begin{minipage}{.45\linewidth}
\begin{flushright}  
CP3-19-18
\end{flushright} 
\end{minipage}

{\flushright 
\vspace{1.8cm}}
\begin{center}
{\Large \textbf{Relativistic and spectator effects in leptogenesis with heavy sterile neutrinos}}\\
\bigskip
\vspace{5mm}
{ \large \bf Bj\"orn Garbrecht$^1$, Philipp Klose$^{2}$, Carlos Tamarit$^1$}\\
\vskip5mm
$^1$Physik-Department T70,  Technische Universität München,\\ James-Franck Straße 1, 85748 Garching, Germany\\
\vskip5mm
$^2$Centre for Cosmology, Particle Physics and Phenomenology - CP3, Université catholique de
Louvain, Chemin du Cyclotron 2, 1348 Louvain-la-Neuve, Belgium
\end{center}
\vspace{1cm}
\begin{abstract}
For leptogenesis with heavy sterile neutrinos above the electroweak scale, asymmetries produced at early times (in the relativistic regime) are relevant, if they are protected from washout. 
This can occur for weak washout or when the asymmetry is partly protected by being transferred to spectator fields. 
We thus study the relevance of relativistic effects for leptogenesis in a minimal seesaw model with two sterile neutrinos in the strongly hierarchical limit. 
Starting from first principles, we derive a set of momentum-averaged fluid equations to calculate the final $B-L$ asymmetry as a function of the washout strength and for different initial conditions at order one accuracy. 
For this, we take the leading fluid approximation for the relativistic $CP$-even and odd rates. 
Assuming that spectator fields remain in chemical equilibrium, we find that for weak washout, relativistic corrections lead to a sign flip and an enhancement of the asymmetry for a vanishing initial abundance of sterile neutrinos. 
As an example for the effect of partially equilibrated spectators, we consider bottom-Yukawa and weak-sphaleron interactions in leptogenesis driven by sterile neutrinos with masses $\gtrsim \unit[5\times10^{12}]{GeV}$. 
For a vanishing initial abundance of sterile neutrinos, this can give rise to another flip and an absolute enhancement of the final asymmetry in the strong washout regime by up to two orders of magnitude relative to the cases either without spectators or with fully equilibrated ones. 
These effects are less pronounced for thermal initial conditions for the sterile neutrinos. 
The $CP$-violating source in the relativistic regime at early times is important as it is proportional to the product of lepton-number violating and lepton-number conserving rates, and therefore less suppressed than an extrapolation of the nonrelativistic approximations may suggest.
\end{abstract}

\begin{center}
\vfill\flushleft
\noindent\rule{6cm}{0.4pt}\\
{\small  E-mail addresses: \tt garbrecht@tum.de, philipp.klose@uclouvain.be, carlos.tamarit@tum.de}

\end{center}

\bigskip


\newpage
\tableofcontents
\section{Introduction}
\label{sec:intro}

The origins of the primordial baryon asymmetry and neutrino masses remain fundamental open questions in cosmology and particle physics.
Both are  simultaneously addressed by the mechanism of leptogenesis~\cite{Fukugita:1986hr}, in which the asymmetry is generated from the interplay between lepton number ($L$) violating out-of-equilibrium  {reactions}  of sterile neutrinos
and nonperturbative sphaleron processes that reprocess part of the lepton charge into a net baryon {number ($B$)} asymmetry.  
As the early universe was filled with a hot, dense plasma, finite temperature effects can be of importance for leptogenesis.%
Generically, this is expected if leptogenesis occurs at least partly at temperatures either comparable or larger than the mass scale of the dynamically important sterile neutrinos, i.e. when the neutrinos are relativistic. 
This is the case in  ``weak washout'' scenarios, where the relevant washout processes always remain out of equilibrium, so that the asymmetries produced at early times/high temperature survive until after the washout processes freezeout%
\footnote{Recall that departure of equilibrium is one of the conditions for baryogenesis pointed by Sakharov~\cite{Sakharov:1967dj}}.
On the other hand, in ``strong washout'' scenarios the washout processes do equilibrate at intermediate temperatures, 
and the standard lore of leptogenesis holds that the final asymmetry will be predominantly generated around the time of their freezeout. 
In this case the sterile neutrinos are nonrelativistic, so that finite-temperature effects are sub-leading and can be calculated with simplifying nonrelativistic approximations, 
as applied for example in Refs.~\cite{Salvio:2011sf,Laine:2011pq,Biondini:2013xua,Biondini:2015gyw,Biondini:2016arl,Bodeker:2017deo}.

Nonetheless, reactions occurring at early times can in fact affect the final asymmetry even in the strong washout regime due to spectator effects%
\footnote{Recall that spectator fields are standard model fields that do not couple to the sterile neutrinos directly. 
They can carry $B$ or $L$ and thus affect the lepton asymmetries via  the various Standard Model interactions.}.
If the spectator interactions become faster than the Hubble rate during leptogenesis, their net effect is to enforce chemical equilibrium constraints among the abundances of the participating particles~\cite{Barbieri:1999ma,Buchmuller:2001sr,1404.2915}.
However, if spectator interactions only equilibrate partially  during leptogenesis, one has to account for the nontrivial dynamics of the Standard Model interactions. 
In particular, there is a possibility that part of the lepton asymmetry could be transferred to spectator fields, where it could remain shielded from washout if the latter is only active before the spectators equilibrate and the asymmetry is transferred back to the leptons.
The Standard Model processes equilibrate successively as the universe cools down, so that it is natural to study scenarios in which some interactions are partially equilibrated. 
While gauge and top-Yukawa interactions can be assumed to always reach equilibrium for reheating temperatures below the scale of grand unification, strong sphalerons equilibrate at $T\sim 2.4\times 10^{13}$~GeV, 
while bottom Yukawa and weak sphaleron interactions do so at $T_b\sim 4.2\times 10^{12}$~GeV and $T_{\rm ws}\sim 1.8\times 10^{12}$~GeV, respectively. 
The remaining Yukawa interactions equilibrate in succession at lower temperatures ~\cite{1404.2915}.

A study of the impact of partially equilibrated spectator fields was presented in Ref.~\cite{1404.2915}, but only for the strong washout scenario with thermal initial abundance of the sterile neutrinos. 
In that case, relativistic effects can be ignored due to the small deviation of the sterile neutrinos from equilibrium  at high temperature, implying suppressed  violation of baryon number at early times.
Since the equilibration of the sterile neutrinos between the epochs of reheating and leptogenesis is not always guaranteed, 
non-thermal or vanishing initial conditions for sterile neutrinos remain well motivated, and a reappraisal of the impact of spectator fields accounting for relativistic effects remains to be done. 
 
For such a reappraisal, a proper understanding of finite temperature effects on the sterile neutrino reaction rates is paramount, and in recent years there has been substantial progress on this front. 
To put the current work into context, we therefore briefly summarize the theoretical status of the relativistic and intermediate rates~(cf. Ref.~\cite{Biondini:2017rpb} for a recent review):

\begin{description}
\item[1. Reaction rates in low-scale leptogenesis:]
In low-scale scenarios, such as those with GeV-scale sterile neutrinos \cite{Akhmedov:1998qx,Asaka:2005pn}, the asymmetry can be generated in the fully relativistic regime, 
in which plasma effects such as scatterings, soft gauge boson emission, and thermal masses dominating the kinematics of one-to-two processes become important, all of which go beyond the simple description in terms of decays and inverse decays of the sterile neutrinos.
For this regime, the sterile neutrino relaxation rates have been studied extensively throughout the recent years~\cite{1012.3784,1202.1288,1302.0743,1303.5498}. 
As relativistic interactions are not helicity blind, it is necessary to distinguish processes that either conserve and flip lepton helicity.  
When assigning lepton number to the different helicity states of the sterile neutrinos, this distinction becomes equivalent to differentiating between lepton-number conserving (LNC) and violating (LNV) processes. 
For scenarios with GeV-scale sterile neutrinos the importance of LNV processes has been pointed out in Ref.~\cite{Hambye:2017elz,Hambye:2016sby}, while the full the interplay with the LNC rates
has been established in Refs.~\cite{Ghiglieri:2017gjz,Eijima:2017anv,Antusch:2017pkq,Ghiglieri:2017csp,Eijima:2017cxr,Eijima:2018qke,Ghiglieri:2018wbs}. \
\item[2. Reaction rates in high-scale leptogenesis:]
In high-scale leptogenesis scenarios with sterile neutrino masses way above the electroweak scale~\cite{Fukugita:1986hr,Buchmuller:2005eh},  
even when the asymmetry is dominantly produced in the relativistic regime, the washout processes are generally active throughout the transition to the nonrelativistic regime. 
The pertinent relaxation rates of the sterile neutrinos at finite temperature have been calculated for arbitrary masses in Refs.~\cite{1302.0743,Ghisoiu:2014ena} by summing over LNC and LNV processes. 
However, in order to accurately calculate the washout effects and the $CP$-violating source for both the relativistic and intermediate regimes, it becomes necessary to distinguish the two types of  processes,
and obtaining estimates of their rates applicable to high-scale leptogenesis scenarios is one of the main goals of the present paper.
For this one has to drop the approximations of the fully relativistic regime, 
but we may simplify the calculation of the $CP$-violating source by considering the hierarchical limit in which the heavier sterile neutrinos decouple.

There have been previous efforts to include thermal effects and to calculate the production and the washout of the asymmetry in the relativistic regime of leptogenesis with heavy sterile neutrinos, 
but two-by-two scattering processes (for which the leading-logarithmic contribution can be captured when including the finite width of the Standard Model leptons) have not been accounted for~\cite{Kiessig:2011fw,Kiessig:2011ga}, 
and in addition the quantum-statistical factors were not included accurately in some cases~\cite{Giudice:2003jh}. 
Refs.~\cite{Pilaftsis:2003gt,Pilaftsis:2005rv,hep-ph/0401240} included relativistic corrections for the washout arising from scatterings with Standard Model particles, yet these were not considered for the $CP$-violating source.
Thermal corrections to the latter are known even at the NLO level, but only in the nonrelativistic regime~\cite{Biondini:2015gyw,Biondini:2016arl,Bodeker:2017deo}.
We are therefore also interested here in updating the analytic description of Ref.~\cite{hep-ph/0401240} for the weak washout regime in view of the recent advancements in the calculation of the relativistic reaction rates.
\item[3. Equilibration rate of spectator fields:]
Interactions that change the charge density
of spectator fields are mediated by Yukawa interactions.
In the symmetric electroweak phase, these rates include at leading
order the effect from gauge bosons, which is why methodically, they
can be obtained in a similar manner as the LNC conserving rates for
the sterile neutrinos in the relativistic regime. In Ref.~\cite{1404.2915},
the rates from Ref.~\cite{1303.5498} have thus been applied to
the case of spectator fields. We note that there is a recent
detailed analysis of the equilibration of right-handed electrons~\cite{Bodeker:2019ajh} that reports a rate that is $2.8$
times higher than what one would obtain from  Ref.~\cite{1303.5498}.
While work on a resolution of this discrepancy is certainly in order,
we note that an enhancement in the rate for spectator processes assumed
in this present work can be absorbed within choosing a larger values $M_1$
such as to remain in the regime of partial equilibration for a particular
spectator field.
\end{description}

In view of the above, the aim of this paper is two-fold. 
First, we seek to obtain consistent evolution equations for the particle asymmetries {in high-scale leptogenesis} that remain valid in the transition from the relativistic through the intermediate to the nonrelativistic regime,
and to compute the associated source, equilibration and backreaction rates, including the leading effects from scatterings in the plasma. 
Second, we study the impact of the relativistic effects on the generation of the asymmetry, either ignoring spectators or accounting for their partial equilibration. 
To illustrate these  effects, we study a simplified one-flavour regime in which the sterile neutrinos couple to a single combination of flavours of Standard Model leptons. 
This is appropriate for high temperatures $\gtrsim 4\times 10^{11}$ GeV, in which the lepton Yukawa couplings are not yet in equilibrium and the leptons are effectively indistinguishable. 
For our study of spectator effects at such high temperatures, we consider scenarios with a lightest right-handed neutrino with a mass around $ 10^{13}$ GeV and with partially equilibrated bottom Yukawa and weak sphaleron interactions.
As mentioned earlier, these interactions equilibrate at  $T\sim10^{12}$ GeV, so that the choice of $ M_1\sim10^{13}$ GeV ensures partial equilibration of the spectators until after the freeze-out of the lepton asymmetry at $T\sim M_1$.
The previously described set-up corresponds to a minimal leptogenesis scenario that satisfies the bound  $M_1\gtrsim 10^9\,{\rm GeV}$ imposed by the observed oscillations of the Standard Model neutrinos~\cite{hep-ph/0202239}.
\footnote{{This bound can be evaded by relying on e.g. resonant enhancement, lepton flavour effects, or the possibility that the freeze out of the asymmetry is controlled by the quenching of sphaleron processes at the electroweak phase transition, rather than by the Maxwell suppression of the sterile neutrinos;
for reviews, see e.g. Refs.~\cite{Buchmuller:2005eh,Davidson:2008bu,Garbrecht:2018mrp}. }}

The remainder of this paper is organized as follows. 
In Sec.~\ref{sec:intromodel}, we specify the seesaw model that we use to illustrate our developments and explain how we account for the expansion of the universe. 
The resulting fluid equations for leptogenesis are presented in Sec.~\ref{sec:boltzmann_eq}, where we qualitatively explain the particular terms in the fluid equations and present approximate forms for the reaction rates that can be used in phenomenological calculations.
In Sec.~\ref{sec:derivation_fluid_eq_rates}, we give the full derivation of the fluid equations based on the closed time-path (CTP) formalism of nonequilibrium quantum field theory~\cite{Schwinger:1960qe,Keldysh:1964ud,Calzetta:1986cq}, 
and give technical details for the calculation of the sterile neutrino reaction rates.
In particular, we study the behaviour of the reaction rates in the relativistic and intermediate regimes.
In Section~\ref{sec:numerical_scans}, we solve the fluid equations numerically for a toy model without spectator interactions and for a realistic model accounting for partially equilibrated bottom Yukawa and weak sphaleron interactions.
For the toy model we present a semianalytical approximation of the asymmetry in the weak washout regime, and in the realistic model we isolate the effects from partially equilibrated spectator fields by comparing with the results based on the assumption of fully equilibrated spectators. 
Our conclusions can be found in Sec.~\ref{sec:conclusions}.
\section{Simplified Model for minimal High-Scale leptogenesis}
\label{sec:intromodel}

We consider a minimal type I seesaw model with two sterile Majorana neutrinos $N_1$ and $N_2$ that couple to the Higgs boson $\phi$ and a single linear combination ${l_\smallpara }$ of the left-handed Standard Model lepton doublets,
\begin{equation}
\mathcal{L} = \mathcal{L}_{SM} +  \frac{1}{2} \bar{N}_i \i \slashed{\partial} N_{i}  - \frac{1}{2 }  \bar{N}_i \widetilde M_{ij} N_{j}  - F^*_{ i} \,\,\overline{l}_{\smallpara} \widetilde{\phi}^{*}P_R N_{ i} - F_{i}\, \bar{N}_{i} \widetilde\phi  \,P_L l_{\smallpara}
\ .
\label{eq:minimal_seesaw_def}
\end{equation}
In this Lagrangian, we use the shorthand notation $\widetilde\phi_a\equiv \epsilon_{ab}\phi_b$, where $a,b$ are indices of the fundamental representation of ${\rm SU}(2)$ and $\epsilon_{ab}$ is the $2\times 2$ antisymmetric tensor with $\epsilon_{12}=1$. 
We take the sterile neutrino mass matrix to be real and diagonal, $\tilde M_{ij} = \text{diag} \big(\tilde M_1, \tilde M_2 \big) $, and assume strongly hierarchical masses, $\tilde M_1 \ll \tilde M_2$.
Focusing on energy scales $E \ll \tilde M_2$, we may then construct an effective theory by integrating out the heavy $N_2$.
To leading order in an expansion in  $\nicefrac{E}{\tilde M_2}$, the effects of virtual $N_2$ exchanges are captured by the two the Weinberg-like operators
\begin{align}
\label{eq:Weinberg_ops}
O_{L} 	&\equiv\frac{ F_2^2 }{2\tilde M_2} \big( \bar{l^c_{\smallpara}}\, \widetilde{\phi} \big) \p_L ( \tilde{\phi}^\top {l_\smallpara} \big) \ ,
&
O_{R}  	&\equiv O_L^\dagger =  \frac{(F^*_{2 })^2}{2\tilde M_2} \big( \bar{{l_\smallpara}} \widetilde{\phi}^* \big) \p_R \big( \widetilde{\phi}^\dagger {l^c_\smallpara} \big) \ ,
\end{align}
that have to be added to the part of Lagrangian in Eq.~\eqref{eq:minimal_seesaw_def} that remains after neglecting the terms involving $N_2$. For phenomenological models of the light neutrino masses,
one may want add more than two sterile neutrinos to the
Standard Model, but in order to study leptogenesis
in the hierarchical limit, the present model with
two new neutrinos is sufficient.

The expansion of the universe is captured by following the approach in \cite{1002.1326,Garbrecht:2018mrp}, where it was assumed that leptogenesis takes place in a spatially flat Friedman-Robertson-Walker (FRW) cosmology during the radiation dominated period. 
In this case, one can work in a comoving Minkowski frame in which the effect of the expansion is captured by masses that grow with the scale factor. 
In this frame one has comoving three-momenta $\mathbf{k}$ related to physical momenta $\mathbf{k}_{\rm phys}$ as $\mathbf{k} \equiv a(t) \, \mathbf{k}_{\rm phys}$, 
while the on-shell relation  for the sterile neutrinos reads  $\eta_{\mu \nu} k^\mu k^\nu = a(t)^2 \tilde M_i^2 \equiv M^2_i$. 
Within this setup, the temperature $T$ and entropy $s$ are taken to be comoving quantities as well, and we choose to normalise the scale factor $a(t)$ such that $T = T_\text{phys}$ for $a(t) = 1$, giving $T=T_{\rm phys} \cdot a(t),\, s=s_{\rm phys}\cdot a(t)^3$.
Finally, we define the time variable $z \equiv  \nicefrac{\tilde M_1}{ T_\text{phys}} = \nicefrac{M_1}{T}$, which is normalised such that $z = 1$ for $T = M_1$. In terms of $z$ the Hubble rate is given as $H = \nicefrac{\tilde M_1^2}{T z^2}$. \\

In general, we denote the (anti-)particle number densities in the early universe as $n_X^\pm (t)$, where the subscript $X$ specifies the particle species and the $\pm$ distinguishes particles and antiparticles. 
We count each member of a gauge multiplet as a separate particle species, which leads to occasional explicit factors of $N_c = 3$ and $g_w = 2 $ when summing over quark color or electroweak isospin.
For each species, the particle number yield is given by
\begin{equation}
\label{eq:charge_def}
Y_X \equiv \frac{1}{s} \,(n^+_X - n^-_X)
\ .
\end{equation}
Since the sterile neutrinos $N_i$ are Majorana particles, one has $n_{N_i}^+ = n_{N_i}^-$, and the charge density as in Eq.~\eqref{eq:charge_def} is not a useful in order to quantify asymmetries in these particles. 
Instead, we use the yields for even and odd combinations of helicity,
\begin{equation}
Y_{N_i \, \text{even}/\text{odd} } =  \frac{1}{s} (n_{N_i \, +} \pm n_{N_i \, -} ), 
\end{equation}
where $n_{N_{i\, h}}$ is the number density of sterile neutrinos with a given helicity $h$. 

Finally, we assume that the Standard Model particles remain in kinetic equilibrium throughout leptogenesis. 
This is a good approximation for physical temperatures well below the equilibration temperature of the electroweak gauge interactions in the Standard Model, $T_\text{phys} \ll T_\text{eq} \sim \unit[10^{16}]{GeV}$. 
In this regime the time evolution of the Standard Model charges is determined by their respective chemical potentials and vice versa. 
For small chemical potentials and relativistic species, one has 
\begin{align}
\mu_X &\approx \frac{3 g_s \, s}{T^2} Y_X\,,
\end{align}
with $g_s = 2$ for fermionic degrees of freedom and $g_s = 1$ for scalars. 
In chemical equilibrium, which is approached when reaction rates are much faster than the Hubble rate, the rates for a given reaction and its inverse match, which enforces the equality of the total chemical potentials of reactants and products.
Crucially however, chemical equilibrium may or may not hold throughout the temperature range that we consider in this work: 
For instance, the thermally averaged rates of Yukawa interactions are given as $h^2 \gamma^{h} T_{\rm phys}$, 
where $\gamma^{h}$ is a numerical constant of order $10^{-3}$-$10^{-2}$, depending on the representations of the fields coupled through the Yukawa interaction~\cite{1404.2915,1303.5498,Bodeker:2019ajh}, and $h$ is the Yukawa coupling in question.
In comparison, the Hubble rate  $H\propto T^2_{\rm phys}$ drops faster with decreasing temperature. 
Therefore, depending on the temperature at which the $B-L$-violating interactions mediated by $N_1$ freeze out, we may be allowed to ignore the spectator processes or assume fully equilibrated spectator fields, or we may need to treat them as partially equilibrated. 
Notably, among the spectator degrees of freedom, one has to consider the Standard Model leptons and their associated flavour effects. 
In the case of full chemical equilibration, leptogenesis has to be treated in the fully flavoured regime~\cite{Endoh:2003mz,Pilaftsis:2005rv,Abada:2006fw,Nardi:2006fx} 
by using the individual lepton flavour asymmetries instead of those in the linear combination $l_\parallel$ of flavours that appear in the Lagrangian~(\ref{eq:minimal_seesaw_def}).
In the intermediate regime, besides the partial equilibration of spectator fields, one has to account for the decoherence dynamics of the correlations among the lepton flavours~\cite{DeSimone:2006nrs,1007.4783}. 
To separate flavour decoherence from spectator effects, we therefore decide to focus on temperatures above $10^{12} {\rm GeV}$, where the $\tau$ leptons reach chemical equilibrium. 
Methodologically it should be straightforward to combine both, partial flavour decoherence and partial equilibration of spectator fields in one single calculation.

\section{Relativistic fluid equations for leptogenesis}
\label{sec:boltzmann_eq}

The dynamics of leptogenesis is usually described in terms of momentum averaged fluid equations for the yields $Y_X$. 
We will follow this approach here, but it should be noted that, when relativistic corrections are of importance for the sterile neutrinos, the leading fluid approximation discards potentially relevant information on the distribution functions. 
In this case, working directly with the distribution functions is expected to give order one corrections for the final $B-L$ asymmetry, as can be inferred from dedicated studies of leptogenesis with ${\rm GeV}$ scale sterile neutrinos~\cite{Asaka:2011wq,Ghiglieri:2017csp}. 
In contrast, the reduction to fluid equations is a very good approximation when leptogenesis occurs in the strong washout regime and the partial equilibration of spectator fields can be neglected~\cite{Basboll:2006yx,HahnWoernle:2009qn}.

Although the correct form of the fluid equations can be inferred by physical intuition when balancing the scattering processes in the thermal plasma while accounting for the expansion of the universe, 
we perform a derivation from  first principles by means of the Schwinger-Keldysh CTP formalism~\cite{Schwinger:1960qe,Keldysh:1964ud} applied to nonequilibrium quantum-field theory~\cite{Calzetta:1986cq}. 
The advantage of this approach is that we may compute the $CP$-violating source terms as well as the relaxation rates toward equilibrium throughout the transition from relativistic to nonrelativistic sterile neutrinos in a way that is self-consistent without overly relying on physical assumptions.
In comparison to $S$-matrix approaches, the CTP formalism allows to automatically incorporate quantum statistical factors and enforce the constraints from unitarity and invariance under charge conjugation, parity and time-reversal. %
In the $S$-Matrix formalism this requires the added complication of real-intermediate-state subtraction, as the sterile neutrinos are not proper asymptotic states. 
The CTP approach has been amply used in studies of leptogenesis~ \cite{Buchmuller:2000nd,DeSimone:2007gkc,Garny:2009rv,Garny:2009qn,Garny:2010nj,1002.1326,1007.4783,Garny:2010nz,Anisimov:2010dk,Garbrecht:2011aw} and
particularly in resonant leptogenesis~\cite{Garny:2011hg,1007.4783}
and in the context of leptogenesis from GeV-scale sterile neutrinos~\cite{Drewes:2016gmt}, where the small masses involved require a fully relativistic treatment.
In full detail, this computation is presented in Section~\ref{sec:derivation_fluid_eq_rates}. 
Here, we only give a summary of the resulting fluid equations and the relevant equilibration rates, which are used to perform the numerical parameter scan in Section~\ref{sec:numerical_scans}. \\

To connect with more familiar treatments of leptogenesis, it is instructive to start by considering the nonrelativistic regime $z=M_1/T\gg1$, 
which is entirely driven by $N_1 \leftrightarrow l_\smallpara \phi$ and $N_1 \leftrightarrow \overline l_\smallpara \phi^\dagger$ decay and inverse decay processes. 
In the strongly hierarchical limit $\nicefrac{\tilde M_1}{\tilde M_2} \to 0$, there are only two free parameters. 
The first is the zero temperature decay width of the lightest sterile neutrino
\begin{align}
\Gamma_D (z\rightarrow\infty) 
&\equiv \Gamma_D (N_1 \to l_\smallpara \phi) + \Gamma_D (N_1 \to \overline l_\smallpara \phi^\dagger \big) %
= g_w \frac{\abs{F_{1}}^2 \tilde{M}_1}{16 \pi } 
\ ,
\end{align}
and the second 
is the decay asymmetry \cite{Covi:1996wh,hep-ph/9710460} 
\begin{align}
\label{eq:epsilon0}\epsilon_0 
&\equiv \frac{ \Gamma_D (N_1 \to l_\smallpara \phi) -\Gamma_D (N_1 \to \overline l_\smallpara \phi^\dagger )}{ \Gamma_D (N_1 \to l_\smallpara \phi) + \Gamma_D (N_1 \to \overline l_\smallpara \phi^\dagger) } %
= \frac{1+g_w}{16 \pi} \frac{\Im [ F_{1}^{*2} F_{2}^{2} ] }{| F_{1} |^2 } \frac{\tilde M_1}{\tilde M_2}
\ .
\end{align}

Using these two coefficients, it is possible to write down a simple set of fluid equations that is valid in the nonrelativistic regime:
\begin{subequations}
\begin{align}
\label{eq:n_even_boltzmann_nr}
\dz Y_{N_1 \text{even}} &= - \Gamma \big(Y_{N_1 \text{even}} - Y_{N_1 \text{eq}}\big), \\
\label{eq:y_bl_boltzmann_nr}
\dz Y_{\Delta_\smallpara} &= - \epsilon_0 \Gamma \big(Y_{N_1 \text{even}} - Y_{N_1 \text{eq}}\big) + \eta_{N_1} \Gamma \big(Y_{l_\smallpara} + \frac{1}{2} Y_\phi \big)
\ .
\end{align}
\end{subequations}
In these equations, $Y_{l_\smallpara}$ and $Y_\phi$ are the yields of the individual members of the $l_\smallpara$ and $\phi$ gauge doublets%
\footnote{We assume that initially, there are no charges associated with diagonal generators of a gauge group. 
Since the Standard Model gauge symmetries are intact at the temperatures considered, all members of a gauge multiplet share a common yield.}, 
$Y_{N_1 \text{eq}}$ is the value of the sterile neutrino yield in thermal equilibrium, and $\eta_{N_1}$ is a correction factor that captures the Boltzmann suppression of the inverse decays at late times. 
$Y_{N_1,\rm eq}$ and $\eta_{N_1}$ are given as
\begin{subequations}
\begin{align}
\label{eq:Yeq_eta}
Y_{N_1\rm eq}(z) &=\, \frac{T^3}{s \pi^2} \mathcal{I}(z) \ ,
&
\eta_{N_1}(z) &= \frac{6}{\pi^2}\mathcal{J}(z) \ ,
\end{align}
where
\begin{align}
\label{eq:I_exp}
\mathcal{I}(z) 
&\equiv \int_z^\infty\!\! dy\,\frac{y\sqrt{y^2-z^2}}{e^y+1} = z^2 \sum_{n=1}^{\infty} \frac{\left( -1 \right)^{n+1}}{n} K_2 \left( n \, z \right) \approx  z^2 K_2 \left( z \right) \ , 
\\
\label{eq:J_exp}
\mathcal{J}(z)
&\equiv\int_z^\infty\!\! dy\,\frac{y\sqrt{y^2-z^2}e^y}{(e^y+1)^2} = z^2 \sum_{n=1}^{\infty} \left(-1\right)^{n-1}K_{2} \left(n z\right)\approx z^2 K_2(z) \ ,
\end{align}
\end{subequations}
and $K_m(z)$ denotes the modified Bessel function of the second kind. 
To obtain the expression for $\eta_{N_1}$, we have linearized the fluid equations in the chemical potentials of leptons and Higgs bosons, see section \ref{sec:boltzmann_deriv} for more details.   

The dimensionless equilibration rate $\Gamma$ in Eqs.~\eqref{eq:n_even_boltzmann_nr}, \eqref{eq:y_bl_boltzmann_nr} is different from $\Gamma_D (z\rightarrow\infty) $ due to us working with $z$ instead of the comoving proper time $t_{\rm com}$.
Explicitly, one has
\begin{subequations}
\label{eq:Gammanr}
\begin{equation}
\Gamma = \frac{\text{d}t_{\rm com}}{\text{d}z} \,\Gamma_D (z\rightarrow\infty)  = a(t) \frac{\text{d}t}{\text{d} z} \Gamma_D (z\rightarrow\infty)  = \frac{T \, z}{\tilde M^2_1} \,\Gamma_D (z\rightarrow\infty) = K \cdot z
\ ,
\end{equation}
where the constant
\begin{equation}
\label{eq:washout_K}
K \equiv \frac{T}{\tilde M^2_1} \,\Gamma_D(z\rightarrow\infty) =\frac{\Gamma_D(z\rightarrow\infty)}{H(z=1)}= T \cdot g_w \frac{\abs{F_{1}}^2 }{16 \pi  \tilde M_1} 
\end{equation}
\end{subequations}
denotes the standard washout parameter defined as in e.g. Refs.~\cite{Fukugita:1986hr,hep-ph/0401240,1404.2915}. 
Phenomenologically, one distinguishes the two parametric regimes of $K\ll 1$ and $K\gg 1$, which are referred to as weak and strong washout, respectively.%
\footnote{It is generally understood that the relations $\ll$ and $\gg$ only need to be moderately saturated, which also the convention we use. }

Finally, we have considered the change in the yield $Y_{\Delta_\smallpara}=Y_{B/3}-g_w Y_{l_\smallpara}$ on the left-hand side of Eq.~\eqref{eq:y_bl_boltzmann_nr}, since it is conserved under anomalous weak sphaleron processes.%
\footnote{Note that the baryon asymmetries stored in each generation must be identical and equal to $B/3$, as weak sphalerons change $B$ by the same amount for each generation.}
In principle, the goal of our calculation is $Y_{B-L}$.
However, as the decays and inverse decays of the $N_1$ are the only relevant $B-L$-violating processes and they involve only the linear combination $l_\smallpara$ of lepton flavours, one finds
\begin{align}
\label{eq:DeltalparequalsB-L}
Y_{\Delta_\smallpara}=Y_{B-L}\ ,
\end{align}
so that we may use $Y_{\Delta_\smallpara}$ instead of $Y_{B-L}$. 

Physically, the collision term on the right-hand side of Eqs.~\eqref{eq:n_even_boltzmann_nr} describes the relaxation of the sterile neutrino toward equilibrium, 
and the first term on the right-hand side of Eq.~\eqref{eq:y_bl_boltzmann_nr} is the source of the asymmetry.
The washout of the lepton asymmetry via the inverse decays is captured by the second term in Eq.~\eqref{eq:y_bl_boltzmann_nr}. 

As a consistency check, we compare the nonrelativistic fluid equations~\eqref{eq:n_even_boltzmann_nr} and~\eqref{eq:y_bl_boltzmann_nr} with those in Refs.~\cite{hep-ph/0401240,1404.2915}.
For this, we neglect the Standard Model spectator fields and set $Y_{\phi}=0$ as well as  $Y_{l_\parallel}=-1/2 Y_{B-L}$.
Doing so, we find a relative prefactor of $\nicefrac{12}{\pi^2}$ in $\eta_{N_1}$ with respect to the results in Ref.~\cite{hep-ph/0401240}. 
However, taking the limit $z^2 K_2(z) \to z\sqrt{ \frac{\pi}{2} z } e^{-z} $ for $z \to \infty$ we reproduce the expression given for the washout term in Ref.~\cite{1404.2915}.
This small difference arises from the fact that the relation between the washout rate and the relaxation rate implied in Ref.~\cite{hep-ph/0401240}, apparently based on the ratio of number densities of sterile neutrinos and Standard Model leptons, is not valid. 
To obtain the factor $\eta_{N_1}$, we have instead performed an expansion in terms of the chemical potentials of the Higgs and lepton fields, as it is appropriate for rates that are proportional to the charge asymmetry.\\

Moving on to arbitrary temperatures, we first note two reasons why the fluid equations \eqref{eq:n_even_boltzmann_nr} and \eqref{eq:y_bl_boltzmann_nr} are valid only in the nonrelativistic regime,
apart from thermal corrections for the sterile neutrino interaction rates being neglected.

First, the equations do not account for the time-dilation of relativistic sterile neutrinos in the plasma frame. 
Following the approach in \cite{hep-ph/0401240}, this effect can be taken into account by multiplying $\Gamma$ with the thermal average of the sterile neutrino time dilation factor, 
\begin{align}
\label{eq:gamma_augmented}
\Gamma &\to K \cdot z \, \frac{1}{\gamma} \ , 
&
\text{where}&
&
\frac{1}{\gamma} \equiv \frac{1}{\mathcal{I}(z)} \int_z^\infty\!\! dy\,\frac{\sqrt{y^2-z^2}}{e^y+1} \approx \frac{K_1(z)}{K_2(z)}
\ .
\end{align}
As discussed below, this augmented version of $\Gamma$ works well for $z \gtrsim 1$, while the expression without the time dilation factor is sufficient only for $z\gtrsim 10$. 
For $z \lesssim 1$, thermal corrections have to be taken into account for a self-consistent computation of the sterile neutrino equilibration rates. 

Second, the nonrelativistic fluid equations make no distinction between the two helicity eigenstates of the sterile neutrino. 
This a reflection of the fact that nonrelativistic interactions become insensitive to spin. 
However, in the relativistic regime, the reaction rates are helicity dependent, so that we have to keep track of two independent sterile neutrino equilibration rates $\Gamma$ and $\tilde \Gamma$, 
which are defined such that $\Gamma$ is helicity even and $\tilde \Gamma$ is helicity odd. 
Assigning positive (negative) lepton number to sterile neutrinos with negative (positive) helicity%
\footnote{This is in keeping with assigning positive lepton number to the left-handed lepton doublets in the Standard Model.}, 
we then encode the finite temperature corrections to $\Gamma$ and $\tilde \Gamma$ in terms of a lepton-number conserving (LNC) rate $\gamma_\text{\rm LNC}$ and a lepton-number violating (LNV) rate $\gamma_\text{\rm LNV}$.
Explicitly, we parametrize
\begin{align}
\label{eq:gamma_av}
\Gamma &= K \frac{1}{2} (\gamma_\text{\rm LNC} + \gamma_\text{\rm LNV} )
&
\text{and}&
&
\tilde{\Gamma} \equiv K \frac{1}{2} (\gamma_\text{\rm LNC} - \gamma_\text{\rm LNV}) \ ,
\end{align}
where the LNC and LNV rates $\gamma_\text{\rm LNC}$ and $\gamma_\text{\rm LNV}$ are defined such that they only depend on Standard Model parameters. 
They are computed in Section \ref{sec:gamma_computation}.

Using the equilibration rates $\Gamma$, $\tilde \Gamma$ and following the derivation in Section \ref{sec:boltzmann_deriv}, we find the momentum-averaged relativistic fluid equations 
\begin{subequations}
\label{eq:Boltzmann:all:toy}
\begin{align}
\label{eq:n_even_boltzmann}
\dz Y_{N_1 \text{even}} &= - \Gamma \cdot \big( Y_{N_1 \text{even}} - Y_{N_1 \text{eq}} \big) \ , \\
\label{eq:n_odd_boltzmann}
\dz Y_{N_1 \text{odd}} &= - \Gamma \cdot Y_{N_1 \text{odd}} - \eta_{N_1} \tilde{\Gamma} \cdot \big(Y_{l_\smallpara} + \frac{1}{2} Y_\phi \big) \ , \\
\label{eq:y_bl_boltzmann}
\dz Y_{\Delta_\smallpara} &=  \tilde{\Gamma} \cdot Y_{N_1 \text{odd}} - \epsilon_\text{eff} \Gamma \cdot \big(Y_{N_1 \text{even}} - Y_{N_1 \text{eq}} \big) + \eta_{N_1} \Gamma \cdot \big(Y_{l_\smallpara} + \frac{1}{2} Y_\phi \big) \ .
\end{align}
\end{subequations}
In these equations, $\epsilon_\text{eff}$ is now the effective $CP$-violating parameter at finite temperature.  
Similar to $\Gamma$ and $\tilde \Gamma$, it can be expressed in terms of $\gamma_\text{\rm LNC}$ and $\gamma_\text{\rm LNV}$,
\begin{align}
\label{eq:epsilon_eff_def}\epsilon_\text{eff} 
&= \epsilon_0 \, \frac{\mathcal{K}(z)}{\mathcal{I}(z)} \frac{ 2 \cdot \gamma_\text{\rm LNC} \cdot \gamma_\text{\rm LNV} }{ z^2 ( \gamma_\text{\rm LNC} + \gamma_\text{\rm LNV} ) },
\end{align}
where 
\begin{align*}
\mathcal{K}(z)&=\int_z^\infty\!\! dy\,\frac{y^2\sqrt{y^2-z^2}}{e^y+1}= \sum_{n=1}^\infty (-1)^{n+1} \left( \frac{z^3}{n} K_1(nz) + \frac{3z^2 }{n^2} K_2(nz) \right)
\\
&\approx z^3 K_1(z) + 3 z^2 K_2(z) 
\, 
\end{align*}
and $\mathcal{I}(z)$ is defined as in Eq.~\eqref{eq:Yeq_eta}. 
As before, $\epsilon_0$ is the nonrelativistic decay asymmetry of Eq.~\eqref{eq:epsilon0}.

For the above fluid equations, we can explicitly verify that $\gamma_\text{\rm LNC}$ conserves the generalized lepton number $\tilde{L}=L-n_{N_1+}+n_{N_1-}$, where $L$ is the usual lepton number in the Standard Model.
Indeed, one has
\begin{align}
\dz Y_{B-\tilde{L}} =&\,\dz Y_{B-L}+\dz Y_{N_1\rm odd} =K \gamma_{\rm LNV} \big(-Y_{N_1\rm odd}+\eta_{N_1} \big(Y_{l_\smallpara} + \frac{1}{2} Y_\phi \big) \big)+(\text{source term})
\ ,
\end{align}
justifying the choice of the nomenclature LNC.
Occasionally, we will refer to $\Gamma$ as the relaxation or equilibration rate and to $\tilde\Gamma$ as the backreaction rate. 
This is because the term involving $\tilde \Gamma$ in Eq.~\eqref{eq:n_odd_boltzmann} physically corresponds to the lepton asymmetry being partially transferred back into a sterile neutrino helicity asymmetry as a result of the washout.\\

The fluid equations~\eqref{eq:n_even_boltzmann}--\eqref{eq:y_bl_boltzmann} differ from the nonrelativistic Eqs.~\eqref{eq:n_even_boltzmann_nr}, \eqref{eq:y_bl_boltzmann_nr} obtained in the standard treatment 
via the presence of the two independent rates $\Gamma$, $\tilde\Gamma$, the separation of the sterile neutrino yield in two contributions $Y_{N_1\,\rm even}$, $Y_{N_1\,\rm odd}$, 
and the $z$-dependence of the effective $CP$-violating parameter $\epsilon_{\rm eff}(z)$. 
For leptogenesis with hierarchical sterile neutrinos, strong enough washout and negligible spectator effects, the asymmetry is generated at late times, where $z\gg1$ and the sterile neutrinos are entirely nonrelativistic.
As we derive in Section~\ref{sec:derivation_fluid_eq_rates}, in this regime the LNC and LNV rates are approximately equal, 
so that $\tilde\Gamma \approx 0$, $Y_{N_1\,\rm odd}$ becomes irrelevant and one does indeed recover the nonrelativistic Eqs.~\eqref{eq:n_even_boltzmann_nr}, \eqref{eq:y_bl_boltzmann_nr}.
In contrast, for leptogenesis driven by ${\rm GeV}$-scale sterile neutrinos~\cite{Akhmedov:1998qx,Asaka:2005pn}, the LNC rate is the most important one and the LNV rate can be ignored in large regions of parameter space, %
so that $\Gamma \approx \tilde \Gamma$ and there is once again only one independent rate appearing in the fluid equations. 
Generically however, both rates may be relevant, which has been explored in Refs.~\cite{Hambye:2017elz,Hambye:2016sby,Ghiglieri:2017gjz,Eijima:2017anv,Antusch:2017pkq,Ghiglieri:2017csp,Eijima:2017cxr,Eijima:2018qke,Ghiglieri:2018wbs} for ${\rm GeV}$-scale sterile neutrinos. 
In that framework, the $CP$-violating effects have been obtained from numerical solutions of the oscillations among the sterile neutrinos. 
In contrast to that, in the present work we consider the interplay of LNC and LNV rates in leptogenesis from decays, where flavour oscillations are negligible and the sterile neutrinos have masses  close to  the natural scale of the type-I seesaw mechanism. 
Since in the relativistic regime the LNC and LNV rates differ and $\epsilon_{\rm eff}(z)$ departs from $\epsilon_0$, we expect nontrivial effects when a sizable asymmetry is generated at early times that does not suffer from strong washout, 
as can happen for nonthermal initial conditions for the sterile neutrinos in the weak washout regime or in the presence of partially equilibrated spectator fields. \\

In the case of partially equilibrated spectators, the Boltzmann equations~\eqref{eq:n_even_boltzmann}--\eqref{eq:y_bl_boltzmann} are no longer sufficient, since one also has to describe the time-evolution of the additional dynamical Standard Model degrees of freedom.
Proceeding as in Ref.~\cite{1404.2915}, we augment the Boltzmann equations with an additional equation for each partially equilibrated spectator. 
As mentioned in the introduction, the bottom-Yukawa and weak sphaleron interactions are expected to be partially equilibrated during leptogenesis for sterile neutrinos with masses $M_{N_1} \sim \unit[10^{13}]{GeV}$, 
since their respective equilibration temperatures are estimated to be $4.2\times10^{12}$ GeV and $\unit[1.8\times10^{12}]{GeV}$~\cite{1404.2915}.
In the relevant temperature range, the first and second generation Yukawa interactions are negligible 
and the corresponding quarks are effectively indistinguishable\footnote{This can also be understood from the conservation of the charges of the approximate $U(2)$ flavour-symmetry due to the small Yukawa couplings of the first two generations.}.
As a result, they have a single, common yield:  $Y_{Q_1}=Y_{Q_2}\equiv Y_Q$ for the left-handed doublets, and  $Y_{u_i}=Y_{d_i}\equiv Y_d,i=1,2$ for the right-handed quarks. 
Analogously, the Standard Model Yukawa interactions of the left-handed lepton doubles are negligible, so that the two flavour combinations $l_{\perp 1}$, $l_{\perp 2}$ not coupling to the sterile neutrinos share a single common yield $Y_{l_\perp 1} = Y_{l_\perp 2} \equiv Y_{l_\perp}$. 
The third generation of quarks is affected by the partially equilibrated bottom-Yukawa interactions, and we denote their yields as $Y_{Q_3}, Y_t, Y_{b}$.  
Adding the fluid equations for the spectator processes as in Ref.~\cite{1404.2915}, and using the definition $Y_{\Delta \rm down}\equiv Y_{ b}-Y_{d}$, one obtains
\begin{subequations}
\label{eq:Boltzmann:all:b:sph}
\begin{align}
\label{eq:Boltzmann_n_even}
\dz Y_{N_1 \text{even}} 		&= - \Gamma \cdot \big(Y_{N_1 \text{even}} - Y_{N_1 \text{eq}} \big) 
\ , \\ 
\dz Y_{N_1 \text{odd}} 		&= - \Gamma \cdot Y_{N_1 \text{odd}} - \eta_{N_1} \tilde{\Gamma} \cdot \big(Y_{l_\smallpara} + \frac{1}{2} Y_\phi \big)
\ , \\
\label{eq:Boltzmann_b_l_charge}
\dz Y_{\Delta_{\smallpara}} 	&=  \tilde{\Gamma} \cdot Y_{N_1 \text{odd}} - \epsilon_\text{eff} \Gamma \cdot \big(Y_{N_1 \text{even}} - Y_{N_1 \text{eq}} \big) + \eta_{N_1} \Gamma \cdot \big(Y_{l_\smallpara} + \frac{1}{2} Y_\phi \big)
\ , \\
\label{eq:Boltzmann_b_Yukawa}
\dz Y_{\Delta_{\text{down}}} 	&= -   \Gamma_{\rm down}  \big( Y_b - Y_{Q_3} + \frac{1}{2}Y_\phi \big) 
\ , \\
\label{eq:Boltzmann_weak_sphal}
\dz Y_{l_{\perp}} 			&= - \Gamma_\text{ws}  \big( 9Y_{Q_3} + 18Y_Q + 3Y_{l_{\smallpara}} + 6Y_{l_{\perp}} \big) \ .
\end{align}
\end{subequations}
At temperatures of about $\unit[10^{12}]{GeV}$, the equilibration rates $\Gamma_{\rm down}$ and $\Gamma_{\rm ws}$ for the bottom Yukawa~\cite{1404.2915} and weak sphaleron~\cite{1303.5498,Moore:2000ara} interactions are given as
\begin{align}
\label{eq:Gammas_b_sph}
\Gamma_{\rm down} &\approx 1.0 \cdot 10^{-2} \frac{h_b^2 T}{ \tilde M_1}
&
\text{and}&
&
\Gamma_\text{ws} &\approx \left(8.24 \pm 0.10 \right) \left( \log \left( \frac{m_D}{g_2^2 T} \right) + 3.041\right) \frac{g_2^2 T^3}{2{m_D}^2 \tilde M_1}\cdot \alpha_2^5 \ ,
\end{align}
where $h_b$ is the bottom-Yukawa coupling, $\alpha_2 \equiv \nicefrac{g_2^2}{4\pi}$ is the coupling strength of the electroweak $\SU(2)_L$ gauge interaction and $m^2_D \approx \frac{11}{6} g_2^2 T^2 $ is the Debye mass of the $\SU(2)_L$ gauge bosons. 

In order to solve the fluid equations~\eqref{eq:Boltzmann_n_even}-\eqref{eq:Boltzmann_weak_sphal}, we need five relations to express the yields $Y_{l_{\smallpara}}$, $Y_{Q_3}$, $Y_b$, $Y_Q$ and $Y_\phi$ appearing on the right-hand side %
in terms of the three yields $Y_{\Delta_{\smallpara}}$, $Y_{l_{\perp}}$ and $Y_{\Delta_\text{down}}$ on the left-hand side. 
For this, we first use the constraints imposed onto the chemical potentials corresponding to the yields by the spectator interactions that are fully equilibrated in the relevant temperature range, $10^{12}\,{\rm GeV}\lesssim T\lesssim 10^{13} \,{\rm GeV}$.
As discussed in Ref.~\cite{1404.2915}, this is the case for top-Yukawa interactions and strong sphaleron processes, yielding two relations.
The remaining three relations can be obtained by enforcing $Y_{\Delta_\perp} \equiv Y_{B/3} - 2 g_w Y_{l_\perp}=0$, equal quark asymmetries for the three generations, and that the weak hypercharge vanishes. 
The first condition follows from the fact that the weak sphalerons are the only source of $B$ violation, as they change $B/3$ and $L$ by the same amount in each generation and hence conserve $Y_{B/3}-2 Y_{l_\perp}$ in absence of lepton-flavour violation.%
\footnote{Recall that there are two  generations of leptons perpendicular to $l_\smallpara$, so that $Y_{B/3} - 2 g_w Y_{l_\perp 1,2}$ are not conserved if lepton-flavour violating interactions are present.}
The second condition is a result of the weak and strong sphalerons being flavour blind;
the additional flavour violation via charged currents for the left-handed quarks then only retains the chemical equilibrium among them.
Finally, the third condition is a consequence of charge-neutral initial conditions and gauge invariance.
Putting everything together results in the relations \cite{1404.2915}
\begin{align}
\label{eq:charge_relations}
\left(\begin{array}{c}
Y_{l_{\smallpara}}\\
Y_{Q_3}\\
Y_b\\
Y_Q\\
Y_\phi
\end{array}
\right)
&=
\begin{pmatrix}
-\frac{1}{2} 	& 1 			& 0 				\\
\frac{1}{23} 	& \frac{1}{2} 	& -\frac{10}{23} 	\\
\frac{1}{46} 	& \frac{1}{2} 	& \frac{18}{23} 	\\
-\frac{1}{46} 	& \frac{1}{2} 	& \frac{5}{23} 		\\
-\frac{7}{23} 	& 0 			& \frac{24}{23} 	
\end{pmatrix}
\cdot 
\left(\begin{array}{c}
Y_{\Delta_{\smallpara}}\\
Y_{l_{\perp}}\\
Y_{\Delta_\text{down}}
\end{array}\right)\ .
\end{align}
This equation can also be used to explicitly verify Eq.~\eqref{eq:DeltalparequalsB-L}, which states $Y_{\Delta_\smallpara} = Y_{B-L}$.

In cases with fully equilibrated bottom Yukawa and weak sphaleron interactions, one obtains two additional chemical equilibrium constraints for the yields, giving
\begin{align}
\label{eq:chemical_b_sph}
Y_{l_\smallpara}=&\,-\frac{13}{30}Y_{\Delta\smallpara}, & Y_\phi=&\,-\frac{1}{5}Y_{\Delta\smallpara}\ .
\end{align}
These relations can be used in Eqs.~\eqref{eq:n_even_boltzmann}--\eqref{eq:y_bl_boltzmann} to obtain a closed system, replacing the final two equations \eqref{eq:Boltzmann_b_Yukawa} and \eqref{eq:Boltzmann_weak_sphal}.\\

\begin{figure}[ht]
\centering
\includegraphics[width=0.49\textwidth, trim = 0 20 0 10]{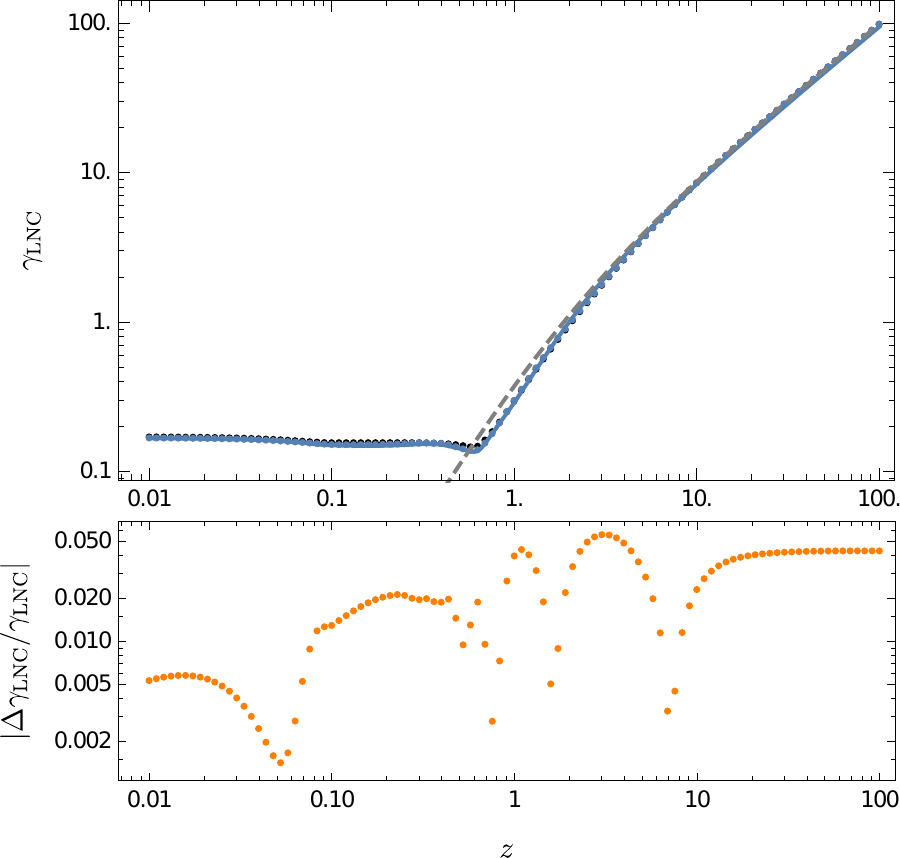}%
\hskip.2cm
\includegraphics[width=0.49\textwidth, trim = 0 20 0 10]{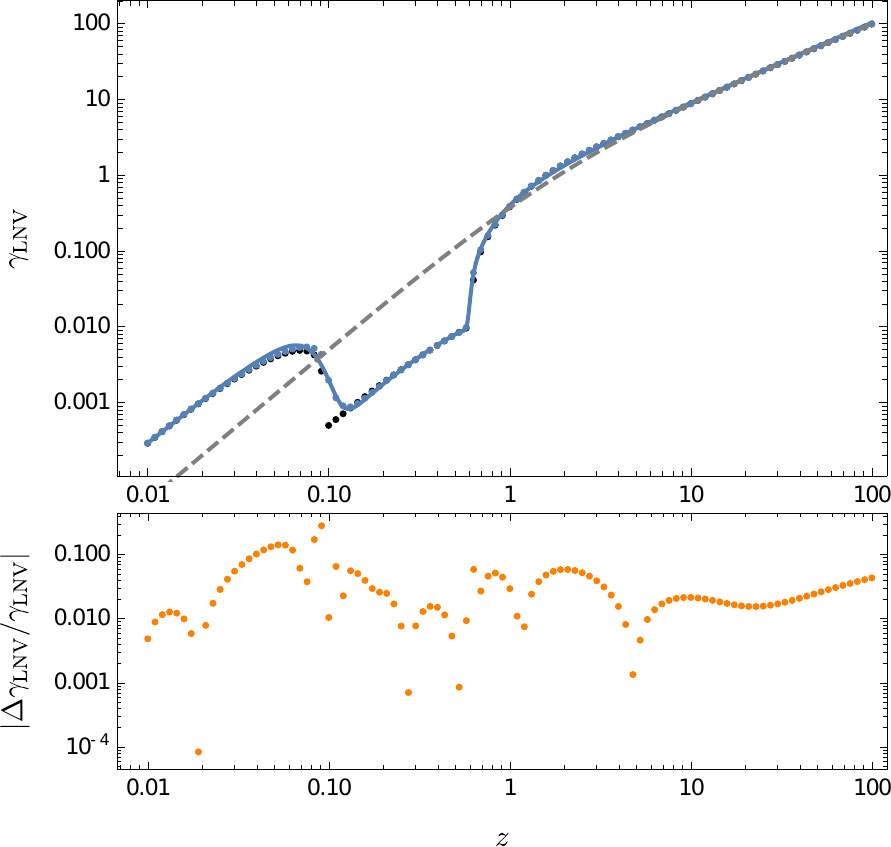}\\
\vskip.2cm
\hskip.15cm
\includegraphics[width=0.485\textwidth]{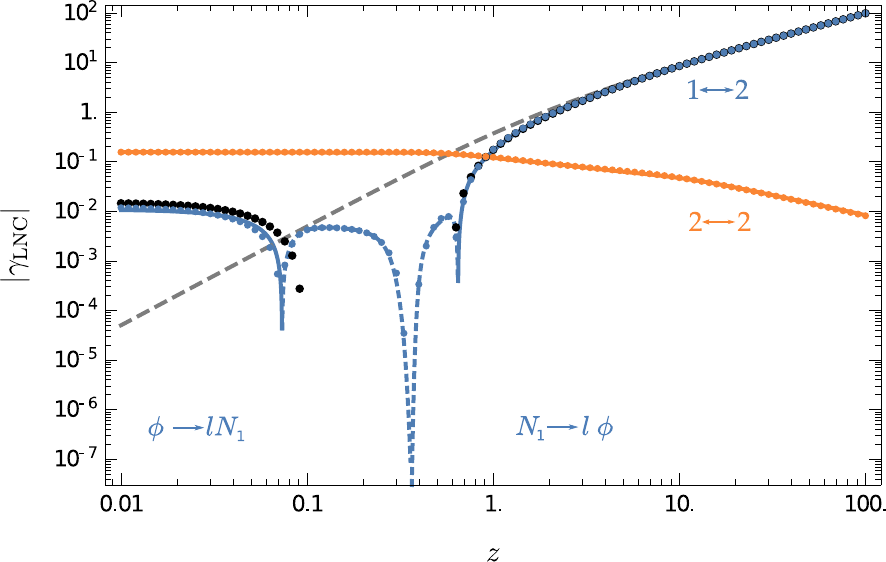}
\hskip.15cm
\includegraphics[width=0.485\textwidth]{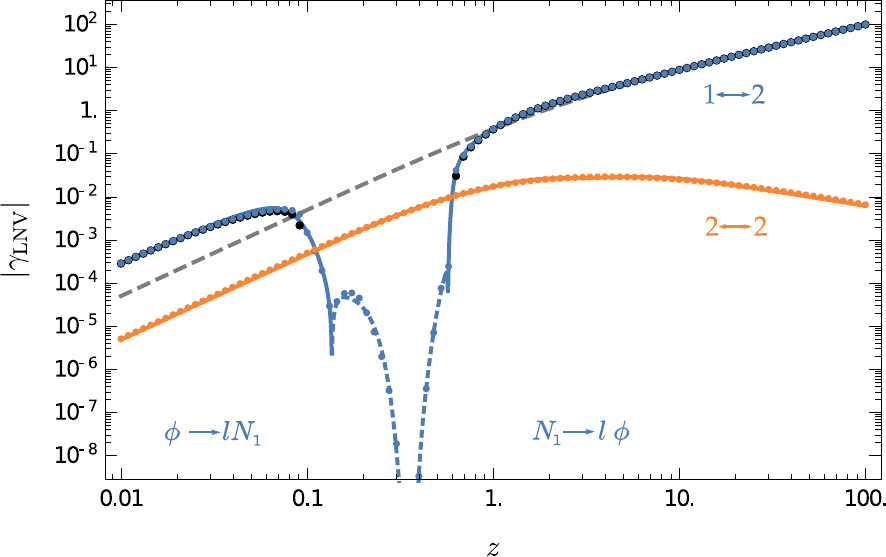}
\vskip-.3cm
\caption{\footnotesize\label{fig:eq_rates_analyt}%
First line: Numerical estimates for $\gamma_\text{LNC/LNV}(z)$ (blue dots) overlaid on exp. \eqref{eq:gammaLNC12}-\eqref{eq:gammaLNV22} (solid blue lines) %
and the result with simplified lepton propagators for the $1\leftrightarrow2$ processes (black dots). 
Second line: The relative deviations between the full numerical results and the approximate formulae (orange dots). 
Third line: Absolute values of the individual $1\leftrightarrow2$ contributions (blue), $1\leftrightarrow2$ contributions with simplified lepton propagators (black), and $2\leftrightarrow2$ contributions (orange) for $\gamma_\text{LNC/LNV}(z)$. 
Numerical results are shown as dots, while the approximate formulae are shown as solid lines (for positive contributions) or dashed lines (for negative contributions, generated by hole contributions).  
In all plots, dashed gray lines show the nonrelativistic $1\leftrightarrow2$ result $\gamma_{\rm LNC/LNV}\approx z K_1(z)/K_2(z)$.
}
\end{figure}
\begin{figure}[ht]
\centering
\includegraphics[width=0.495\textwidth, trim = 0 0 0 40]{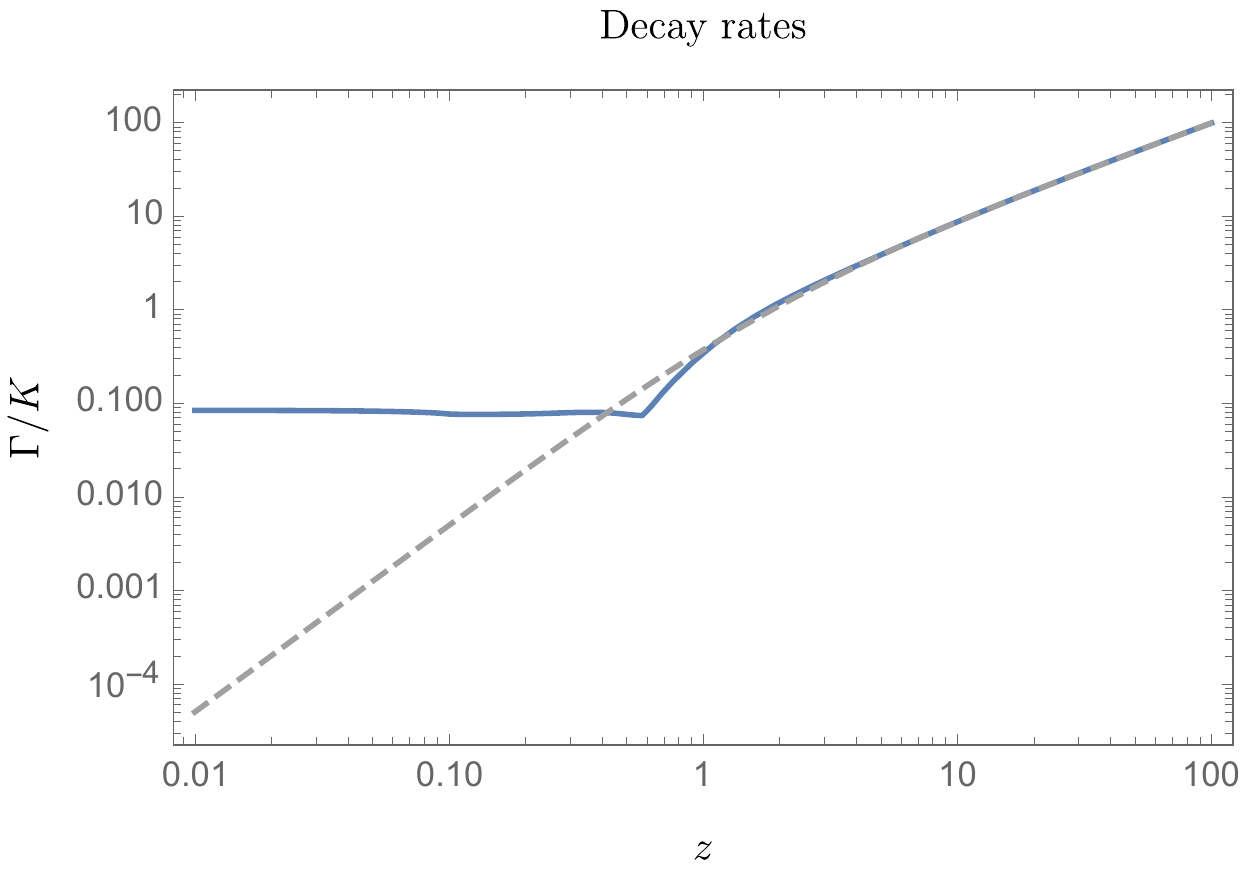}
\includegraphics[width=0.495\textwidth, trim = 0 0 0 40]{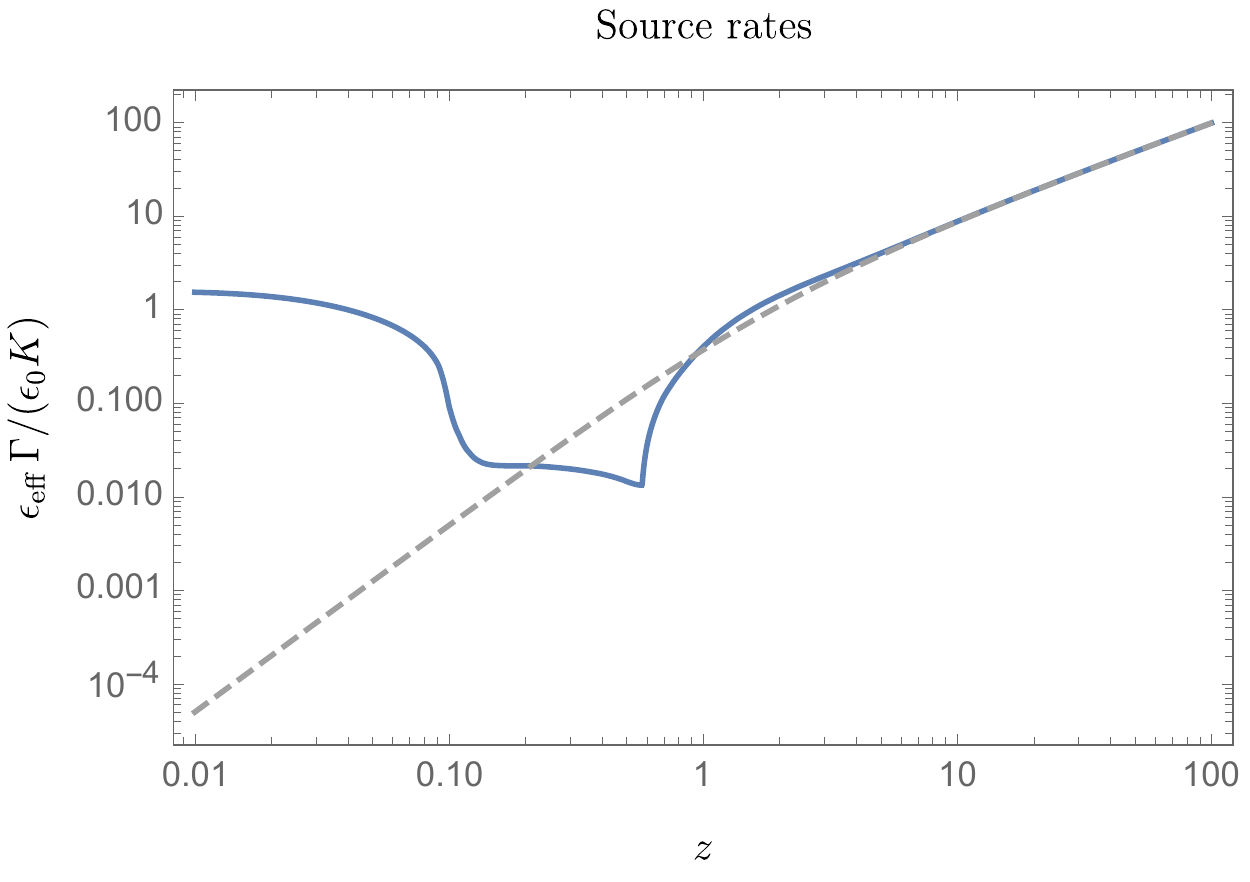}
\\
\includegraphics[width=0.495\textwidth, trim = 0 20 0 0]{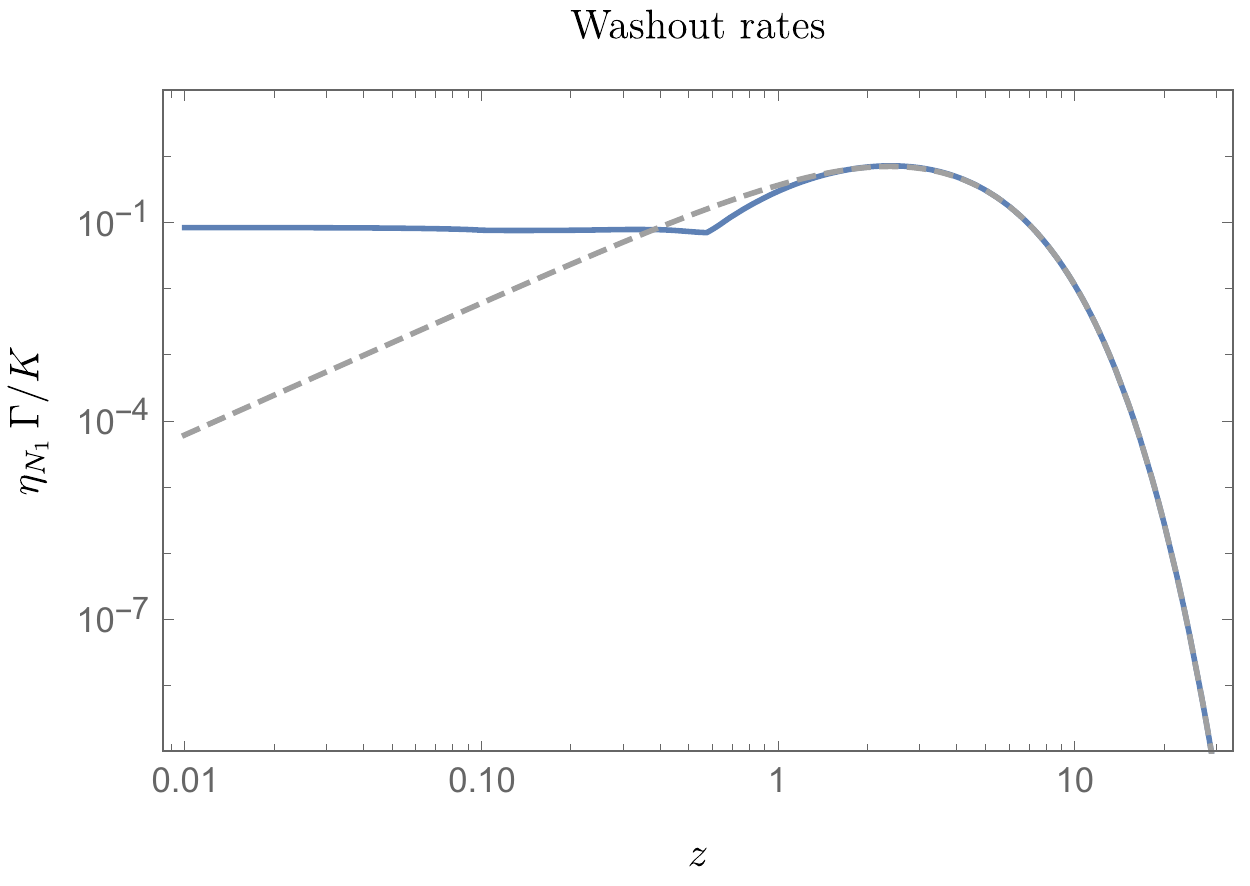}
\includegraphics[width=0.495\textwidth, trim = 0 20 0 0]{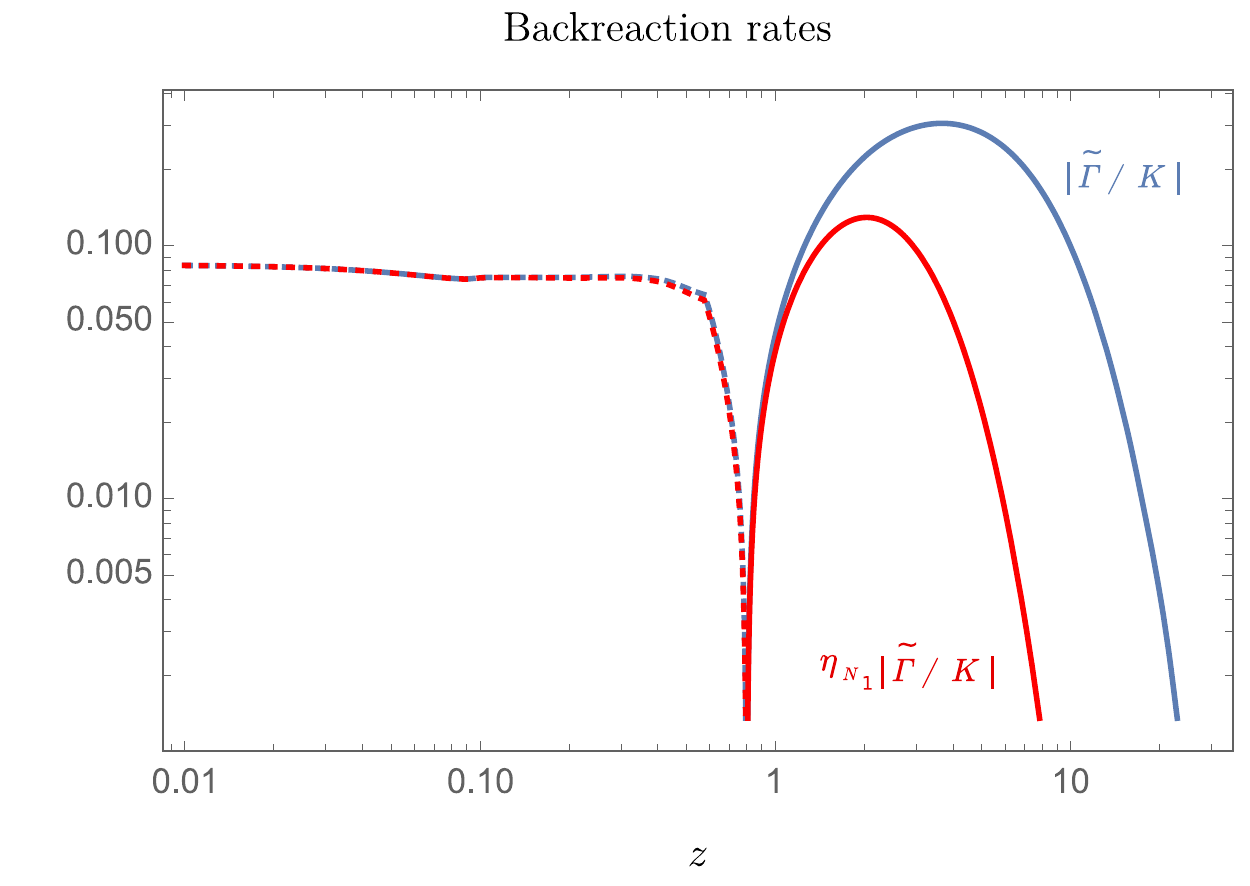}
\caption{\footnotesize\label{fig:rates}%
Rates appearing in the fluid equations, with appropriate constant rescalings. 
The nonrelativistic limits for the decay, source, and washout rates are shown as dashed gray lines. 
Note the enhancement of the source for small $z<0.1$.}
\end{figure}

To conclude this section, we present approximate formulae for the rates $\gamma_{\rm LNC}$ and $\gamma_{\rm LNV}$.
These reproduce the results of the numerical computation in Section~\ref{sec:gamma_computation} with a relative precision that is typically better than 5\%, 
while dropping to 10-30\% only for narrow intervals of $z$ {in the vicinity of kinematic thresholds } in the case of $\gamma_{\rm LNV}$. 
As elaborated in Section~\ref{sec:gamma_computation}, the rates receive contributions which can be interpreted as arising from $1\leftrightarrow2$ processes in the thermal plasma involving $N_1$, $\phi$ and $l_\smallpara$, 
as well as $2\leftrightarrow2$  scatterings involving $N_1$, $\phi$, $l_\smallpara$ and weak gauge bosons. 
For the leptonic states taking part in the $1\leftrightarrow2$ processes, we account for the two branches of the modified plasma dispersion relation corresponding to either particle or hole modes.
Hence, we write $\gamma_{\rm LNC/LNV}=\gamma_{\rm LNC/LNV}^{1\leftrightarrow2}+\gamma_{\rm LNC/LNV}^{1\leftrightarrow2,h}+\gamma_{\rm LNC/LNV}^{2\leftrightarrow2}$, 
with the three different contributions corresponding respectively to $1\leftrightarrow2$ processes with lepton particle-modes, $1\leftrightarrow2$ reactions with lepton hole-modes and $2\leftrightarrow 2$ scatterings. 
Defining $L\equiv \log_{10}z,$ the approximate formulae for the different contributions (valid for scales around $10^{12}$ GeV) are
\begin{subequations}
\begin{align}
\label{eq:gammaLNC12}
&\log_{10} \gamma_{\rm LNC}^{1\leftrightarrow2} \approx 
\left\{
\begin{array}{cc}
\left(0.125 e^{-73.8 (L + 0.994)^2}+1\right) \left(-\frac{1}{\left| L\right| ^{6.51}}-1.79\right),& z\leq 0.3 \ ,
\\
\left(139 e^{-27.9 (L + 0.653)^2}+1\right) \left(0.994 L-\frac{35.2}{\left| L+1.97\right| ^{5.66}}\right),& z> 0.3 \ ,
\end{array}\right.
\\ 
\label{eq:gammaLNC12h}
&\log_{10} -\gamma_{\rm LNC}^{1\leftrightarrow2,h} \approx
\left\{
\begin{array}{ccc}
\left(1.69 e^{-1.01 (L-1.31)^2}+1\right) \left(-\frac{0.002}{\left| L\right| ^{9.53}}-2.29\right),& z\leq 0.36 \ ,
\\
\left(0.733 e^{15.5 (L+0.186)^2}+1\right) \left(-0.631 \left| L-0.578\right| ^{50.2}-1.20\right),& 0.36\leq z< 0.8 \ ,\\
\left(1.16 e^{-5.59 (L-0.607)^2}+1\right) \left(-\frac{0.586}{\left| L-1.42\right| ^{5.25}}-1.98\right),&z\geq 0.8 \ ,
\end{array}\right.
\\
\label{eq:gammaLNC22}
&\log_{10} \gamma_{\rm LNC}^{2\leftrightarrow2} \approx
\begin{multlined}[t]
-0.987 \left(0.417 L+\sqrt{(0.174 L-0.161) L+0.0571}+0.615\right)
\\
\times\left(0.180 e^{-5.00 \left|0.412 - L\right| ^{2.2}}+1\right) \ ,
\end{multlined}
\end{align}
\end{subequations}
\begin{subequations}
\begin{align}
\label{eq:gammaLNV12}
&\log_{10}  \gamma_{\rm LNV}^{1\leftrightarrow2} \approx
\left\{
\begin{array}{cc}
\left(1-0.887 e^{-11.1 (L+0.753)^2}\right) \left(-\frac{0.584}{\left| L+0.179\right| ^{9.06}}+1.90 L+0.256\right) \ ,& z\leq 0.55 \ ,
\\
\left(19.0 e^{-213 (L+0.360)^2}+1\right) \left(1.023 L-\frac{0.117}{\left| L+0.399\right| ^{1.39}}\right) \ ,& z> 0.55 \ ,
\end{array}
\right.
\\
\nonumber
&\log_{10} -\gamma_{\rm LNV}^{1\leftrightarrow2,h} \approx
\\ 
\label{eq:gammaLNV12h}
&\left\{
\begin{array}{ccc}
\left(23.7 e^{-10.6 (L-0.011)^2}+1\right) \left(2.00 L-2.52 -4.44 \times10^{-7} \left| L+4.70\right| ^{8.63}\right)\ ,& z\leq 0.36 \ ,
\\
\left(1-0.947 e^{-1.85 (L+0.191)^2}\right) \left(-1.97 \left| L-0.613\right| ^{60.1}-63.2\right)\ ,& 0.36\leq z< 0.65 \ ,\\
\left(0.302 e^{0.218 (L-0.971)^2}+1\right) \left(-0.400 \left| L+0.931\right| ^{5.41}-2.30\right) \ ,&z\geq 0.65 \ ,
\end{array}\right.\\
\label{eq:gammaLNV22}&\log_{10} \gamma_{\rm LNV}^{2\leftrightarrow2}\approx 
\begin{multlined}[t]
\left(0.648 L-\sqrt{L (1.83 L-0.704)+0.195}-1.03\right)
\\
\times\left(0.291 e^{-4.46(L-0.297)^2}+1\right)
\ .
\end{multlined}
\end{align}
\end{subequations}
In Fig.~\ref{fig:eq_rates_analyt}, we compare the numerical results obtained in Section~\ref{sec:gamma_computation} with the approximate formulae~\eqref{eq:gammaLNC12}--\eqref{eq:gammaLNV22}. 
Note that for large $z$, the $1\leftrightarrow2$ processes dominate, and one has
\begin{align}
\gamma_{\rm LNC}\approx\gamma_{\rm LNV} \approx  z\, \frac{K_1(z)}{K_2(z)} \approx z \left(1 - \frac{3}{2z} \right) \ ,
\quad
z \gg 1 \ ,
\end{align}
and therefore
\begin{align}
\Gamma &\approx z \, \frac{K_1(z)}{K_2(z)} \approx z \big(1 - \frac{3}{2z} \big) \ , 
\quad 
\tilde\Gamma \approx 0 \ ,
\quad
z \gg 1
\ .
\end{align}
Similarly, Eq.~\eqref{eq:epsilon_eff_def} gives
\begin{align}
\epsilon_{\rm eff}(z) \approx \epsilon_0 \frac{K_1(z)(z K_1(z)+3K_2(z))}{z K_2(z)^2}\approx \epsilon_0\left(1 + \frac{3}{2 z^2}\right),
\end{align}
so that we explicitly recover the nonrelativistic fluid equations~\eqref{eq:n_even_boltzmann_nr}, \eqref{eq:y_bl_boltzmann_nr}. 
For lower values of $z$ the rates receive significant corrections from both $1\leftrightarrow2$ and $2\leftrightarrow2$ processes, 
with the latter dominating $\gamma_{\rm LNC}$ for $z\lesssim 1$ and $\gamma_{\rm LNV}$ in the window $0.1\lesssim z \lesssim 0.6$, where the $1\leftrightarrow 2$ processes are suppressed or forbidden
due to the kinematic blocking caused by the thermal masses acquired by the Standard Model particles and holes in the early universe plasma.
This kinematic blocking is the main cause for the features of the $1\leftrightarrow2$ contributions that can be seen in Fig.~\ref{fig:eq_rates_analyt} for $z \ll 1$.
In this regime, the $1\leftrightarrow2$ processes are dominated by $H \leftrightarrow N_1 l$ decays, as the thermal mass of the Higgs overcomes the sum of the masses of $N_1$ and $l_\smallpara$. 
Consequently, the standard $N_1 \leftrightarrow l H$ decays only dominate the $1\leftrightarrow2$ contributions for $z\gtrsim 1$. 
The hole contributions to $1\leftrightarrow2$ processes are negative and only relevant for $0.1\lesssim z \lesssim 1$.
However, in this window the total rate is dominated by the scattering contributions, so that holes do not play an appreciable role in the total rates. 
This is also shown in the lower panels of Fig.~\ref{fig:eq_rates_analyt}, where negative values of $\gamma_{\rm LNC/LNV}^{1\leftrightarrow2}$ due to the hole contributions are drawn as dashed blue lines, while scattering contributions are drawn in orange.
More details are given in Sec.~\ref{sec:gamma_computation}.

In Fig.~\ref{fig:rates}, we also illustrate the derived rates $\Gamma$,$\tilde\Gamma$, $\eta_{N_1}\Gamma$, $\eta_{N_1}\,\tilde\Gamma$, and $\epsilon_{\rm eff}\Gamma$ that appear in the fluid equations together with their respective nonrelativistic approximations. 
A noteworthy feature is the large  enhancement of the source term $\epsilon_{\rm eff}\Gamma$ at $z<0.1$, which is the results of the $H \leftrightarrow N_1 l$ decay processes generating a large contribution.
As will be seen, even when neglecting spectator effects, the enhanced source can result in a large increase of the asymmetry compared to the nonrelativistic approximation, if the generation of this asymmetry is effective at early times, when $z\ll 1$.
This is the case e.g. in the weak washout regime for a vanishing initial abundance of $N_1$.

\section{Derivation of the fluid equations and computation of the equilibration rates in the closed time-path formalism}
\label{sec:derivation_fluid_eq_rates}

In this section, we derive the fluid equations presented in the previous section from first principles, using the standard techniques of the CTP formalism as reviewed in Appendix~\ref{sec:ctp_basics}.
The derivation itself is presented in section~\ref{sec:boltzmann_deriv}, and in Section~\ref{sec:gamma_computation} we then compute relevant LNC and LNV rates appearing in the fluid equations. 

Our calculation follows the approach of Ref.~\cite{1007.4783}. 
The strategy is to recover the particle number densities $n_X^\pm (t)$ from the two-point correlation functions of the corresponding fields. 
The time-evolution of these correlators  is governed by Schwinger-Dyson equations, 
which can be converted into approximate Boltzmann equations for the local number densities after performing a simultaneous expansion in gradients and deviations from thermal equilibrium~\cite{Calzetta:1986cq}.
Averaging over the momenta in the thermal plasma then yields the final set of fluid equations in terms of particle yields.

The computation of the LNC and LNV rates poses no major problem in the nonrelativistic limit $z\gg1$, since one may neglect finite-temperature corrections. 
However, these corrections are of paramount importance in the ultrarelativistic limit $z\ll1$. 
In this regime, an accurate computation becomes more involved, since the presence of the additional temperature scale means that higher-order loops, %
even if suppressed by larger powers of couplings, can be enhanced by powers of $T/m$, with $m$ being a relevant mass or momentum scale.
Hence, one has to resort to resummation schemes that incorporate all the relevant Feynman diagrams. 
The dominant loop corrections may be approximated by so-called ``hard thermal loops'' (HTL) \cite{Braaten:1989mz} that can give enhancement factors of the form $\kappa^2 T^2/m^2$, 
where $\kappa$ denotes either a gauge coupling, a Yukawa coupling, or the square root of a scalar quartic coupling. 
The HTL resummation entails absorbing such loops to all orders into effective propagators and vertices. 
For the LNC rates in leptogenesis, the HTL-resummed calculations presented in Refs.~\cite{1012.3784,1202.1288,1303.5498,Ghisoiu:2014ena,Glowna:2015aos} 
are necessary because the $2\leftrightarrow2$ scattering processes involving $t$-channel lepton exchanges are infrared divergent at tree level.
After the resummation, it is found that these rates go as $g^2 \ln g^{-2}$ (with $g$ standing here schematically for the gauge couplings $g_{2,1}$
for weak isospin or hypercharge, respectively) as opposed to the $g^2$ for $2\leftrightarrow2$ contributions without infrared enhancement. 

In the present work, we are interested in the evolution of both the LNC and LNV rates throughout the whole transition from the ultrarelativistic to the nonrelativistic regime.  
As such,  we cannot make any assumption on the value of the ratio of $M_{1}$ and $T$, which makes obtaining analytic results rather challenging. 
Thus, we will resort to numerical estimates of the rates at leading logarithmic order in the HTL resummation scheme.
A systematic extension of the LNC rates toward the nonrelativistic regime and an improvement of the approximation beyond leading logarithmic order 
has to account for the cancellation of real with virtual divergences from soft and collinear radiation of gauge bosons and has been addressed in Refs.~\cite{1302.0743,Ghisoiu:2014ena}. 
Demonstrating this cancellation explicitly is involved. 
However, when expressing the rates in terms of self energies of the sterile neutrinos in the CTP formalism, 
the HTL resummation for contributions corresponding to wave-function corrections to the lepton and Higgs propagators automatically accounts for both real and virtual effects.~\cite{1302.0743}.
We will, however, not account for vertex corrections in the self-energies of the sterile neutrino, as these do not give rise to enhanced $g^2 \ln g^{-2}$ corrections. 
Therefore, our results will be accurate to leading logarithmic order, i.e. to $g^2 \ln g^{-2}$, what will be refereed to briefly as ``leading-log accuracy''%
\footnote{Note that here, ``leading-log'' refers to logarithms of the gauge coupling that arise from infrared effects, 
as opposed to the logarithms involving the renormalization scale that give rise to running gauge couplings.}.

\subsection{Derivation of the fluid equations}
\label{sec:boltzmann_deriv}

We start by considering the Schwinger-Dyson equations at leading order in the gradient expansion, Eq.~\eqref{eq:SDEs:Wigner},  whose derivation is summarized in Appendix~\ref{sec:ctp_basics}. 
Taking the Hermitian part and assuming the universe to be spatially homogeneous and isotropic (so that spatial gradients may be neglected) results in the so-called ``kinetic equation''~\cite{1007.4783,1606.06690}
\begin{equation}
\label{eq:kadanoff_baym}
\partial_t \i \boldsymbol{S}_{X}^{<,>} =
- \big[ \boldsymbol{\mathcal{H}}_{X,\text{eff}} , \boldsymbol{S}_X^{<, >} \big] + \big[ \boldsymbol{\Sigma}_{X}^{<,>}, \boldsymbol{S}_X^{\mathcal{H} } \big] 
+ \frac{1}{2} \big( \big\{ \boldsymbol{\Sigma}_{X}^{>}, \boldsymbol{S}_X^{<} \big\} - \big\{ \boldsymbol{\Sigma}_{X}^{<}, \boldsymbol{S}_X^{>} \big\} \big)
\end{equation}
with the shorthand notations
\begin{align}
\boldsymbol{S}_{X}^{<,>,{\cal H}} &\equiv \i \gamma_0 S_X^{<,>,\cal{H}} \ , 
&
\boldsymbol{\Sigma}_{X}^{<,>} &\equiv \Sigma_X^{<,>} \gamma_0 \ ,
&
 \boldsymbol{\mathcal{H}}_{X,\text{eff}} &\equiv \big( \slashed{k} - m_X - \Sigma^{\mathcal{H}}_X \big) \gamma_0
\ .
\end{align}
Note that the term involving the anticommutators describes the decay, inverse decay and
scattering processes and is therefore referred to as collision term.
Now, we take the kinetic equation~\eqref{eq:kadanoff_baym} and apply the integrations~\eqref{eq:lep_density_def}, \eqref{eq:neut_density_def} that lead to the particle number densities. 
This yields
\begin{subequations}
\begin{align}
\nonumber
\dz Y_{N_1 \text{even}/\text{odd}} 
&= \dz \frac{1}{s} \big( n_{N_1 +} \pm n_{N_1 -} \big) 
\\
\label{eq:neut_charge_eom}
&= \frac{1}{2 \tilde M_1 s} \int \frac{ \text{d}^4 k }{ (2 \pi)^4 }\, \text{sign} (k_0) \tr \big[ \big( \p_+ \pm \p_- \big) \big( \i \Sigma_{N_1}^{>} (k)\i S_{N_1}^{<}(k) - \i \Sigma_{N_1}^{<}(k) \i S_{N_1}^{>}(k) \big) \big] \ ,
\\
\nonumber
\dz Y_{l_\smallpara} 
&= \dz \frac{1}{s} \big( n_{l_\smallpara}^+ - n_{l_\smallpara}^- \big) 
\\
\label{eq:lep_charge_eom}
&= \frac{1}{\tilde M_1 s} \int \frac{ \text{d}^4 p }{ (2 \pi)^4 }  \tr \big[ \i \Sigma_{l_\smallpara}^{>}(p) \i S_{l_\smallpara}^{<}(p) - \i \Sigma_{l_\smallpara}^{<} (p)\i S_{l_\smallpara}^{>}(p) \big] \ .
\end{align}
\end{subequations}
To proceed, we evaluate these equations by means of a perturbative expansion in the sterile neutrino Yukawa couplings. 
Notice that, while the Schwinger-Dyson Eqs.~\eqref{eq:SDEs:Wigner} form a closed system, 
the number densities on the left-hand side of Eqs.~\eqref{eq:neut_charge_eom} and~\eqref{eq:lep_charge_eom} do not contain the full information on the distributions 
that appear in the propagators and self energies in the collision integrals on the right. 
Furthermore, the above equations do not yield any information on the spectral functions that also enter into the collision integrals and therefore have to be obtained elsewhere. 
As a result, we have to provide further input in the shape of appropriate expressions for the Wightman functions and the CTP self-energies of the Standard Model leptons, the Higgs boson, and the sterile neutrinos, that have to be inserted into the collision term.

First, we consider the Wightman functions of the Standard Model particles and sterile neutrinos. 
Since the Standard Model particles are assumed to remain in kinetic equilibrium, their Wightman functions are given as in Eqs.~\eqref{eq:Wightman_eq} and~\eqref{eq:dist_eq}, 
with the spectral functions $\Delta_X^{\cal A},\,S_X^{\cal A}$ that include the dispersive and absorptive effects from the interactions with the plasma. 
These corrections are necessary for the fluid equations to remain valid in the regime relativistic sterile neutrinos. 
We refer to Section~\ref{sec:gamma_computation} for  more details on the spectral functions of the Standard Model particles.
 
In contrast, the sterile neutrinos may be far away from kinetic equilibrium throughout leptogenesis, so that the parametrization in Eqs.~\eqref{eq:Wightman_eq} and~\eqref{eq:dist_eq} cannot be assumed to be valid for these particles. 
In the nonrelativistic regime, where $|\mathbf k|\ll M_1$ for the typical momentum of a sterile neutrino, 
taking the kinetic equilibrium form for their distribution function does not lead to a sizable error in the result of the calculation~\cite{Basboll:2006yx,HahnWoernle:2009qn}.
In general however, we must assume helicity-dependent distribution functions that are not simply characterized by a chemical potential, and which will be left unspecified in the present work. 
This is the origin of the dominant theoretical uncertainty within the fluid equation approach. 
For the spectral function $S^{\cal A}_{N_1}$, we may simply take the tree-level result of Eq.~\eqref{eq:spectral_eq_tree}, as the smallness of the Yukawa couplings of the  sterile neutrino imply a strong suppression of its width. 
Proceeding in this way, we may write
\begin{subequations}
\label{eq:SN:tree}
\begin{align}
 \i S^{<}_{N_1} (k)
&= - 2 \pi \sum_h \p_h (\slashed{k} + M_{1}) \text{sign} (k_0) \delta ( k^2 - M_{1,}^2 ) f_{N_1 h} (k_0) \ ,
\\
 \i S^{>}_{N_1 } (k)
&= 2 \pi \sum_h \p_h (\slashed{k} + M_{1}) \text{sign} (k_0) \delta ( k^2 - M_{1}^2 ) (1 - f_{N_1 h} (k_0) )
\ .
\end{align}
\end{subequations}
The above helicity decomposition can be justified from the fact that the helicity projection operator commutes with the kinetic equation \cite{1112.5954}. 
The $f_{N_1 h} (k_0)$ with $h=\pm 1$ denote the two sterile neutrino distribution functions, for which we define deviations from equilibrium as
\begin{align}
\delta f_{N_1 h} (k_0) &\equiv f_{N_1 h} (k_0) - f_{N_1 h, \text{eq}} (k_0) \ ,
&
\text{where}&
&
f_{N_1 h, \text{eq}} (k_0) = f_F (k_0) = \frac{1}{e^{\beta k_0} +1}
\end{align} 
is the Fermi-Dirac distribution. 
\begin{figure}[t]
\centering
\raisebox{-0.0cm}{\includegraphics[width=.55\textwidth]{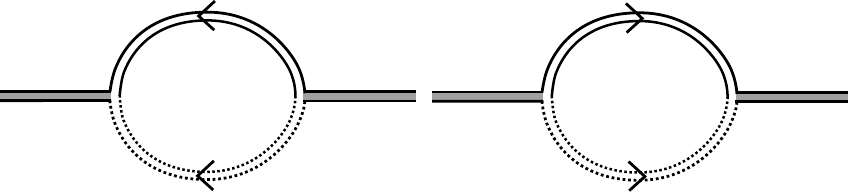}}
\caption{\label{fig:sterile-self-energy}\footnotesize%
2PI diagrams contributing to the self energy of the lightest sterile neutrino. 
The double lines represent resummed propagators.
Scalar lines are dotted, and those corresponding to sterile neutrinos are indicated with a gray shading.
Note that the self energies ${\rm i}\Sigma_{N_1}$ correspond to the amputated diagrams.}
\end{figure}
Next, we consider the sterile neutrino and lepton self-energies. 
The sterile neutrino self-energy is formally defined as in Eq.~\eqref{eq:self_energy}, and to leading order in the Yukawa couplings $F_1$, 
it can be obtained by amputating the 2PI diagrams shown in Fig.~\ref{fig:sterile-self-energy}.
Generically, it can be decomposed as
\begin{align}
\i \Sigma_{N_1}^{>,<}(k)
= g_w \left| F_{1} \right|^2 \left( \p_L \gamma_\mu \i \hat{\Sigma}^{\mu>,<}_{N_1, L}(k) + \p_R \gamma_\mu \i \hat{\Sigma}^{\mu>,<}_{N_1, R}(k) \right) \ ,
\end{align}
where we have implicitly defined the left- and right-chiral reduced self-energies $\hat{\Sigma}^\mu_{N_1,L}$ and $\hat{\Sigma}^\mu_{N_1,R}$. 
In the nonrelativistic limit, these reduced self-energies are approximately temperature-independent, but in general they can receive sizable thermal corrections generated by the Standard Model gauge interactions. 
To one-loop order, when dropping the Yukawa couplings, one has~\cite{1002.1326}
\begin{align}
\label{eq:sigma_n_l_def}
\i \hat{\Sigma}^{\mu >,<}_{N_1,L}(k)
&=\frac{1}{2} \int \frac{\text{d}^4 p}{ (2 \pi)^4 } \tr\big[\gamma^\mu \p_L \i S^{>,<}_{l_\smallpara} (p)  \p_R \big] \i \Delta^{>,<}_{\phi} (k-p)
\intertext{and}
\i \hat{\Sigma}^{\mu >,<}_{N_1,R}(k)
&= \frac{1}{2} \int \frac{\text{d}^4 p}{ (2 \pi)^4 } \tr \big[ \gamma^\mu \p_R C (\i S^{<,>}_{l_\smallpara} (-p))^{\top}C^\dagger\p_L \big] \i \Delta^{<,>}_{\phi} (p-k)
\ ,
\end{align}
where $C$ is the Dirac charge-conjugation matrix. As noted before, the Standard Model Wightman functions are taken to include finite-temperature wave-function type corrections. 
In principle, the self energies receive additional vertex type corrections from electroweak interactions. 
As noted before, these corrections contribute only at order $g^2$ and are therefore negligible at leading log accuracy.  
For more details, we again refer to the discussion in Section~\ref{sec:gamma_computation}. 

Since we assume that the Standard Model particles are in kinetic equilibrium
and have small asymmetries, we may expand the sterile neutrino self energy in $\nicefrac{\mu_X}{T}$, giving~\cite{1606.06690}
\begin{subequations}
\begin{align}
\label{eq:close_to_eq_expansion_LH}
\i \hat{\Sigma}^{\mu >,<}_{N_1,L}(k)
&= \i \hat{\Sigma}_{N_1}^{\mu>,<}(k) - 2\, \frac{\mu_{l_\smallpara} + \mu_H}{T} f_F(k_0) (1 - f_F(k_0) ) \hat{\Sigma}_{N_1}^{\mu}(k) + O \vspace{-2pt} \big( \nicefrac{\mu_{l_\smallpara}^2}{T^2} , \nicefrac{\mu_{\phi}^2}{T^2} \big)
\intertext{and}
\label{eq:close_to_eq_expansion_RH}
\i \hat{\Sigma}^{\mu >,<}_{N_1,R}(k)
&= \i \hat{\Sigma}_{N_1}^{\mu>,<}(k) + 2\, \frac{\mu_{l_\smallpara} + \mu_H}{T} f_F(k_0) ( 1 - f_F(k_0) ) \hat{\Sigma}_{N_1}^{\mu}(k) + O \vspace{-2pt} \big( \nicefrac{\mu_{l_\smallpara}^2}{T^2} , \nicefrac{\mu_{\phi}^2}{T^2} \big) \ .
\end{align}
\end{subequations}
Here we have introduced the reduced self-energies $\hat{\Sigma}_{N_1}^{\mu>,<}$ and $\hat{\Sigma}_{N_1}^\mu $, which are defined via the relations
\begin{align}
\hat{\Sigma}_{N_1}^{\mu>,<}(k)
\equiv &\,\eval{\hat{\Sigma}^{\mu >,<}_{N_1,L}(k)}_{\mu_{l_\smallpara}, \mu_{\phi} = 0} = \eval{\hat{\Sigma}^{\mu >,<}_{N_1,R}(k)}_{\mu_{l_\smallpara},\mu_H = 0} 
\end{align}
and
\begin{equation}
\label{eq:hatSigmaN}
\begin{aligned}
\hat{\Sigma}_{N_1}^\mu(k) 
& \equiv \,\frac{\i}{2} \big( \hat{\Sigma}_{N_1}^{\mu >}(k)- \hat{\Sigma}_{N_1}^{\mu <}(k) \big) \\
& = f^{-1}_F(k_0)\int \frac{\text{d}^4 p}{(2\pi)^4} f_F(p_0) f_B(k_0 - p_0) \Delta_\phi^{\mathcal{A}}(k-p) \tr \big[ \p_L \gamma^\mu S_{l}^{\mathcal{A}}\left(p\right) \p_R \big] 
\ ,
\end{aligned}\end{equation}
where
\begin{align}
f_B(k_0)=\frac{1}{e^{\beta k_0}-1}
\end{align}
is the Bose-Einstein distribution.
\begin{figure}[t]
\centering
\raisebox{-.19cm}{\includegraphics[width=.27\textwidth]{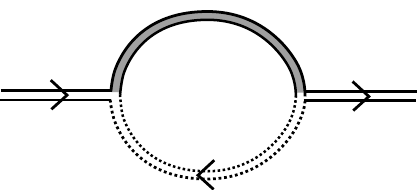}}
\includegraphics[width=.3\textwidth]{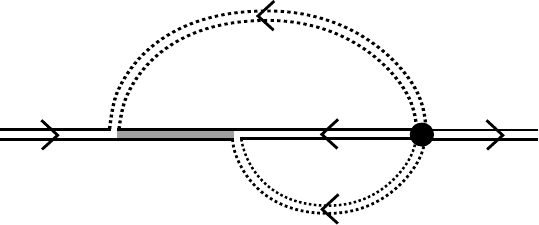}
\includegraphics[width=.3\textwidth]{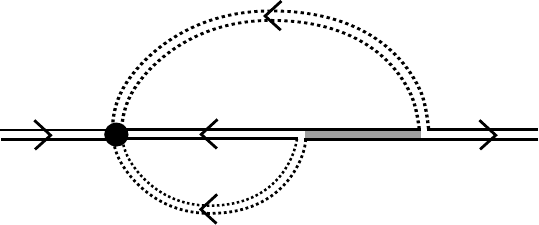}
\caption{\label{fig:lepton-self-energy}\footnotesize%
2PI diagrams contributing to the leptonic self-energies. 
The first diagram is the LO contribution, and the two-loop diagrams generate the $CP$-violating source term. The conventions for the lines are as in Fig.~\ref{fig:sterile-self-energy}.
The dotted vertices correspond to the effective interaction of Eq.~\eqref{eq:Weinberg_ops}. The self energy ${\rm i}\Sigma_{l_\smallpara}$ results
when amputating these diagrams.}
\end{figure}

In the case of the Standard Model leptons, only the $B-L$ violating contributions to the self-energies enter explicitly into the fluid equations~\eqref{eq:lep_charge_eom}.
When restricting to these pieces, the leading order contribution in the sterile neutrino Yukawa couplings is given by the first diagram in Fig.~\ref{fig:lepton-self-energy},
\begin{align}
\label{eq:sigma_lep_LO}
\i \Sigma_{l_\smallpara, LO}^{>,<}(k)
&= \left| F_{1} \right|^2 \int \frac{\text{d}^4 p}{ (2 \pi)^4 } \p_R \i S^{>,<}_{N_1} (p) \p_L \i \Delta^{<,>}_\phi (p-k)
\ , 
\end{align}
where as before $\i \Delta^{<,>}_\phi (p-k)$ refers to the full propagator, including in particular the wave-function type corrections for spectral function of the Higgs boson.  
Again, we neglect vertex-type corrections to the lepton self-energy, since they do not contribute at leading-log accuracy in the electroweak coupling. 
Further, we note that $\i S^{<,>}_{N_1}$ in Eq.~\eqref{eq:sigma_lep_LO} is understood to be given by the tree-level expression~\eqref{eq:SN:tree}, since we expand the lepton selfenergy to order $|F_1|^2$. 
In principle, the above 2PI expression for $\i \Sigma_{l_\smallpara}^{<,>}$ at one-loop order does include $CP$-violating effects from mixing of the sterile neutrinos, 
but these are encoded in terms of wave-function type loop corrections for $\i S^{<,>}_{N_1}$, which only enter into the expression in Eq.~\eqref{eq:sigma_lep_LO} at order $ O(F^6)$.
It is therefore important to note that the self energy~\eqref{eq:sigma_lep_LO} with $\i S^{<,>}_{N_1}$ as in Eq.~\eqref{eq:SN:tree} is $CP$ conserving: 
it generates the washout and backreaction terms in the fluid equations but does not give rise to the $CP$-violating source term.
Indeed, the leading order $CP$-violating term appears at $O (F^4)$ in the sterile neutrino Yukawa couplings. 
The relevant contributions to the source term are generated by the second and third diagram in Fig.~\ref{fig:sterile-self-energy}, which include the vertices of the effective operators of Eq.~\eqref{eq:Weinberg_ops}. 
These diagrams give
\begin{align}
\label{eq:sigma_lep_source}
\i \Sigma_{l_\smallpara , S}^{>,<}(k)
&= \left| F_{1} \right|^2 \int \frac{\text{d}^4 p}{ (2 \pi)^4 } \p_R \i S^{>,<}_{S} (p) \p_L \i \Delta^{<,>}_\phi (p-k)
\ , 
\end{align}
where we have defined the shorthand
\begin{equation}\begin{aligned}
 \i S^{>,<}_{S}(p) \equiv
&\, \pm \i \frac{3F_{1}^{*2} F_{2}^2}{M_{2} \left| F_{1 }\right|^2 } \Big( \p_R \i S^{>,<}_{N_1}(p) \p_R \i \gamma_\mu \hat{\Sigma}^{\mu \ T, \overline{T} }_{N_1,R}(p) - \p_R \i S^{\overline{T},T}_{N_1}(p) \p_R \i \gamma_\mu \hat{\Sigma}^{\mu \ >,<}_{N_1,R}(p) \Big)
\\
&\mp\i \frac{3F_{1}^{2} F_{2}^{*2}}{M_{2} \left| F_{1 }\right|^2 } \Big(\i \gamma_\mu\hat{\Sigma}^{\mu \ \overline{T},T }_{N_1,R}(p)\p_L \i S^{>,<}_{N_1}(p) \p_L -\i \gamma_\mu\hat{\Sigma}^{\mu \ >,<}_{N_1,R}(p)\p_L  \i S^{T,\overline{T}}_{N_1}(p) \p_L  \Big).
\end{aligned}\end{equation}

Having now specified the necessary expressions for the Wightman functions and self energies, we can proceed with deriving the fluid equations. 
To this end, we first use expressions \eqref{eq:sigma_lep_LO} and \eqref{eq:sigma_lep_source} to rewrite Eq. \eqref{eq:lep_charge_eom} as
\begin{align}
\label{eq:lep_charge_eom_v2}
\dz g_w Y_{l_\smallpara}
&= \frac{1}{\tilde M_1 s} \int \frac{ \text{d}^4 k}{ (2 \pi)^4 }  \tr \big[ \big( \i S_{N_1}^{>} + \i S^{>}_{S} \big) (k) \p_L \i \Sigma_{N_1}^{<} (k) - \big( \i S_{N_1}^{<} + \i S^{<}_{S}\big) (k) \p_L \i \Sigma_{N_1}^{>} (k) \big] \ . 
\end{align}
Next, we make use of the assumption that the sterile neutrino remains close to equilibrium and that the Standard Model asymmetries remain small.
If this is true, we may expand the right-hand side of Eqs. \eqref{eq:neut_charge_eom} and \eqref{eq:lep_charge_eom} to linear order in $\delta f_{N_1 h}$ and $\mu_{l,\phi}$. {Now, in view of the applications in Sec.~\ref{sec:numerical_scans}, where we assume 
vanishing initial conditions and hence  $\delta f_{N_1 h}$ is initially large, this implies an order one
inaccuracy for the washout rate as well as for the backreaction rate producing a helicity asymmetry in
the sterile neutrinos. For the scenario with partially
equilibrated spectator fields in the strong washout regime, this may not be problematic
because the washout and backreaction at early times have a subdominant effect with respect to the reactions occurring at later times when
$\delta f_{N_1 h}$ is indeed small. For weak washout,
$\delta f_{N_1 h}$ may remain large throughout all times
prior to Maxwell suppression, such that we do not accurately capture the Pauli-blocking factors from the sterile neutrinos that appear in the washout and the backreaction terms. While it would be interesting to
account for this, the rates would then be no longer simply functions of $z$ but also of $\delta f_{N_1 h}$. We therefore relegate this matter to future work, where
the fluid approximation is not made and $\delta f_{N_1 h}$ is calculated accurately enough to address this matter.}

First we  consider the terms generated by the Standard Model asymmetries.
Taking $\delta f_{N_1 h} \to 0$ and inserting the expansions \eqref{eq:close_to_eq_expansion_LH}, \eqref{eq:close_to_eq_expansion_RH} into Eqs. \eqref{eq:neut_charge_eom} and \eqref{eq:lep_charge_eom_v2}, we find
\begin{subequations}
\label{eq:boltzman_washout}
\begin{align}
\dz \eval{Y_{N_1 \text{even}}}_{\delta f_{N_1 h} \to 0}	&= 0 \ ,
&
\dz\eval{ Y_{N_1 \text{odd}}}_{\delta f_{N_1 h} \to 0}	&= \widetilde W \ ,
&
\dz g_w \eval{Y_{l_\smallpara}}_{\delta f_{N_1 h} \to 0} 	&= W \ , 
\end{align}
\text{with}
\begin{align}
\label{eq:washout_def}
\widetilde W	&= - \frac{12K}{T^3}\, \left( Y_{l_{\smallpara}} + \frac{1}{2} Y_{\phi} \right)  \left. \int \frac{\text{d}^3 \mathbf{k}}{(2\pi)^3}( 1 - f_F(k_0) ) f_F(k_0) \,\tilde\gamma(k) \,\right|_{k_0 = \sqrt{\mathbf{k}^2 + M_{1}^2} } \ ,
\\
W			&= - \frac{12K}{T^3} \, \left( Y_{l_{\smallpara}} + \frac{1}{2} Y_{\phi} \right)\left. \int \frac{\text{d}^3 \mathbf{k}}{(2\pi)^3} ( 1 - f_F(k_0) ) f_F(k_0)\,\gamma(k) \,\right|_{k_0 = \sqrt{\mathbf{k}^2 + M_{1}^2} } \ ,
\end{align}
\end{subequations}
where  the washout parameter $K$ was defined in Eq.~\eqref{eq:washout_K}, and we further introduced
\begin{align}
\label{eq:ktilde_gamma_def}
    \tilde{k}^\mu \equiv&\, \frac{1}{2} h \tr [ \p_h \gamma^5 \gamma^\mu \slashed{k}] = \big(\abs{\mathbf{k}}, k_0 \hat{\mathbf{k}} \big), &
    \gamma(k) \equiv& \frac{32 \pi}{T} \frac{k_\mu \hat{\Sigma}_{N_1}^{\mathcal{A} \mu}(k) }{k_0},
&
\tilde{\gamma}(k) \equiv&\, \frac{32 \pi}{T} \frac{\tilde{k}_\mu \hat{\Sigma}_{N_1}^{\mathcal{A} \mu}(k) }{k_0} \ .
\end{align}

Next, we consider the contributions generated by the sterile neutrino being out of equilibrium. 
Taking $\mu_{l,\phi} \to 0$, and defining  $f_{N_1, \text{even}/\text{odd}} \equiv f_{N_1 +} \pm f_{N_1 -}$, the result is
\begin{subequations}
\label{eq:boltzman_neutrino}
\begin{align}
\dz \eval{Y_{N_1 \text{even}}}_{ \mu_X\to 0}	&= D \ ,
&
\dz \eval{Y_{N_1 \text{odd}}}_{ \mu_X\to 0}	&= B \ ,
&
\dz \eval{g_w Y_{l_\smallpara}}_{\mu_X\to 0} 	&= \widetilde B + S \ ,
\end{align}
\text{with}
\begin{align}
\label{eq:decay_def}
D
&= - \frac{K}{s} \left. \int \frac{\text{d}^3 \mathbf{k}}{(2\pi)^3}\, \delta f_{N_1, \text{even}} (k_0) \gamma(k) \right|_{k_0 = \sqrt{\mathbf{k}^2 + M_{1}^2} } \ , 
\\
B
&= -  \frac{K}{s}\left. \int \frac{\text{d}^3 \mathbf{k}}{(2\pi)^3}\, \delta f_{N_1, \text{odd}} (k_0) \gamma(k) \right|_{k_0 = \sqrt{\mathbf{k}^2 + M_{1}^2} } \ ,
\\
\widetilde B
&= - \frac{K}{s} \left. \int \frac{\text{d}^3 \mathbf{k}}{(2\pi)^3}\, \delta f_{N_1, \text{odd}} (k_0) \tilde\gamma(k) \right|_{k_0 = \sqrt{\mathbf{k}^2 + M_{1}^2} } \ , 
\\
\label{eq:source_term}
S
&=  \epsilon_0 \frac{K}{s} \hspace{-5pt} \left. \int \frac{\text{d}^3 \mathbf{k}}{(2\pi)^3}\, \delta f_{N_1, \text{even}} (k_0)  \bigg( \frac{(32 \pi)^2}{T} \frac{ \hat{\Sigma}_{N_1 \mu} \hat{\Sigma}_{N_1}^\mu(k)}{k_0} \bigg) \right|_{k_0 = \sqrt{\mathbf{k}^2 + M_{1}^2} } \ ,
\end{align}
\end{subequations}
where $\epsilon_0$ is the zero-temperature decay-asymmetry defined as in Eq.~\eqref{eq:epsilon0}.
The source term~\eqref{eq:source_term} is consistent with the calculation in Ref.~\cite{Covi:1996wh}, provided that one there takes the limit $\nicefrac{M_1}{M_2} \to 0$.

To further simplify the source term, we note that since we work in the plasma frame the spatial part of the reduced self-energy, $\hat{\Sigma}_{N_1}^i(k)$, has to be proportional to $k^i$, as it is the only nonvanishing spatial momentum available.
Also, $k^\mu$ and the vector $\tilde k^\mu$ defined in Eq.~\eqref{eq:ktilde_gamma_def} are orthogonal in the four-dimensional sense, while having spatial components parallel to $k^i$. 
Thus, they span the two-dimensional sub-space of four vectors with spatial part proportional to $k^i$, so that we may use the decomposition
\begin{equation}
\hat{\Sigma}_{N_1}^\mu = \frac{1}{k^2} \big[ k^\mu (\hat{\Sigma}_{N_1}^\alpha k_\alpha ) - \tilde k^\mu (\hat{\Sigma}_{N_1}^\alpha \tilde k_\alpha) \big] \ .
\end{equation}
Using this relation and the definitions of $\gamma(k),\tilde\gamma(k)$ in Eq.~\eqref{eq:ktilde_gamma_def},
the scalar product appearing in the $CP$-violating source term \eqref{eq:source_term} can be cast as
\begin{align}
\frac{(32 \pi)^2}{T} \frac{ \hat{\Sigma}_{N_1 \mu}(k) \hat{\Sigma}_{N_1}^\mu(k)}{k_0}
&= \beta k_0  \frac{1}{z^2} ( \gamma^2(k) - \tilde \gamma^2(k) ) \ .
\end{align}

Finally, to obtain the fluid equations presented in Section~\ref{sec:boltzmann_eq}, we approximate $\gamma(k)$, $\tilde\gamma(k)$  by their thermal average over the equilibrium distribution of the sterile neutrino. 
For a general momentum-dependent variable $X(k)$, we define the thermal average as
\begin{align}
\label{eq:thermal_average}
\langle X \rangle =\frac{2}{n_{N_1,\rm eq}}\,\,\int\,\frac{d^3\mathbf{k}}{(2\pi)^3}\,X(k) f_F(k) \ , 
\end{align}
where $n_{N_1,\rm eq}$ is the equilibrium number density of the sterile neutrinos, 
\begin{align}
n_{N_1,\rm eq}  =\left.2 \int \frac{ \text{d}^3 \mathbf{k} }{(2\pi)^3} f_F(k)\right|_{k^0=\sqrt{\mathbf{k}^2 + M_{1}^2}}  = \frac{T^3}{ \pi^2} z^2 \sum_{n=1}^{\infty} \frac{\left( -1 \right)^{n+1}}{n} K_2 \left( n \cdot z \right) \approx \frac{T^3}{\pi^2} z^2 \cdot K_2 \left( z \right)
\ .
\end{align}
Using the definition~\eqref{eq:thermal_average}, we now implement the thermal averaging by replacing
\begin{align}
\gamma(k) &\to \langle \gamma \rangle \ ,
&
\tilde \gamma(k) &\to \langle \tilde \gamma \rangle \ ,
&
\text{and}&
&
\beta k_0 ( \gamma^2(k) - \tilde \gamma^2(k)) &\to \langle \beta k_0 ( \gamma^2 - \tilde \gamma^2 ) \rangle \ .
\end{align} 
From the averaged rates $\langle\gamma\rangle,\,\langle\tilde\gamma\rangle$, we finally introduce the dimensionless LNC and LNV rates of Sec.~\ref{sec:boltzmann_eq}:
\begin{align}\label{eq:gammas_LNC_LNV}
\gamma_\text{\rm LNC} &= \langle \gamma + \tilde \gamma \rangle \ ,
&
\gamma_\text{\rm LNV} &= \langle \gamma - \tilde \gamma \rangle \ .
\end{align}
The average of the source term in terms of $\gamma_{\rm LNC},\,\gamma_{\rm LNV}$  is given as
\begin{equation}
 \langle \beta k_0 ( \gamma^2 - \tilde \gamma^2 ) \rangle \approx \langle \beta k_0 \rangle \langle \gamma + \tilde \gamma \rangle \langle \gamma - \tilde \gamma \rangle = \langle \beta k_0 \rangle \, \gamma_\text{\rm LNC} \cdot \gamma_\text{\rm LNV} \ .
\end{equation}
At this point, it is prudent to reiterate that the momentum averaging leads to the order one uncertainty
corresponding to the leading order fluid approximation in the relativistic regime.
Once $\gamma(k),\,\tilde\gamma(k)$ are substituted by their averages, 
the remaining momentum integrals can be cast in terms of the integrals ${\cal I}(z),{\cal J}(z), {\cal K}(z)$ that have been defined back in Eqs.~\eqref{eq:I_exp}, \eqref{eq:J_exp}, and~\eqref{eq:epsilon_eff_def}.
Indeed, we find for the integrals appearing in Eqs.~\eqref{eq:boltzman_neutrino} and~\eqref{eq:boltzman_washout}
\begin{align}
W			&= - 12 \Gamma \, \left( Y_{l_{\smallpara}} +\frac12 Y_\phi\right) \frac{{\cal I}(\bar M_1/T)}{2\pi^2} \ ,
\end{align}
where $\Gamma$ is given by Eq.~\eqref{eq:gamma_av},
such that $\eta_{N_1}$ follows as given in Eq.~\eqref{eq:Yeq_eta}. Further,
\begin{align}
\label{eq:etaN1}
\langle\beta k_0\rangle=&\,\frac{{\cal K}(z)}{{\cal I}(z)}
\end{align}
and
\begin{align}
\frac{1}{s}\int \frac{\text{d}^3 \mathbf{k}}{(2\pi)^3} \,\delta f_{N_1,\rm even}(k_0)=&\,Y_{N_1,\rm even}-Y_{N_1,\rm eq}  \ , 
\\
\frac{1}{s}\int \frac{\text{d}^3 \mathbf{k}}{(2\pi)^3} \,\delta f_{N_1,\rm odd}(k_0)=&\,Y_{N_1,\rm odd} \ ,
\end{align}
where we have also used that $Y_{N_1,\rm odd}$ vanishes in equilibrium.
Putting everything together, we finally obtain the momentum-averaged fluid equations for leptogenesis:
\begin{align}
\label{eq:Boltzmann}
\begin{aligned}
\dz Y_{N_1 \text{even}}
&= D \\
&\approx - \Gamma \cdot (Y_{N_1 \text{even}} - Y_{N_1 \text{eq}}) 
\ , \\
\dz Y_{N_1 \text{odd}}
&= B + \widetilde W \\
&\approx - \Gamma \cdot Y_{N_1 \text{odd}} - \eta_{N_1} \tilde{\Gamma} \cdot \left(Y_{l_\smallpara} + \frac{1}{2} Y_\phi \right) 
\ , \\
\dz g_w Y_{l_\smallpara} 
&= \widetilde B + S + W \\
&\approx  - \tilde{\Gamma} \cdot Y_{N_1 \text{odd}} + \epsilon_\text{eff} \Gamma \cdot (Y_{N_1 \text{even}} - Y_{N_1 \text{eq}}) - \eta_{N_1} \Gamma \cdot \left(Y_{l_\smallpara} + \frac{1}{2} Y_\phi \right)
\ ,
\end{aligned}
\end{align}
where the two independent rates $\Gamma,\tilde\Gamma$ are given by Eq.~\eqref{eq:gamma_av}, while the effective sterile neutrino decay asymmetry is
\begin{align}
\epsilon_\text{eff} = \epsilon_0 \langle \beta k_0 \rangle \frac{2 \, \gamma_\text{\rm LNC} \cdot \gamma_\text{\rm LNV} }{z^2 ( \gamma_\text{\rm LNC} + \gamma_\text{\rm LNV} ) } \ .
\end{align}
The factor of $\eta_{N_1}$ can be understood as accounting for the fact that $\Gamma$ and $\tilde\Gamma$ are defined as averages with respect to the sterile neutrino momentum distribution, 
rather than the momentum distributions of the Standard Model lepton and Higgs boson.
Above, we have derived it from the absorptive part of the self energy of the Standard Model lepton.
To obtain its value in a more intuitive way based on a collision term that one would set up in a Boltzmann equation, 
we consider the yield corresponding to the difference between the net-numbers of leptons and anti-leptons turned into sterile neutrinos via the Yukawa interactions per given unit $z$-interval. 
Since the Standard Model particles are in kinetic equilibrium, this difference is given as
\begin{align}
W
&\equiv - \Gamma \frac{2}{s} \int \frac{ \text{d}^3 \mathbf{k} }{(2\pi)^3} \Big( \frac{1}{ e^{ \beta (k_0 - \mu_{l_\smallpara} - \mu_H) } + 1 } - \frac{1}{ e^{\beta (k_0 + \mu_{l_\smallpara} + \mu_H) } + 1 } \Big)
\\ 
&= \Gamma \frac{12}{T^3} \left(Y_{l_\smallpara} + \frac{1}{2} Y_\phi \right) \int \frac{ \text{d}^3 \mathbf{k} }{(2\pi)^3}  \frac{ e^{ \beta k_0 } }{( e^{ \beta k_0 } + 1 )^2 } +  O \vspace{-2pt} \big( \nicefrac{\mu_X^2}{T^2} \big)
\ , 
\end{align}
and therefore
\begin{equation}
\eta_{N_1}
\equiv \frac{12}{T^3} \int \frac{ \text{d}^3 \mathbf{k} }{(2\pi)^3}  \frac{ e^{ \beta k_0 } }{( e^{ \beta k_0 } + 1 )^2 } =\frac{6}{\pi^2} \int_{z}^{\infty} \hspace{-5pt} \text{d}y \hspace{5pt} y (y^2 - z^2)^{\frac{1}{2}} \frac{ e^y }{( e^y + 1 )^2 }
=\frac{6}{\pi^2}\, {{\cal J}(z)}
\ .
\end{equation}

To close this section, we note that the fluid equations~\eqref{eq:Boltzmann} do not account for $Y_{l_\smallpara}$-violating Standard Model interactions. 
However, all of these rates are $B-L$ conserving, so that the evolution of $Y_{B-L}$ must follow directly from the equation for $Y_{l_\smallpara}$ in~\eqref{eq:Boltzmann}. 
As argued in Sec.~\ref{sec:boltzmann_eq}, even with partially equilibrated sphaleron reactions, chemical equilibrium constraints can be seen to imply that the baryon number charge is equally distributed among flavours, 
while nonzero values of $B-L$ are restricted to the flavour that couples to the sterile neutrinos. 
This implies  that 
\begin{align}
Y_{B-L}=\frac{Y_B}{3}-g_w Y_{l\smallpara}\equiv Y_{\Delta_\smallpara}\Rightarrow \dz Y_{\Delta_\smallpara}=-\dz g_w Y_{l\smallpara}.
\end{align}
Taking this into account, one recovers the fluid equations \eqref{eq:n_odd_boltzmann}-\eqref{eq:y_bl_boltzmann} from Eq.~\eqref{eq:Boltzmann}.
\subsection{Computation of the rates $\gamma_\text{\rm LNC}$ and $\gamma_\text{\rm LNV}$}
\label{sec:gamma_computation}

To compute $\gamma_\text{\rm LNC}$ and $\gamma_\text{\rm LNV}$, which are defined by Eqs.~\eqref{eq:gammas_LNC_LNV}, \eqref{eq:thermal_average} and~\eqref{eq:ktilde_gamma_def}, we use the expression for $\hat{\Sigma}_{N_1}^\mu$ given in Eq.~\eqref{eq:hatSigmaN}:
\begin{align}
\label{eq:sigma_n_def}
\hat{\Sigma}_{N_1}^\mu(k)
&= f^{-1}_F(k_0)\int \frac{\text{d}^4 p}{(2\pi)^4} f_F(p_0) f_B(k_0 - p_0) \Delta_\phi^{\mathcal{A}}(k-p) \tr \big[\gamma^\mu \p_L S_{l}^{\mathcal{A}}\left(p\right) \p_R \big] 
\ ,
\end{align}
where $f_F$ and $f_B$ are the Fermi-Dirac and Bose-Einstein distributions and $S_l^\mathcal{A}$ and $\Delta_\phi^\mathcal{A}$ are the Standard Model lepton and Higgs boson spectral functions. 
We recall that we have defined $\hat{\Sigma}_{N_1}^\mu$ as the reduced spectral self-energy of the sterile neutrinos with the Standard Model fields in kinetic equilibrium and with vanishing chemical potentials (see Eqs.~\eqref{eq:close_to_eq_expansion_LH},\eqref{eq:close_to_eq_expansion_RH}).
Therefore, we approximate $S_l^\mathcal{A}$ and $\Delta_\phi^\mathcal{A}$ as the one-loop resummed equilibrium spectral functions, which solve the Schwinger-Dyson equations in the absence of gradients.
Neglecting the tree-level masses of the Higgs boson, and with vanishing tree-level masses for the leptons in the phase of restored electroweak symmetry, these resummed spectral functions are given by~\cite{Garbrecht:2008cb,1007.4783}
\begin{align}
\label{eq:resummed_spectral}
S_{l}^{\mathcal{A}}\left(p\right) &= \p_L \left[\left(\slashed{p} -{\Sigma}^\mathcal{H}_{l} \left(p \right) \right) \cdot \frac{\Gamma_{l}}{\Omega_{l}^2 + \Gamma_{l}^2} %
- {\Sigma}^\mathcal{A}_{l}\left(p \right) \frac{\Omega_{l}}{\Omega_{l}^2 + \Gamma_{l}^2} \right]\, \p_R \ ,
&
\Delta_\phi^{\mathcal{A}}(q) &= \frac{\Gamma_{\phi}}{\Omega_{\phi}^2 + \Gamma_{\phi}^2} \ ,
\end{align}
with 
\begin{align}\begin{aligned}
\Gamma_{\phi}\left(q \right) 
&= \Pi^\mathcal{A}_{\phi} \ ,
&	
\Gamma_{l}\left(p\right)	
&= 2 \left( p_{\mu} - \Sigma^\mathcal{H}_{l, \mu} \right) \cdot \Sigma^{\mathcal{A},\mu}_{l} \ ,
\\
\Omega_{\phi}\left(q \right)	
&= q^2 - \Pi^\mathcal{H}_{\phi} \ ,
&
\Omega_{l}\left(p\right)
&= \left( p_{\mu} - \Sigma^\mathcal{H}_{l, \mu} \right)^2 - \left(\Sigma^{\mathcal{A}}_{l,\mu}\right)^2 
\ ,
\end{aligned}\end{align} 
where $\Pi^{\mathcal{H}/\mathcal{A}}_\phi$ and $\Sigma_{l}^{\mathcal{H}/\mathcal{A}}$ are finite-temperature Standard Model self-energies of the active lepton and the Higgs-boson.

To compute the LNC and LNV rates at leading-log accuracy for a range of temperatures covering the relativistic and nonrelativistic regimes, 
we keep only the HTL contributions within the spectral functions in the one-loop self-energy~\eqref{eq:sigma_n_def}. 
As explained at the beginning of Sec.~\ref{sec:derivation_fluid_eq_rates}, these may be enhanced by factors of $T^2/m^2$ compared to the tree-level result for $\hat\Sigma_{N_1}$, where $m$ is a mass or momentum scale. 
In fact, the HTL contributions must be resummed in order to get accurate results in the relativistic regime. 
In our case, the HTL expressions can also be recovered by evaluating the full one-loop self-energies close to $p^2, q^2 \to 0$. 
In this limit, one has \cite{Weldon:1982bn,Bellac:2011kqa}
\begin{align}
\label{eq:HTL_self_energies}
\begin{aligned}
\Pi^{\mathcal{H},\text{HTL}} &= m^2_\phi \ ,
&
\Sigma_l^{\mathcal{H},\text{HTL}}(p) &= \frac{m_l^2}{4} \frac{ \slashed{\tilde{p}} }{ \abs{\mathbf{p}}^2 } \ln \left| \frac{p_0 + \abs{\mathbf{p}} }{p_0 - \abs{\mathbf{p}} } \right| - \frac{m_l^2}{2} \frac{ \slashed{\hat{p}} }{ \abs{\mathbf{p}}^2 } \ , 
\\
\Pi^{\mathcal{A},\text{HTL}} &= 0 \ ,
&
\Sigma_l^{\mathcal{A},\text{HTL}}(p) &= \frac{m_l^2}{4} \frac{\slashed{\tilde{p}}}{|\mathbf{p}|^2} 2 \pi \, \theta(-p^2)
\ .
\end{aligned}
\end{align}
Here, we have again used $\tilde{p}^\mu \equiv (\abs{\mathbf{p}}, p_0 \frac{ \mathbf{p} }{ \abs{\mathbf{p}} } )$ and further defined $\hat{p}^\mu \equiv (0, \mathbf{p} )$. 
The parameters $m^2_\phi$ and $m^2_l$ denote the thermal masses of the Higgs boson and lepton,
\begin{align}
\label{eq:m_thermal}
m^2_l &= \frac{1}{16} ( 3 g_2^2 + g_1^2 ) T^2 
&
\text{and}&
&
m^2_\phi &= \frac{1}{16} ( 3 g_2^2 + g_1^2 + 4 h_t^2 + 8 \lambda_\phi ) T^2 \ ,
\end{align}
where $h_t$ is the top-quark Yukawa-coupling and $\lambda_\phi$ the quartic self-coupling of the Higgs boson.
Note that within the HTL approximation, the width of the Standard Model particles is zero for time-like momenta $p^2 >0$.
As a result, the resummed spectral propagators of Eq.~\eqref{eq:resummed_spectral} then take the same shape as the tree-level results of Eq.~\eqref{eq:spectral_eq_tree}, except for a modified on-shell condition,
\begin{align}
\label{eq:spectral_schannel}
\begin{aligned}
 p^2>0:\quad S_l^{{\cal A},\rm HTL}(p)=&\,\pi\,\text{sign}(p^0)\,\delta\left((p_\mu-\Sigma_{l,\mu}^{{\cal H},\rm HTL}(p))^2\right)\p_L(\slashed{p}-\slashed{\Sigma}^{\mathcal{H},\rm HTL}_{l} \left(p \right) )\p_R \ , \\
\Delta_\phi^{{\cal A},\rm HTL}(p)=&\,\pi\,\text{sign}(p^0)\,\delta(p^2-m^2_\phi)
\ .
\end{aligned}\end{align}

\begin{figure}
\centering
\includegraphics[width=.8\textwidth, height = 8cm]{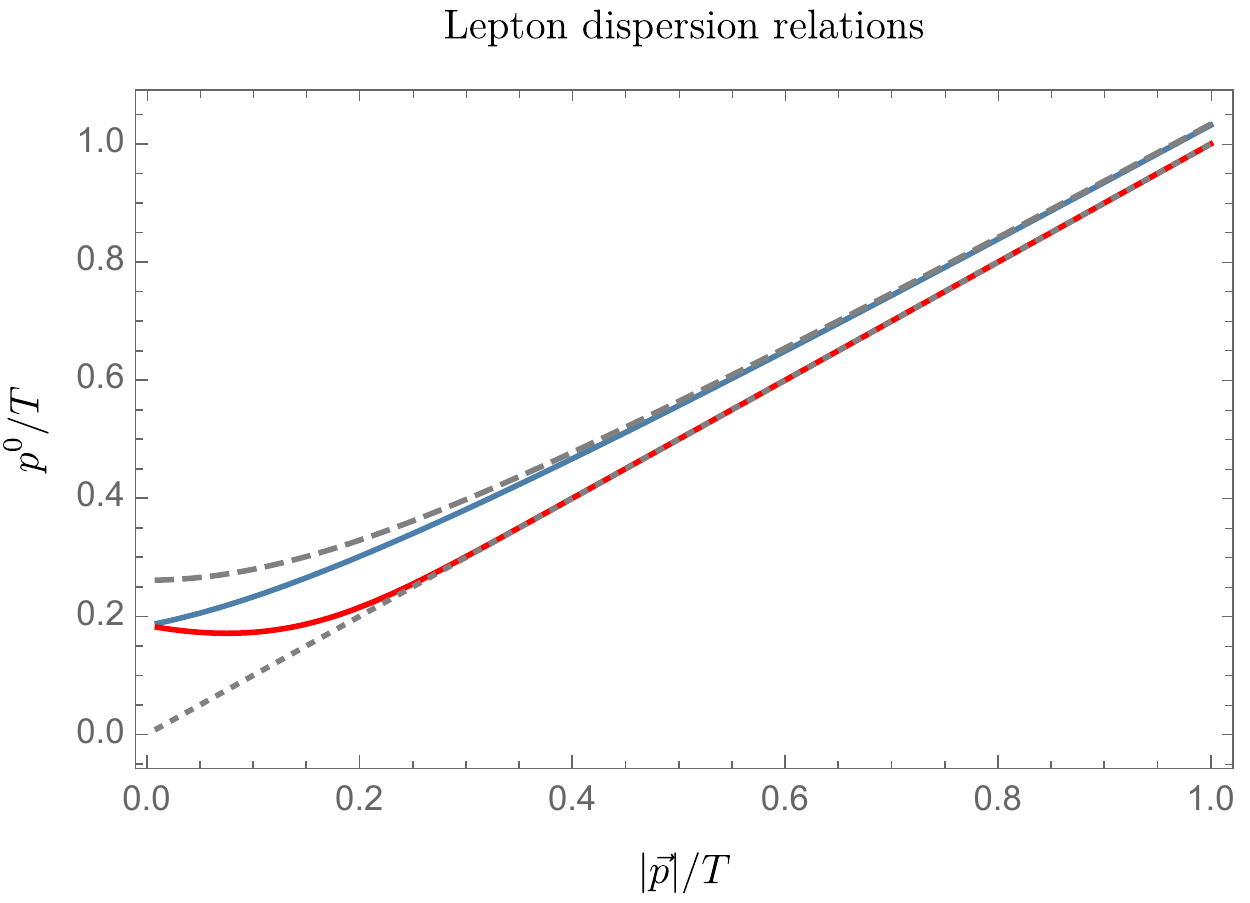}
\caption{\label{fig:dispersion}\footnotesize%
The HTL dispersion relations for leptonic pseudoparticles (solid blue) and holes (solid red), plotted against the dispersion relation of a massless particle (dotted gray) and a particle with mass $m_l$ (dashed gray).}
\end{figure}

When substituting the above spectral propagators into expression~\eqref{eq:sigma_n_def} for $\hat{\Sigma}_{N_1}^\mu$, 
the on-shell conditions enforced by the $\delta$-functions in Eq.~\eqref{eq:spectral_schannel} allows for the interpretation of the resulting contributions to $\hat{\Sigma}_{N_1}^\mu$ as arising from on-shell $1\leftrightarrow2$ processes in the thermal plasma. 
These pieces, that we label $\hat{\Sigma}_{N_1,1\leftrightarrow2}^\mu$, can be computed analytically when simplifying the HTL dispersion relation, which enforces
\begin{align}
\label{eq:lepton_dispersion}
(p_\mu-\Sigma_{l,\mu}^{{\cal H},\rm HTL})^2&= 0 \ ,
\end{align}
by using the lepton thermal mass instead of the full Hermitian self-energy.
Physically, this corresponds to neglecting contributions generated by collective excitations in the plasma, i.e. the ``holes''.
Explicitly, taking
\begin{align}
\label{eq:simplified_spectral}
S_{l}^{\mathcal{A},\rm HTL,}\left(p\right) &\approx \pi \, \text{sign} (p_0) \, \delta (p^2 - m_l^2) \p_L \slashed{p} \p_R , \quad (p^2>0),
\end{align}
one can approximate  $\hat{\Sigma}_{N_1,1\leftrightarrow2}^\mu$ as
\begin{align}
\label{eq:sigma_1to2}
\hat{\Sigma}^\mu_{N_1, 1\leftrightarrow2}(k)
&\approx \frac{T}{32 \pi \abs{\mathbf{k}}} \Big[ \frac{\abs{A}^2}{M_1^2} {I}_{0} \Big(\frac{A^2}{M_1^2}\frac{k_0}{T},\frac{B^2}{M_1^2}\frac{ \abs{\mathbf{k}} }{T},\frac{k_0}{T} \Big) \cdot k^\mu \\
\nonumber
&\hspace{80pt} + \text{sgn}(A^2) \frac{T}{\abs{\mathbf{k}}} {I}_{1} \Big(\frac{A^2}{M_1^2}\frac{k_0}{T},\frac{B^2}{M_1^2}\frac{ \abs{\mathbf{k}} }{T},\frac{k_0}{T} \Big) \cdot \tilde{k}^\mu \Big] \ ,
\end{align}
where $\tilde{k}$ is defined as in Eq.~\eqref{eq:ktilde_gamma_def}, and the quantities $A,B$ are defined as
\begin{align}
A^2 &\equiv M_{1}^2 + m_l^2 - m_\phi^2,
&
B^2 &= \sqrt{A^4 - 4 m_l^2 M_{1}^2}
\ .
\end{align} 
The dimensionless integrals ${I}_{0/1} (\alpha, \beta, y)$ are given by
\begin{align}
\label{eq:I_0_bar_def}
\begin{aligned}
{I}_{0} (\alpha,\beta,y)
&= \beta + \ln \bigg| \frac{ 1 + e^{-\frac{1}{2} (\alpha + \beta) } }{ 1 + e^{-\frac{1}{2} (\alpha - \beta) } } \bigg| + \ln \bigg| \frac{ 1 - e^{\frac{1}{2} (\alpha - \beta) - y } }{ 1 - e^{\frac{1}{2} (\alpha + \beta) - y } } \bigg|
\ , \\
{I}_{1} (\alpha,\beta,y)
&= \beta \left[ \ln \! \big| \big( 1 + e^{-\frac{1}{2} (\alpha + \beta) } \big) \big( 1 + e^{-\frac{1}{2} (\alpha - \beta) } \big) \big| \right.- \left. \ln \! \big| \big( 1 - e^{\frac{1}{2} (\alpha - \beta) - y} \big) \big( 1 - e^{\frac{1}{2} (\alpha + \beta) - y } \big) \big| \right] \\
&\hspace{10pt} - 2 \left[ \text{Li}_2 (-e^{-\frac{1}{2} (\alpha - \beta) }) - \text{Li}_2 (-e^{-\frac{1}{2} (\alpha + \beta) }) \right.\left. + \text{Li}_2 (e^{\frac{1}{2} (\alpha - \beta) - y }) - \text{Li}_2 (e^{\frac{1}{2} (\alpha + \beta) - y }) \right]
\ ,
\end{aligned}
\end{align}
where ${\rm Li}_2(x)$ denotes the polylogarithmic function of second order. 
Using the result in Eq.~\eqref{eq:sigma_1to2}, the rates $\gamma^{1\leftrightarrow2}_{\rm LNC/LNV}$ can be obtained by averaging over sterile neutrino momentum distribution 
in thermal equilibrium as per Eqs.~\eqref{eq:gammas_LNC_LNV} and \eqref{eq:ktilde_gamma_def}, which can be done numerically.
In order to check the validity of the approximation Eq.~\eqref{eq:simplified_spectral} for calculating the $1\leftrightarrow2$ contributions, 
we have to substitute the propagators~\eqref{eq:spectral_schannel} into $\hat{\Sigma}_{N_1,1\leftrightarrow 2}^\mu$ and resort to numerics when evaluating this object as well as for the momentum averaging. 
Due to the $\delta$-functions in Eq.~\eqref{eq:spectral_schannel}, the calculation of $\hat{\Sigma}_{N_1}^\mu$ requires identifying the propagating modes that solve Eq.~\eqref{eq:lepton_dispersion} for $p^2>0$. 
These are the well-known pseudoparticle and hole excitations,  whose dispersion relations are illustrated in Fig.~\ref{fig:dispersion}. For large $|\mathbf{p}|$, the dispersion relation of the pseudoparticles is indeed captured by the effect of the thermal mass, 
as it is used in the calculations leading to the approximation in Eq.~\eqref{eq:sigma_1to2}. In the same limit, the holes approach a massless dispersion relation, yet they also decouple. 
This can be seen by expressing the spectral lepton propagator in Eq.~\eqref{eq:spectral_schannel} as a sum over contributions proportional to $\delta((p^0)^2- E^2_{\text{part/hole}}(\mathbf{p}))$, 
with ``$\rm part, hole$'' denoting the particle and hole branches, respectively. 
For the hole contributions, the Jacobian that arises from expressing $\delta((p_\mu-\Sigma_{l,\mu}^{{\cal H},{\rm HTL}}(\mathbf{p}))^2)$ 
in terms of $\delta((p^0)^2- E^2_{\text{hole}}(\mathbf{p}))$ becomes exponentially suppressed for large $|\mathbf{p}|$, capturing the decoupling of the holes.

\begin{figure}[h]
\begin{align*}
\left(\raisebox{-0.6cm}{\includegraphics[width=.2\textwidth]{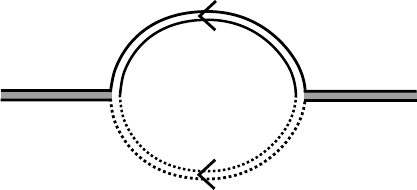}}\right)^{\cal A}=&\,\raisebox{-0.9cm}{\includegraphics[width=.2\textwidth]{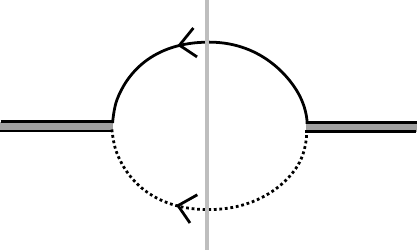}}+\raisebox{-0.9cm}{\includegraphics[width=.2\textwidth]{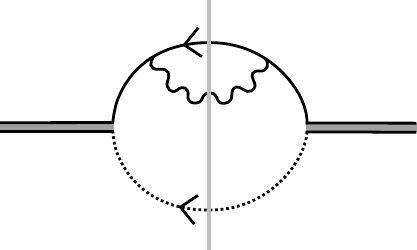}}+\cdots\\
&\hspace{1.25cm}1\rightarrow2\hspace{2.9cm}2\rightarrow2
\end{align*}
\caption{\label{fig:scatterings}\footnotesize Illustration of the origin of the $1\leftrightarrow2$ and  $2\leftrightarrow2$ contributions to the spectral self-energy of the lightest right-handed neutrino. 
Double lines represent resummed propagators, while single lines denote  tree-level propagators with thermal masses. Fermions, scalars and vector fields correspond to solid, dotted or wavy lines, respectively. Sterile-neutrino lines are indicated with a gray shading.}
\end{figure}

We also evaluate the contributions to $\hat{\Sigma}_{N_1}^\mu$ from the region with spacelike lepton propagators, i.e. for $p^2<0$ in Eq.~\eqref{eq:sigma_n_def}, numerically. 
In this case the scalar spectral energy is given by Eq.~\eqref{eq:spectral_schannel}, while the HTL-resummed lepton spectral self-energy in Eq.~\eqref{eq:resummed_spectral} with  nonzero $\Gamma_l$  can no longer be approximated as being proportional to a delta function. 
In this case, with only the scalar propagator forced to be on shell, the contributions to $\hat{\Sigma}_{N_1}^\mu$ can be interpreted as arising from $2\leftrightarrow2$ processes in the thermal plasma involving off-shell lepton exchanges. 
These are the contributions for which earlier works found the $g^2\log g^{-2}$ enhancement from $t$-channel lepton exchanges in the relativistic regime  \cite{1012.3784,1202.1288,1303.5498,Glowna:2015aos}, and which we now evaluate beyond the relativistic approximation. 
For a schematic illustration of the separation of $\hat{\Sigma}_{N_1}^\mu$ into $1\leftrightarrow2$ and $2\leftrightarrow2$ processes,  see Fig.~\ref{fig:scatterings}.

For the numerical computation of the $1\leftrightarrow2$ contributions to the LNC and LNV rates beyond the approximation of Eq.~\eqref{eq:simplified_spectral}, as well as to obtain the $2\leftrightarrow2$ contributions,
we keep the full momentum dependence of the  HTL self-energies of the Standard Model in Eq.~\eqref{eq:HTL_self_energies}, and we proceed to a numerical evaluation of the integration in Eq.~\eqref{eq:sigma_n_l_def}, 
as well as of the momentum averaging necessary to obtain $\gamma_{\rm LNC/LNV}$ from Eqs.~\eqref{eq:gammas_LNC_LNV}, \eqref{eq:thermal_average} and \eqref{eq:ktilde_gamma_def}. 
We employ an adaptive Monte-Carlo integration method, choosing 100 values of $z$ between $10^{-2}$ and $10^{2}$ with even logarithmic spacing. 
In order to determine the values of the couplings, we have chosen a renormalization scale $\mu=T_\text{phys} \sim \unit[10^{12}]{GeV}$ 
and solved the two-loop renormalization group equations of the Standard Model~\cite{Luo:2002ey} imposing initial conditions at low energy that reproduce collider measurements. 
In the matching to experimental results we include two-loop threshold corrections for the Higgs parameters \cite{Degrassi:2012ry}, 
and one-loop electroweak \cite{Hempfling:1994ar} and three-loop QCD corrections \cite{Chetyrkin:1999qi,Melnikov:2000qh} for the determination of the top Yukawa in terms of the top pole mass. 
We use PDG data \cite{Tanabashi:2018oca} except for the Higgs and top masses, which are taken from recent ATLAS measurements: $m_\phi=124.97 $ GeV \cite{Aaboud:2018wps}, and $m_t=172.69$ GeV \cite{Aaboud:2018zbu}. 
This gives as, at the scale $\mu=T_\text{phys}=10^{12}$ GeV,
\begin{align} 
\label{eq:sm_couplings}
\frac{1}{2} g_1^2 + \frac{3}{2} g_2^2 &= 0.546
\ ,
&
h_t &= 0.485
&
\text{and}&
&
\lambda_\phi &= -0.00187
\ .
\end{align}

\begin{figure}[ht]
\centering
\includegraphics[width=0.48\textwidth, trim = 0 10 0 0]{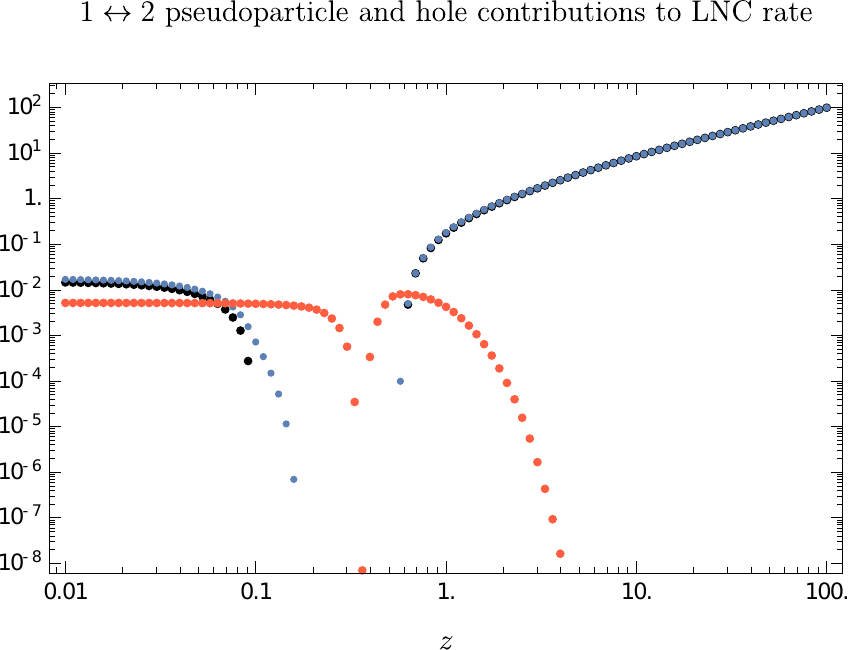}%
\includegraphics[width=0.48\textwidth, trim = 0 10 0 0]{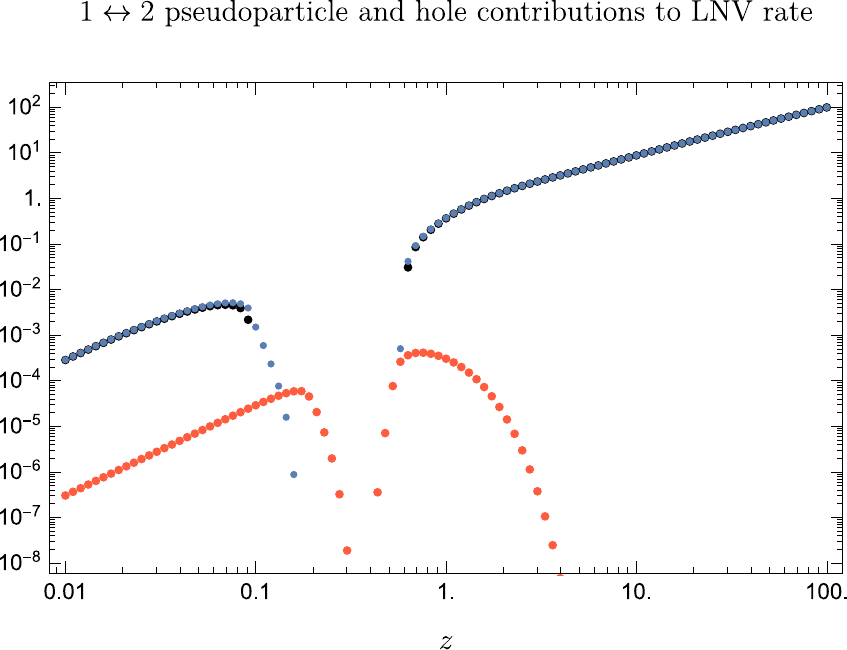}
\caption{\footnotesize\label{fig:12separate}%
$1\leftrightarrow2$ pseudoparticle and hole contributions to the LNC (left) and LNV (right) rates. 
Blue dots give pseudoparticle contributions calculated with the full HTL dispersion relation. 
Black dots give the pseudoparticle contributions obtained with a simplified dispersion relation in terms of the thermal mass $m^2_l$  in Eq.~\eqref{eq:m_thermal}. 
Orange dots show the absolute values of the negative contributions due to holes, obtained with the full HTL dispersion relation.
}
\end{figure}

The final results for the equilibration rates are shown in Figure~\ref{fig:eq_rates_analyt}.
For early times (small $z$),  $\gamma_\text{\rm LNC}$ is dominated by the contributions from $2\leftrightarrow2$ scatterings via lepton exchange and $\gamma_\text{\rm LNV}$ by $1\leftrightarrow2$ processes.
Since the source term $\epsilon_\text{eff} \Gamma$ is proportional to $\gamma_\text{\rm LNV} \cdot \gamma_\text{\rm LNC}$, 
this implies that when $1 \leftrightarrow 2 $ processes are neglected in the relativistic regime, the $CP$-violating source can be underestimated by up to two orders of magnitude. 
This is critical for ultrarelativistic sterile neutrinos, so that the corresponding correction to the source term may be relevant for implementations leptogenesis with ``light'' GeV-scale sterile neutrino, 
 as it has been explored in Refs.~\cite{Hambye:2017elz,Hambye:2016sby,Ghiglieri:2017gjz,Eijima:2017anv,Antusch:2017pkq,Ghiglieri:2017csp,Eijima:2017cxr,Eijima:2018qke,Ghiglieri:2018wbs}.
For both $\gamma_\text{\rm LNC}$ and $\gamma_\text{\rm LNV}$, the rates for $1\leftrightarrow2$ processes become negative and suppressed for intermediate values of $z$ and end up becoming positive again and dominating the rates for $z\gtrsim 1$. 
The suppression is a kinematic effect, taking place when no $1\leftrightarrow2$ processes involving lepton pseudoparticles  can occur on shell. 
For $z\ll1$ the thermal masses of Higgs and leptons in Eq.~\eqref{eq:m_thermal} dominate over the mass $M_1$ of the sterile neutrino, 
so that inverse decays like $\phi \rightarrow lN_1$ are kinematically allowed in the plasma, while decays $N_1\rightarrow \phi l$ are forbidden. 
As the temperature drops, and with it the thermal masses of Higgs bosons and leptons, both Higgs decays and inverse Higgs decays end up becoming blocked. 
However, $\phi \rightarrow N_1 l$ processes involving lepton holes are still allowed, because holes have a lower effective mass than pseudoparticles, as can be seen from the dispersion relation in Figure~\ref{fig:dispersion}. 
Moreover, the contributions of the holes to the rates are negative, which can be understood from the fact that the holes carry negative lepton number, 
so that their creation or destruction in $1\leftrightarrow2$ processes gives rise to changes in $L$ with the opposite sign than those sourced by analogous processes involving lepton pseudoparticles. 
This implies that when hole effects dominate, the $1\leftrightarrow2$ contribution to the rates changes sign, as seen in Fig.~\ref{fig:eq_rates_analyt} for $0.1\lesssim z \lesssim 1$. 
As $z$ increases, decays of $\phi$ into $N_1$ and holes become forbidden, while for $z\gtrsim 0.4$ the decays of $N_1$ into lepton holes and Higgs bosons open up kinematically. 
This explains the drop in the absolute value of the $1\leftrightarrow2$ contribution to the rate for $z\sim 0.4$. 
For $z$ between 0.4 and 1 the decays into holes still dominate the $1\leftrightarrow2$ rates, until for $z\gtrsim1$ the holes decouple, 
as they are produced with very large momentum, and the $N_1\rightarrow \phi l$ decays into lepton pseudoparticles take over, so that the $1\leftrightarrow2$ rate becomes positive again. 
The individual contributions from holes and pseudoparticles to the $1\leftrightarrow2$ rates are shown in Fig.~\ref{fig:12separate}, where the decoupling of holes for $z>1$ becomes apparent.

In Figs.~\ref{fig:eq_rates_analyt} and~\ref{fig:12separate},  we also show the effect of using the full HTL dispersion relation for the lepton pseudoparticles versus simply using the thermal mass $m^2_l$ in Eq.~\eqref{eq:m_thermal}. 
The nontrivial momentum dependence of the Hermitian self-energy of the leptons results in a softening of the kinematic suppression near $z=0.1$, when the inverse decays involving lepton pseudoparticles become blocked.

The relevant rates that appear in the Boltzmann equations \eqref{eq:n_even_boltzmann}--\eqref{eq:y_bl_boltzmann} are shown in Fig.~\ref{fig:rates} together with their nonrelativistic approximations, if applicable. 
As was emphasized at the end of Sec.~\ref{sec:derivation_fluid_eq_rates}, the source rate experiences a large enhancement at $z<0.1$, 
which, as will be seen in the next section, yields a sizable increase in the asymmetry in the weak-washout regime for a vanishing initial abundance of sterile neutrinos. 
Also, the washout rate is somewhat smaller than its nonrelativistic approximation for $z\gtrsim0.2$, which will give rise to a mild enhancement of the asymmetry even in the strong washout regime.

\section{Impact of relativistic and spectator effects on the final $B-L$ asymmetry}
\label{sec:numerical_scans}

Having derived a set of fluid equations and equilibration rates that remain valid throughout the transition from the relativistic to the nonrelativistic regime, %
we now study the impact of early-time relativistic effects and partially equilibrated spectators in models of leptogenesis with heavy sterile neutrinos. 
For this purpose, we perform a parametric study of the final $B-L$ asymmetry, considering both a toy setup without spectators and a realistic scenario for $\tilde M_1$ in the neighbourhood of $\unit[10^{13}]{GeV}$. 

In the toy setup, we neglect all $B-L$ conserving interactions of the Standard Model. 
As a result, there is no way to transfer any net charge from $Y_{l\smallpara}$ to any of the other Standard Model yields. 
Therefore, we can set $Y_{B-L} = Y_{\Delta\smallpara} = -2 Y_{l\smallpara}$, and, for simplicity, we set  $Y_{\phi}=0$, i.e. we ignore the spectator role of the Higgs bosons. 
Doing so, the fluid equations~\eqref{eq:n_even_boltzmann}--\eqref{eq:y_bl_boltzmann} become a closed system that depends only on two free parameters: %
the zero-temperature sterile neutrino decay asymmetry $\epsilon_0$ of Eq.~\eqref{eq:epsilon0} and the washout parameter $K$ of Eq.~\eqref{eq:gamma_av}.
In particular, there is no additional dependence on the mass $\tilde M_1$ of the lightest sterile neutrino, since the $\gamma_\text{\rm LNC}$ and$\gamma_\text{\rm LNV}$ rates depend only on Standard Model parameters.  
In the nonrelativistic limit, we also recover the fluid equations used in Ref.~\cite{hep-ph/0401240}, albeit with the additional prefactor $12/\pi^2$ in the washout rate.

For the realistic scenario, the mass range is chosen such that the sterile neutrinos decay before the weak sphaleron and bottom-Yukawa interactions can fully equilibrate, which happens at temperatures of about $\unit[4\times10^{12}]{GeV}$. 
In this case, one has to solve Eqs.~\eqref{eq:Boltzmann:all:b:sph}, which become a closed system after imposing the chemical constraints of Eq.~\eqref{eq:charge_relations}. 
In addition to being a function of $K$, the asymmetry now depends on $\tilde M_1$ itself, since the spectator rates in Eq.~\eqref{eq:Gammas_b_sph} are functions of $\tilde M_1$ as well as $z$. 
As we are interested in how the partial equilibration of spectator interactions may affect the baryon asymmetry, 
we will compare the asymmetry obtained in the realistic setup with another setup, in which the bottom-Yukawa and weak sphaleron interactions are approximated as fully equilibrated. 
In the latter case the asymmetry is obtained by solving Eqs.~\eqref{eq:Boltzmann:all:toy} with the chemical constraints of Eq.~\eqref{eq:chemical_b_sph}.

In both scenarios, we assume the $B-L$ asymmetry to be vanishing at some initial $z=z_0$ and further take $Y_{N_1\rm odd}(z_0)=0$.

\subsection{Toy setup without spectators}

In order to isolate the effect of the asymmetries generated at early times, when the sterile neutrinos are relativistic, 
we first consider the toy setup without spectator effects, with a particular focus on the weak washout regime. 
Comparing our fluid equations~\eqref{eq:Boltzmann:all:toy} with those used in Ref.~\cite{hep-ph/0401240}, 
we expect to obtain significant corrections in the final $B-L$ asymmetry when thermal corrections to the rates $\gamma_\text{\rm LNC}$ and $\gamma_\text{\rm LNV}$ are of importance. 
As noted before, this should be the case for $K \ll 1$ (i.e. in the weak-washout regime), since then the asymmetries produced at early times are not destroyed by the by the washout. 
In contrast, the final $B-L$ asymmetry obtained when $K \gg 1$ (i.e. the strong-washout regime) is expected to receive no significant corrections compared to the nonrelativistic approximation, %
since the washout destroys all asymmetries existing at early times, and is active until lepton-number violating interactions decouple in the nonrelativistic regime, where the thermal corrections are negligible. 
That being said, we still obtain order one corrections in the strong washout regime due to the extra factor of $12/\pi^2$ in the washout rate compared to the fluid equations used in Ref.~\cite{hep-ph/0401240}.\\

Formally, the general solution to the fluid equations~\eqref{eq:Boltzmann:all:toy} with initial conditions $Y_{N_1\rm odd}(z_0)= Y_{B-L}(z_0) = 0$ can be cast as
\begin{subequations}
\begin{align}
\label{eq:yn_even_sol}
Y_{N_1 \text{even} }(z) 
&= e^{ - I (z,z_0)} Y_{N_1 \text{even} }(z_0) + \int_{z_0}^z \text{d}z' e^{ - I (z,z')} \Gamma (z') Y_{N_1 \text{eq} }(z') \ , 
\\
\label{eq:Y_b_l_sol}
Y \left(z \right) 
&= \int\displaylimits_{z_0}^z \text{d}z'\mathcal{T}  e^{ \int_{z'}^z \text{d}z'' W (z'') } \cdot S ( z' ) \big( Y_{N_1\text{even}} (z') - Y_{N_1, \text{eq}} (z') \big) \ , 
\end{align}
\end{subequations}
where $\mathcal{T}e^{\int dz''W(z'')}$ is a time-ordered matrix exponential and we write in shorthand
\begin{subequations}
\begin{align}
Y &\equiv \left(Y_{N_1 \mathrm{odd}}, \: Y_{B-L}\right)^T \ , 
&
S &\equiv \left(0 , \: -\epsilon_\text{eff} \Gamma \right)^T \ ,
&
W &\equiv 
\begin{pmatrix}
- \Gamma 			&  \frac{\eta_{N_1}}{2} \tilde{\Gamma} \\
 \tilde{\Gamma} 	& - \frac{\eta_{N_1}}{2} \Gamma
\end{pmatrix} 
\ ,
\end{align}
and
\begin{align}
I(z,z') &\equiv \int_{z'}^z \text{d}z'' \Gamma (z'')
\ .
\end{align}
\end{subequations}
To recover the nonrelativistic approximation, one has to take $\tilde \Gamma \to 0$, $\epsilon_\text{eff} \to \epsilon_0$ and use the augmented $\Gamma$ as in Eq.~\eqref{eq:gamma_augmented}. 
Using that the sterile neutrino yield should then be helicity-symmetric, one also finds the constraints $Y_{N_1\rm odd}=0$ and $Y_{N_1\rm even}=Y_{N_1\rm +}=Y_{N_1\rm -}\equiv Y_{N_1}$. 

When parametrizing the final $B-L$ asymmetry, we follow Ref.~\cite{Barbieri:1999ma} and define the efficiency factor $\kappa_f$ via
\begin{align}
\label{eq:kappa_f_def}
Y_{B-L} (z \to \infty) 
&= - \epsilon_0 \ Y_{N_1, \text{eq}} (z_0) \cdot \kappa_f \ .
\end{align} 
Combining the formal solution in Eq.~\eqref{eq:Y_b_l_sol} with Eq.~\eqref{eq:n_even_boltzmann} then implies
\begin{align}
\label{eq:kappa_exp}
\kappa_f \equiv \kappa (z \to \infty),\quad\kappa (z) &\equiv - \frac{1}{Y_{N_1, \text{eq} }(z_0)} \int\displaylimits_{z_0}^z \text{d}z' \left[ \mathcal{T} e^{ \int_{z'}^z \text{d}z'' W (z'') } \right]_{22} \cdot \frac{\epsilon_\text{eff} (z')}{\epsilon_0} \frac{\text{d}}{\text{d} z'}  Y_{N_1 \text{even}} (z')
\ .
\end{align}
Notice that in general $\kappa (z)$ depends on the washout parameter $K$ but not on $\epsilon_0$, since $\epsilon_\text{eff} \propto \epsilon_0$. 
The normalization of $\kappa (z)$ is chosen such that one recovers $\kappa_f = 1 + O(K)$ for initially equilibrated sterile neutrinos in the nonrelativistic approximation.
Explicitly, 
\begin{align}
\label{eq:kappanr}
\kappa^\text{nr} (z) &\equiv - \frac{1}{Y_{N_1, \text{eq} }(z_0)} \int\displaylimits_{z_0}^z \text{d}z' \frac{\text{d}}{\text{d} z'}  Y_{N_1 } (z') e^{ -1/2\int_{z'}^z \text{d}z''\eta_{N_1}(z'')\Gamma (z'')} 
\ .
\end{align}
In the two limiting cases $K\ll 1 $ and $K \gg 1$, it is possible to approximate $\kappa_f$ analytically. 
For this, we consider both the general solution~\eqref{eq:kappa_exp} as well as its nonrelativistic approximation~\eqref{eq:kappanr}.
As we will reiterate below, one should keep in mind that the nonrelativistic approximation is only valid for the strong washout regime, 
so that its application to weak washout is only done here in order to assess the impact of the relativistic corrections. \\

For the nonrelativistic case, we follow the derivation in Ref.~\cite{hep-ph/0401240} while keeping track of the additional prefactor $\nicefrac{12}{\pi^2}$ in front of the washout rate.
As it is done there, we first write the efficiency factor as a sum of contributions from production and decays, 
\begin{align}
\kappa^\text{nr}_f = \kappa^\text{nr}_+ + \kappa^\text{nr}_-
\ .  
\end{align}
Defining $z_{\rm eq}$ to the value of $z$ at which $Y_{N_1}(z)=Y_{N_1\rm eq}(z)$, one finds $ \kappa^\text{nr}_+ =  \kappa_f^\text{nr} - \kappa^\text{nr} (z_\text{eq}) $ and $ \kappa^\text{nr}_- =  \kappa_f^\text{nr} (z_\text{eq}) $.
To estimate $z_{\rm eq}$, we substitute the nonrelativistic rate $\Gamma$ from Eq.~\eqref{eq:Gammanr} and solve
\begin{align}
\label{eq:zeq}
  (z_{\rm eq})^2 K_2(z_{\rm eq}) \overset{!}{=} K\int_0^{z_{\rm eq}} dz' {z'}^3  K_1(z').
\end{align}

In the strong washout regime, we further proceed as in Ref.~\cite{hep-ph/0401240} by noting that the largeness of the equilibration rate $\Gamma \propto K$ ensures
that the sterile neutrino yield will reach and track the equilibrium distribution at very early times, so that $z_\text{eq} \ll 1$.  
Similarly, the large washout rate will erase any asymmetry produced before equilibration, so that the asymmetry arises largely from decays of the sterile neutrinos occurring after the washout becomes Boltzmann suppressed. 
Thus, starting from the nonrelativistic efficiency factor of Eq.~\eqref{eq:kappanr}, and assuming that $Y_{N_1} \approx Y_{N_1, \text{eq}}$, one may write
\begin{align}
\kappa(z)^\text{nr} \approx\frac{1}{2}\int_0^z dz' {z'}^2 K_1(z') e^{-3K/\pi^2\int_{z'}^z dz''{z''}^3K_1(z'')}\equiv \frac{1}{2}\int_0^z dz'e^{-{\cal I}(z')} \ .
\end{align}
As in Ref.~\cite{hep-ph/0401240}, we may approximate the integrand around a saddle point $z'=\bar{z}$ of ${\cal I}(z')$ satisfying
\begin{align}
\label{eq:saddle}
\frac{3}{\pi^2} \bar{z}^3K_1(\bar z)+\frac{3}{\bar z}-\frac{K_2(\bar z)}{K_1(\bar z)}=0 \ .
\end{align}
One can then get a simple estimate of the integral as follows,
\begin{align}
\kappa(z)\approx\frac{1}{2\bar z}\int_0^z dz' {z'}^3 K_1(z') e^{-3K/\pi^2\int_{z'}^z dz''{z''}^3K_1(z'')}=\frac{\pi^2}{6K\bar z}\left(1-e^{-3K/\pi^2\int_{z'}^z dz''{z''}^3K_1(z'')}\right),
\end{align}
so that the strong washout asymmetry, regardless of initial conditions, is given by
\begin{align}
\label{eq:efficiency_strong_washout}
\kappa_f^\text{nr} \approx \kappa^\text{nr}_+ \approx \frac{\pi^2}{6 K\bar z},\quad K\gg1 \ .
\end{align}

In the weak washout regime, the only contribution to $\kappa_f$ that is not suppressed by factors of the washout rate $K$ is 
\begin{align}
\label{eq:efficiency_Y0_nr}
\kappa_f^\text{nr} \approx\frac{Y_{N_1 \text{even} }(z_0) }{Y_{N_1, \text{eq} }(z_0)} + O(K) \ ,
\end{align}
so that there is no final $B-L$ asymmetry for a vanishing initial abundance when taking $K\to 0$. 
To obtain the leading-order expression for $\kappa_f^\text{nr}$ when $Y_{N_1 \text{even} }(z_0) = 0$, we again proceed as in Ref.~\cite{hep-ph/0401240}.
First, we consider $z < z_{\rm eq}$. 
For these values of $z$, the neutrino yield is small and one can ignore it on the right-hand side of the fluid equation~\eqref{eq:n_even_boltzmann_nr} for $Y_{N_1}$. 
Using the expression for $Y_{N_1\rm eq}$ in Eq.~\eqref{eq:Yeq_eta} together the approximation on the right-hand side of Eq.~\eqref{eq:I_exp} for $\mathcal{I}(z)$, one finds 
\begin{align}
\label{eq:YNzminus}
\frac{Y_{N_1}(z)}{Y_{N_1\rm eq}(z_0)}\approx\frac{1}{2}\int_0^z dz' \Gamma(z'){z'}^2 K_2(z'),\quad z<z_{\rm eq} \ .
\end{align}
Inserting \eqref{eq:YNzminus} into Eq.~\eqref{eq:kappanr} and using the expression for $\eta_{N_1}$ in \eqref{eq:Yeq_eta} together the approximation on the right-hand side of Eq.~\eqref{eq:J_exp} for $\mathcal{J}(z)$, one obtains
\begin{equation}
\label{eq:kappaminus}
\begin{aligned}
\kappa^{0,\text{nr}}(z)\approx&\,-\frac{1}{2}\int_0^z\,\Gamma(z'){z'}^2K_2(z')e^{-3/\pi^2\int_{z'}^zdz''\Gamma(z''){z''}^2K_2(z'')}
\\
\approx&\,-\frac{\pi^2}{6}(1-e^{-6/\pi^2 Y_{N_1}(z)/Y_{N_1\rm eq}(z_0)}),\quad z<z_{\rm eq}
\ .
\end{aligned}\end{equation}
Next, we exploit that for small $K$ one has $z_{\rm eq}\gg1$ (see Eq.~\eqref{eq:zeq}), so that the washout rate can be ignored for $z>z_{\rm eq}$. 
As a result, one finds
\begin{align}
\kappa^{0,\text{nr}} (z) - \kappa^{0,\text{nr}} (z_\text{eq}) \approx\frac{1}{Y_{N_1\rm eq}(z_0)}(Y_{N_1}(z)-Y_{N_1}(z_{\rm eq})),\quad z>z_{\rm eq},
\end{align}
where $\kappa(z_{\rm eq}) = \kappa^{0,\text{nr}}_-$ is given by Eq.~\eqref{eq:kappaminus}. 
Thus, the efficiency factor $\kappa(z\rightarrow\infty)$ becomes
\begin{align}
\label{eq:efficiency_weak_washout_nr}
\kappa_f^{0,\rm nr}\approx \frac{Y_{N_1}(z_{\rm eq})}{Y_{N_1\rm eq}(z_0)}-\frac{\pi^2}{6}\left(1-e^{-6/\pi^2 Y_{N_1}(z_{\rm eq})/Y_{N_1\rm eq}(z_0)}\right)\approx \frac{3 }{\pi^2}\left(\frac{Y_{N_1}(z_{\rm eq})}{Y_{N_1\rm eq}(z_0)}\right)^2\approx 1.65\, K^2
\ .
\end{align}
In the last equality, we have used a numerical fit to Eq. \eqref{eq:zeq} to find ${Y_{N_1}(z_{\rm eq})}/{Y_{N_1\rm eq}(z_0)}\approx 2.33K$ for $K\ll1$.\\

Now we move on to the semianalytical approximation of the fully relativistic efficiency factor $\kappa_f$ given in Eq.~\eqref{eq:kappa_exp}.
We only need to consider the weak washout regime, since relativistic corrections are expected to be negligible for $K \gg 1$. 
For this purpose, we start by setting $K \to 0$ in Eq.~\eqref{eq:kappa_exp}.
Then, one recovers the same result as in the nonrelativistic case, 
\begin{align}
\label{eq:efficiency_Y0}
\kappa_f \approx\frac{Y_{N_1 \text{even} }(z_0) }{Y_{N_1, \text{eq} }(z_0)} + O(K) \ .
\end{align}
To otain the leading-order expression for $\kappa_f$ when $Y_{N_1 \text{even}} = 0$, we set $W\to0$, giving
\begin{align}
\label{eq:kappa_weak_washout}
\kappa (z)
&\approx - \frac{1}{Y_{N_1, \text{eq} }(z_0)} \bigg[
Y_{N_1\rm even}(z) %
+ \int\displaylimits_{z_0}^z \hspace{-3pt} \text{d}z' \hspace{3pt} \Big( \frac{\epsilon_\text{eff} (z')}{\epsilon_0} - 1 \Big) \frac{\text{d}}{\text{d} z'} Y_{N_1 \text{even} }(z') \bigg]
\ .
\end{align}
This expression
can be simplified by noticing that the integrand in the second term is dominated by values of $z' \lesssim 1$ because $\epsilon_\text{eff} \to \epsilon_0$ when relativistic corrections are small. 
To approximate $\nicefrac{\text{d} Y_{N_1 \text{even} }(z')}{\text{d} z'}$, we use Eq.~\eqref{eq:n_even_boltzmann} and reinsert the formal solution \eqref{eq:yn_even_sol} for $Y_{N_1\rm even}(z')$ on the right-hand side. 
Expanding around $\Gamma (z')=0$, one finds
\begin{align}
\label{eq:dYz_approx}
\frac{\text{d}}{\text{d} z'} Y_{N_1 \text{even} }(z')
\approx \Gamma (z') 
Y_{N_1 \text{eq} }(z') 
\ .
\end{align}
Inserting Eq.~\eqref{eq:dYz_approx} into Eq.~\eqref{eq:kappa_weak_washout}, we obtain
\begin{equation}
\label{eq:kappa_weak_washout_2}
\kappa (z) \approx 
- \frac{Y_{N_1 \text{even} }(z)}{Y_{N_1, \text{eq} }(z_0)} %
- \int\displaylimits_{z_0}^z \hspace{-3pt} \text{d}z' \hspace{3pt} \Big( \frac{\epsilon_\text{eff} (z')}{\epsilon_0} - 1 \Big) \Gamma (z') \frac{Y_{N_1 \text{eq} }(z')}{Y_{N_1, \text{eq} }(z_0)}
\ .
\end{equation}
To recover the final efficiency factor, we now have to take $z \to \infty$, for which only the integral in the second term contributes. 
To estimate this integral, we use expressions \eqref{eq:gamma_av}, \eqref{eq:epsilon_eff_def} and our numerical results for the rates $\gamma_{\rm LNC/LNV}$. 
Choosing $z_0=0.01$ as the lower bound of the integral, 
we find
\begin{align}
\label{eq:efficiency_weak_washout}
\kappa_f^0\equiv \kappa_f|_{Y_{N_1}(z_0)=0}\approx - \int\displaylimits_{z_0}^\infty \hspace{-3pt} \text{d}z \hspace{3pt} \Gamma (z) \frac{Y_{N_1, \text{eq}}(z)}{Y_{N_1, \text{eq} }(z_0)} \Big( \frac{\epsilon_\text{eff} (z)}{\epsilon_0} - 1 \Big)
\approx - 0.32 \cdot K.
\end{align}
There are two main reasons for our choice of $z_0$: firstly, it corresponds to the minimum value of $z$ for which we have numerically evaluated the rates, 
and secondly, it is consistent with our choice of initial $z$ for carrying out the numerical integration of the fluid equations.
That being said, the dependence of the integral on $z_0$ is very mild. 
For example, extrapolating our numerical rates to lower values of $z$ and choosing $z_0=0$, the $\kappa^0_f$ coefficient increases only by about 4\%.
Notice that this result is qualitatively different from its nonrelativistic counterpart given in Eq.~\eqref{eq:efficiency_weak_washout_nr} for $Y_{N_1 \text{even}} = 0$ initial conditions. 
First, we have fully neglected the washout, setting $W \to 0$. 
Within the nonrelativistic approximation, this would have resulted in a vanishing efficiency factor. 
To see this, recall that the source of the $B-L$ asymmetry is proportional to $\text{d}Y_{N_{\rm even}}/\text{d}z$, so that the $B-L$ asymmetry created during the production of the sterile neutrinos is exactly cancelled out by an opposite-sign asymmetry created during the decay of the sterile neutrinos. 
If the washout is taken into account, a part of the symmetry created during the production of the sterile neutrinos is already washed out by the time the sterile neutrinos start to decay. 
As a result, one is left with a small net efficiency factor of positive sign once the sterile neutrinos have decayed.
Quantitatively, this net efficiency factor is at least quadratic in $K$, with one factor of $K$ resulting from $\text{d} Y_{N_{\rm even}}/\text{d}z \propto K$ another factor of $K$ coming from the nonvanishing washout that needs to be included to obtain a finite result.
In contrast, the efficiency factor \eqref{eq:efficiency_weak_washout} for the initial conditions $Y_{N_1 \text{even}} = 0$ in the fully relativistic description is nonvanishing even when $W\to0$. 
This is due to the relativistic corrections to $\epsilon_\text{eff}$ resulting in $\epsilon_\text{eff} > \epsilon_0$ for small $z$, as illustrated in Fig.~\ref{fig:rates}.
As a result, the asymmetry sourced during the production of sterile neutrinos is larger than the opposite-sign asymmetry generated during their decay, giving a finite efficiency factor of negative sign. 
In addition to this sign flip, the asymmetry is generated at first order in the washout parameter, with the sole factor of $K$ coming from $\text{d} Y_{N_{\rm even}}/\text{d}z \propto K$ as $\nicefrac{\epsilon_\text{eff}}{\epsilon_0}$ is independent of $K$. \\

For the evaluation of $\kappa_f$ in between the weak and strong washout regimes, we resort to a numerical integration of the fluid equations. 
From a practical point of view, we need to compute $\kappa (z)$  for some $z > z_f$, where $T_f = \nicefrac{M_1}{z_f}$ is the is the so-called ``freeze-out'' temperature, below which the asymmetry becomes approximately constant in time. 
In the strong washout regime, we may keep using the nonrelativistic result of Ref.~\cite{hep-ph/0401240}, 
\begin{align}
\label{eq:freeze_out_temp}
z_f  \approx  1 + \frac{1}{2} \log \left( 1 + \frac{\pi K^2}{1024} \log^5 \left(\frac{3125 \pi K^2}{1024} \right) \right) \lesssim 10^2 \ .
\end{align}
In the weak washout regime, we need a different approach to estimate $z_f$. 
Generically, freeze-out occurs for $z_f \gg 1$ due to the long lifetime $\tau \sim \nicefrac{1}{T \cdot K}$ of the sterile neutrinos.
A quantitative statement can be made by considering the relative deviation $\delta(z)$ of $\kappa(z)$ from $\kappa_f=\kappa(z \to \infty)$, 
\begin{align}
\label{eq:delta_f_def}
\delta(z) \equiv \frac{\kappa_f - \kappa (z)}{\kappa_f}
\ .
\end{align}
Using $\delta(z)$, we define the freeze-out time $z_f$ to be the smallest $z$ which fulfills the condition
\begin{equation}
\abs{\delta (z_f) } < r \ ,
\end{equation}
where $r \ll 1$ is chosen relative accuracy at which one wishes to evaluate $\kappa_f$.
For example, for $r \approx 10^{-2}$, we need to integrate at least up to the corresponding $z_f$ 
in order to achieve a one-percent accuracy in evaluating the fluid equations, apart from the additional theoretical uncertainties. 
Our aim is now to give an upper bound on the order of magnitude of $z_f$. 
Using expression \eqref{eq:kappa_exp} and neglecting the washout as before, we find
\begin{align}
\label{eq:delta_f_exp}
\delta(z)
&\approx \frac{1}{\kappa_f} \bigg[ \frac{Y_{N_1 \text{even} }(z)}{Y_{N_1, \text{eq} }(z_0)} 
- \int\displaylimits_{z}^\infty \hspace{-3pt} \text{d}z' \hspace{3pt} \Big( \frac{\epsilon_\text{eff} (z')}{\epsilon_0} - 1 \Big) \frac{\text{d}}{\text{d} z'} \frac{Y_{N_1 \text{even} }(z')}{Y_{N_1, \text{eq} }(z_0)} \bigg] 
\ .
\end{align}
Inserting the solution~\eqref{eq:yn_even_sol} for $Y_{N_1 \text{even}}(z)$ and expanding to first order in $K$, this gives
\begin{align}
\label{eq:delta_f_exp_2}
\delta(z)
&\approx \frac{1}{\kappa_f} \bigg[ 
e^{ - I (z,z_0)} \frac{Y_{N_1 \text{even} }(z_0)}{Y_{N_1, \text{eq} }(z_0)} + \int_{z_0}^z \text{d}z' e^{ - I (z,z')} \Gamma (z') \frac{Y_{N_1 \text{eq} }(z')}{Y_{N_1, \text{eq} }(z_0)} \\
\nonumber
&\hspace{50pt} 
+ \int\displaylimits_{z}^\infty \hspace{-3pt} \text{d}z' \hspace{3pt} \Big( \frac{\epsilon_\text{eff} (z')}{\epsilon_0} - 1 \Big) \Gamma(z') \Big( e^{ - I (z',z_0)} \frac{Y_{N_1 \text{even} }(z_0)}{Y_{N_1, \text{eq} }(z_0)} - \frac{Y_{N_1 \text{eq} }(z')}{Y_{N_1, \text{eq} }(z_0)} \Big)
\bigg]
\ .
\end{align}
Even though the exponential suppression factor $I(z,z_0)$ is formally of order $K$, it may not be neglected, since it is necessary to ensure that all terms go to zero as $z \to \infty$. 
In fact, one finds
\begin{align}
\label{eq:I_approx}
I (z, z_0) &= \int_{z_0}^z \text{d}z' \Gamma (z') = \frac12 K z^2 \Big( 1 + O(\nicefrac{1}{z}) \Big) \overset{z \to +\infty}{\longrightarrow} + \infty \ .
\end{align}
Since we expect $z_f\gg1$ in the weak washout regime, we drop the next-to-leading order corrections to $I (z, z_0) $.

\begin{figure}
\includegraphics[width = 0.49\textwidth]{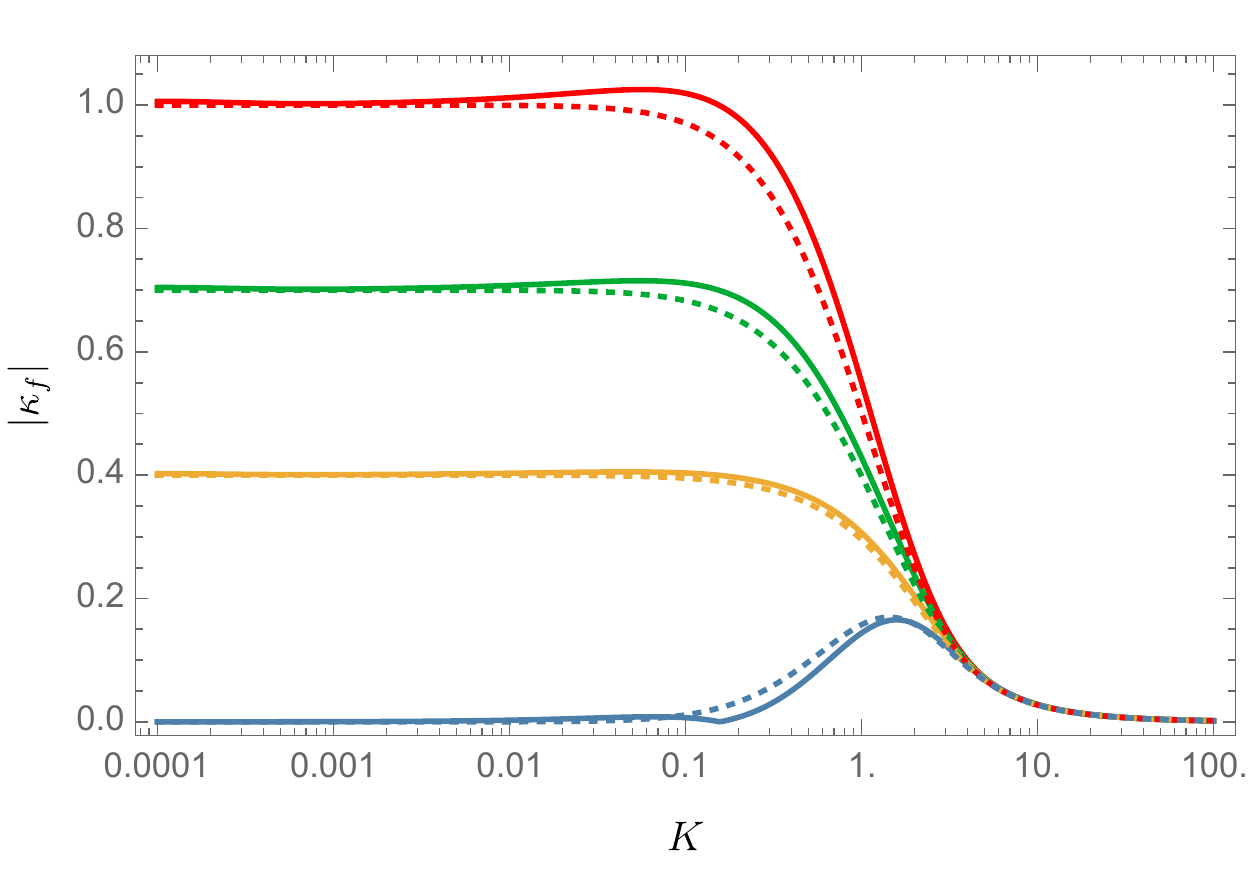}
\includegraphics[width = 0.5\textwidth]{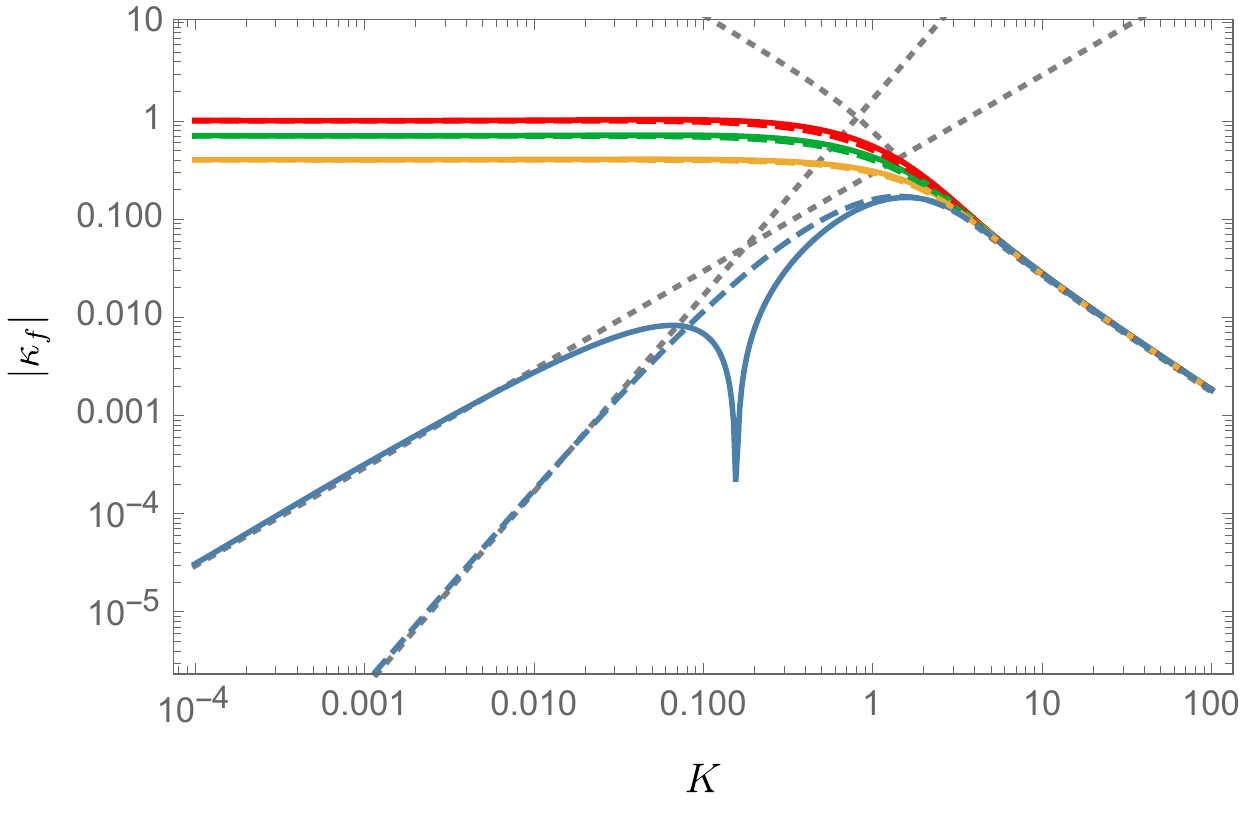}
\caption{\footnotesize\label{fig:eff_scan}%
Numerical scan of $\kappa_f$ vs $K$ for various initial conditions, comparing the nonrelativistic approximation (coloured dashed lines) with the fully relativistic result (solid lines).  
The initial conditions at $z=0.01$ are $Y_{N_1}/Y_{N_1\rm eq}=1$ (red), 0.7 (green), 0.4 (orange) and 0 (blue). 
On the left: single logarithmic plot; on the right: double logarithmic plot. 
On the right, the dashed gray  lines  correspond to the weak washout estimate in Eqs.~\eqref{eq:efficiency_weak_washout}, \eqref{eq:efficiency_weak_washout_nr} 
and the strong washout estimate of Eqs.~\eqref{eq:efficiency_strong_washout}, \eqref{eq:saddle}.}
\end{figure}

To proceed, we evaluate the right-hand side of Eq.~\eqref{eq:delta_f_exp_2} term by term. 
Using the approximation~\eqref{eq:I_approx}, the first term immediately becomes
\begin{equation}
\label{eq:term_1_est}
\frac{1}{\kappa_f} e^{ - I (z,z_0)} \frac{Y_{N_1 \text{even} }(z_0)}{Y_{N_1, \text{eq} }(z_0)} 
\approx \frac{1}{\kappa_f} e^{ - \frac12 K z^2} \frac{Y_{N_1 \text{even} }(z_0)}{Y_{N_1, \text{eq} }(z_0)} \ , 
\end{equation}
so that $z_f \gtrsim \nicefrac{1}{\sqrt{K}} \gg 1$, at least for nonvanishing initial conditions for the sterile neutrinos. 
To find an upper bound on $z_f$, it is sufficient to evaluate the remaining terms in this regime. 
For the second term, taking $z_f \gtrsim \nicefrac{1}{\sqrt{K}}$ enables us to split the integral into two regions with $z' < \tilde z$ and $\tilde z < z'$, where $\tilde z$ is chosen such that $1 \ll \tilde z \ll \nicefrac{1}{\sqrt{K}}$.
In the first region, we may bound the integral as
\begin{align}
\int_{z_0}^{\tilde z} \text{d}z' e^{ - I (z,z')} \Gamma (z') \frac{Y_{N_1 \text{eq} }(z')}{Y_{N_1, \text{eq} }(z_0)}
&\approx e^{- \frac12 K z^2} \int_{z_0}^{\tilde z} \text{d}z' \Gamma (z') \frac{Y_{N_1 \text{eq} }(z')}{Y_{N_1, \text{eq} }(z_0)}
\label{eq:term_2_1_est}
\lesssim 2.65 \, K \, e^{- \frac12 K z^2} \ . 
\end{align}
In the second region, we may use the nonrelativistic approximation to evaluate the integrand since $\tilde z \gg 1$. 
Explicitly, we take
\begin{align}
\label{eq:nr_approx}
\Big( \frac{\epsilon_\text{eff} (z)}{\epsilon_0} - 1 \Big) \Gamma (z) &\approx \frac{3}{2z} K \ ,
&
\frac{Y_{N_1\rm eq}(z)}{Y_{N_1\rm eq}(0)} &\approx \frac{1}{2}\sqrt{\frac{\pi}{2}} z^{3/2}e^{-z} \ .
\end{align}
Inserting these expressions into the integral, we find
\begin{align}
\int_{\tilde z}^{z} \text{d}z' e^{ - I (z,z')} \Gamma (z') \frac{Y_{N_1 \text{eq} }(z')}{Y_{N_1, \text{eq} }(z_0)}
&\approx \frac{1}{2}\sqrt{\frac{\pi}{2}} K \int_{\tilde z}^{z} \text{d}z' {z'}^{1/2} e^{- \frac12 K (z^2 - {z'}^2) - z'}
\label{eq:term_2_2_est}
\lesssim \frac{1}{2} K \sqrt{\frac{\pi z}{2}} e^{- \frac12 K z^2}
\ .
\end{align}
For the third term, we again take $z \gg 1$ to use the approximations~\eqref{eq:I_approx} and~\eqref{eq:nr_approx} to estimate the integral. 
Inserting Eqs.~\eqref{eq:I_approx} and~\eqref{eq:nr_approx} into the both parts of the integrand, one obtains
\begin{subequations}
\begin{align}
\nonumber
\int\displaylimits_{z}^\infty \hspace{-3pt} \text{d}z' \hspace{3pt} \Big( \frac{\epsilon_\text{eff} (z')}{\epsilon_0} - 1 \Big) \Gamma(z') e^{ - I (z',z_0)} \frac{Y_{N_1 \text{even} }(z_0)}{Y_{N_1, \text{eq} }(z_0)}
&\approx \frac32 K \int\displaylimits_{z}^\infty \hspace{-3pt} \text{d}z' \hspace{3pt} \frac1{z'} e^{ - \frac12 K z'^2} \frac{Y_{N_1 \text{even} }(z_0)}{Y_{N_1, \text{eq} }(z_0)} \\
\label{eq:term_3_est}
&\approx \frac32 \frac1{z^2} e^{- \frac12 K z^2} \frac{Y_{N_1 \text{even} }(z_0)}{Y_{N_1, \text{eq} }(z_0)} \ ,
\\
\nonumber
\int\displaylimits_{z}^\infty \hspace{-3pt} \text{d}z' \hspace{3pt} \Big( \frac{\epsilon_\text{eff} (z')}{\epsilon_0} - 1 \Big) \Gamma(z') \frac{Y_{N_1 \text{eq} }(z')}{Y_{N_1, \text{eq} }(z_0)}
&\approx \frac34 K \, \sqrt{\frac{\pi}{2}} \int\displaylimits_{z}^\infty \hspace{-3pt} \text{d}z' \hspace{3pt} z'^{1/2}e^{-z'} \\
\label{eq:term_4_est}
&\approx \frac34 K \, \sqrt{\frac{\pi z}{2}} e^{-z} 
\ .
\end{align}
\end{subequations}
Substituting the estimates~\eqref{eq:term_1_est}, \eqref{eq:term_2_1_est} , \eqref{eq:term_2_2_est}, \eqref{eq:term_3_est} and~\eqref{eq:term_4_est} into expression \eqref{eq:delta_f_exp_2} for $\delta (z)$, we find
\begin{subequations}
\begin{align}
\label{eq:delta_f_estimate}
\abs{\delta(z)}
&\lesssim \frac{1}{\kappa_f} \bigg[ 
\frac{Y_{N_1 \text{even} }(z_0)}{Y_{N_1, \text{eq} }(z_0)} e^{ - \frac12 K z^2} \big( 1 + \frac32 \frac1{z^2} \big) + K e^{- \frac12 K z^2} \big( \frac{1}{2}\sqrt{\frac{\pi z}{2}} +  2.65 \big) + \frac34 K \, \sqrt{\frac{\pi z}{2}} e^{-z} 
\bigg] \\
&\approx \frac{1}{\kappa_f} \bigg[ 
\frac{Y_{N_1 \text{even} }(z_0)}{Y_{N_1, \text{eq} }(z_0)} e^{ - \frac12 K z^2} + \frac{1}{2} K \sqrt{\frac{\pi z}{2}} e^{- \frac12 K z^2} \bigg] 
\ ,
\end{align}
\end{subequations}
where we have again used that $z \gg 1$.
For nonvanishing initial conditions and $r \approx 10^{-2}$, only the first term is relevant, and we obtain $z^2_f \lesssim \nicefrac{10}{K}$. 
For vanishing initial conditions and $r \approx 10^{-2}$, we obtain $z_f \lesssim \nicefrac{1}{K}$.\\
\begin{figure}
\includegraphics[width = 0.5\textwidth]{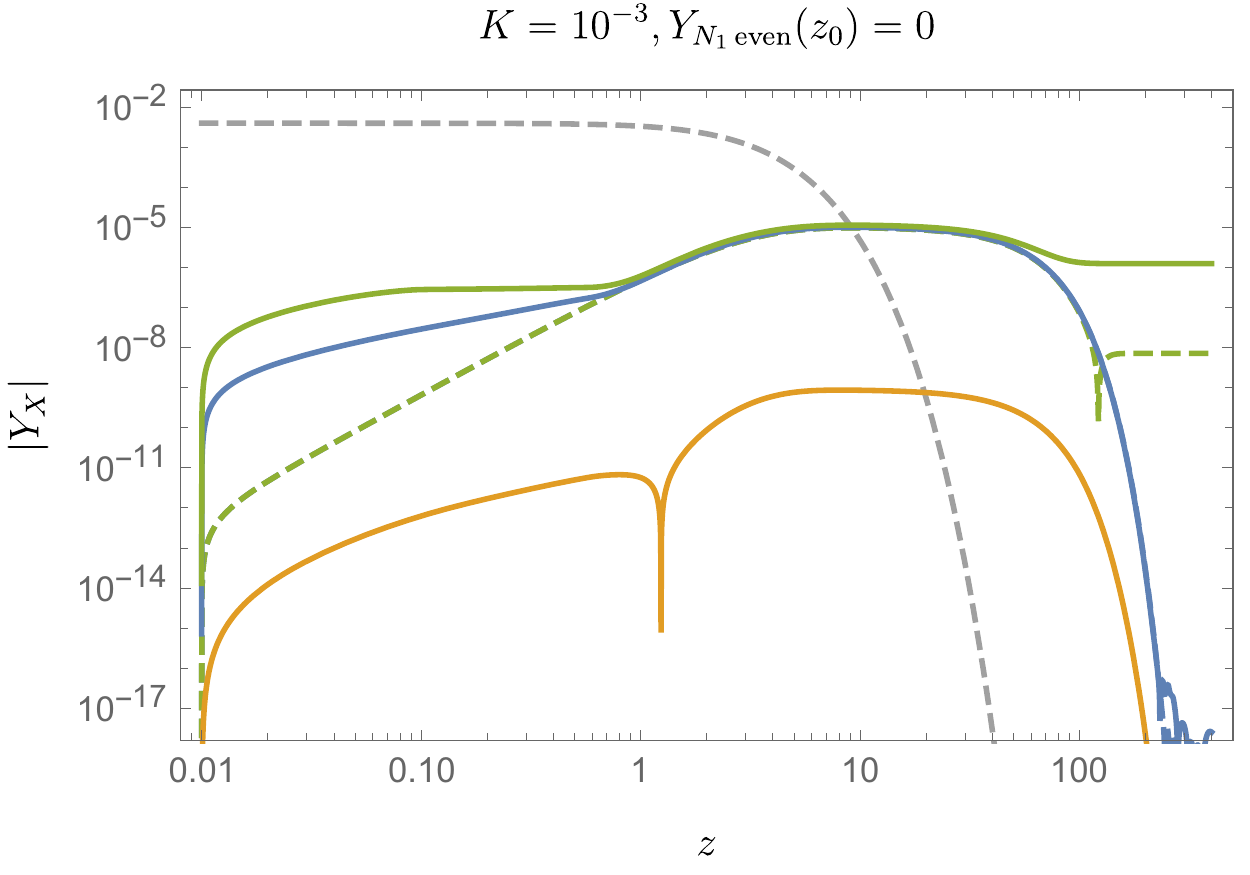}
\includegraphics[width = 0.5\textwidth]{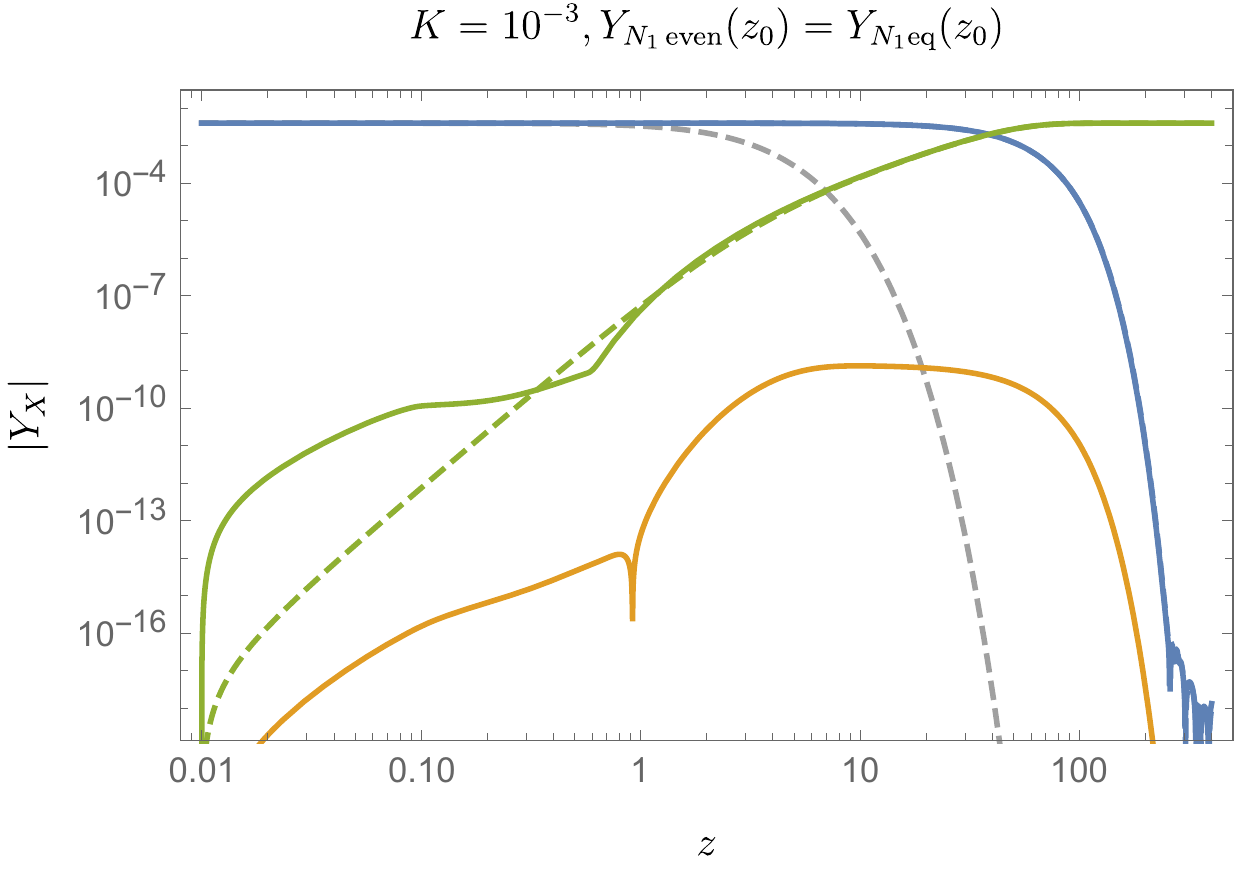}\vskip0.1cm
\includegraphics[width = 0.5\textwidth]{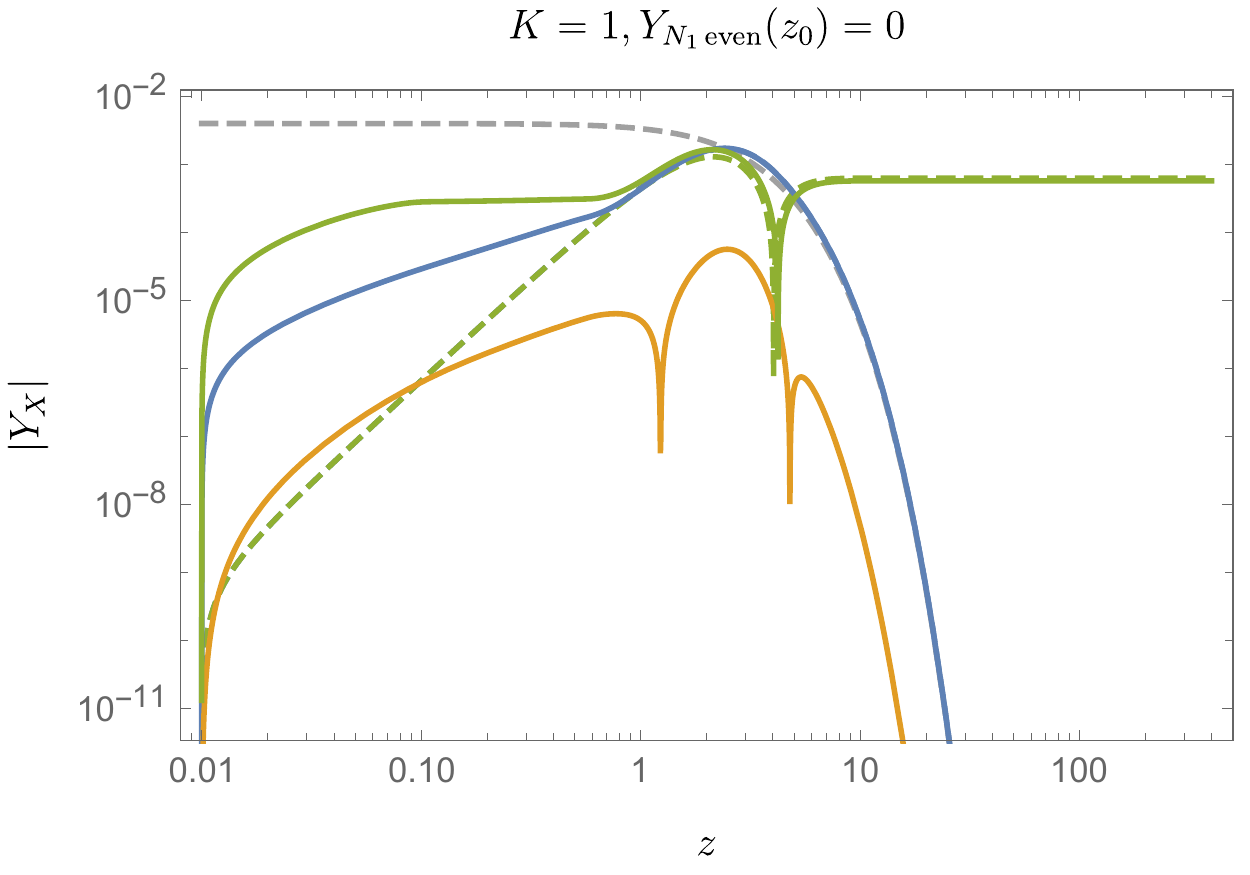}
\includegraphics[width = 0.5\textwidth]{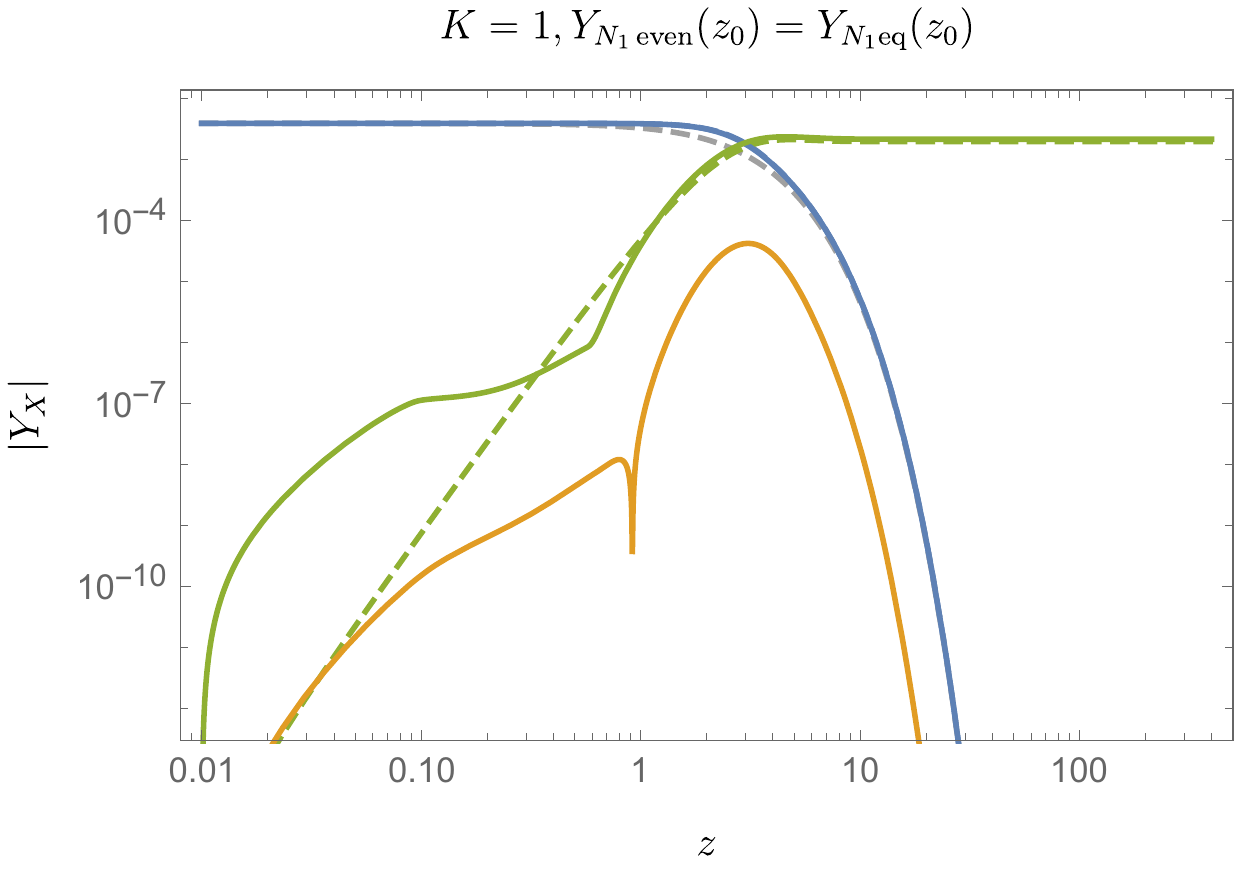}\vskip0.1cm
\includegraphics[width = 0.5\textwidth]{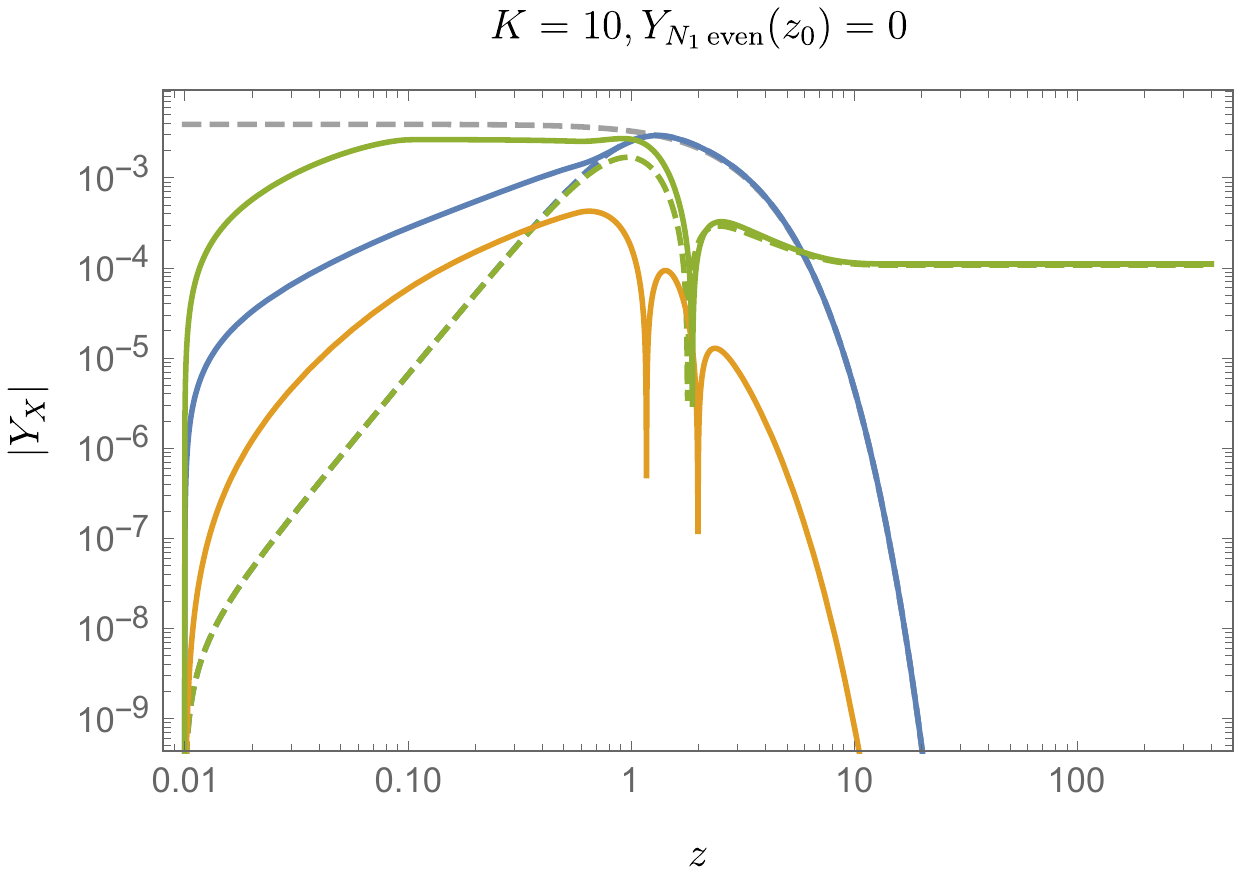}
\includegraphics[width = 0.5\textwidth]{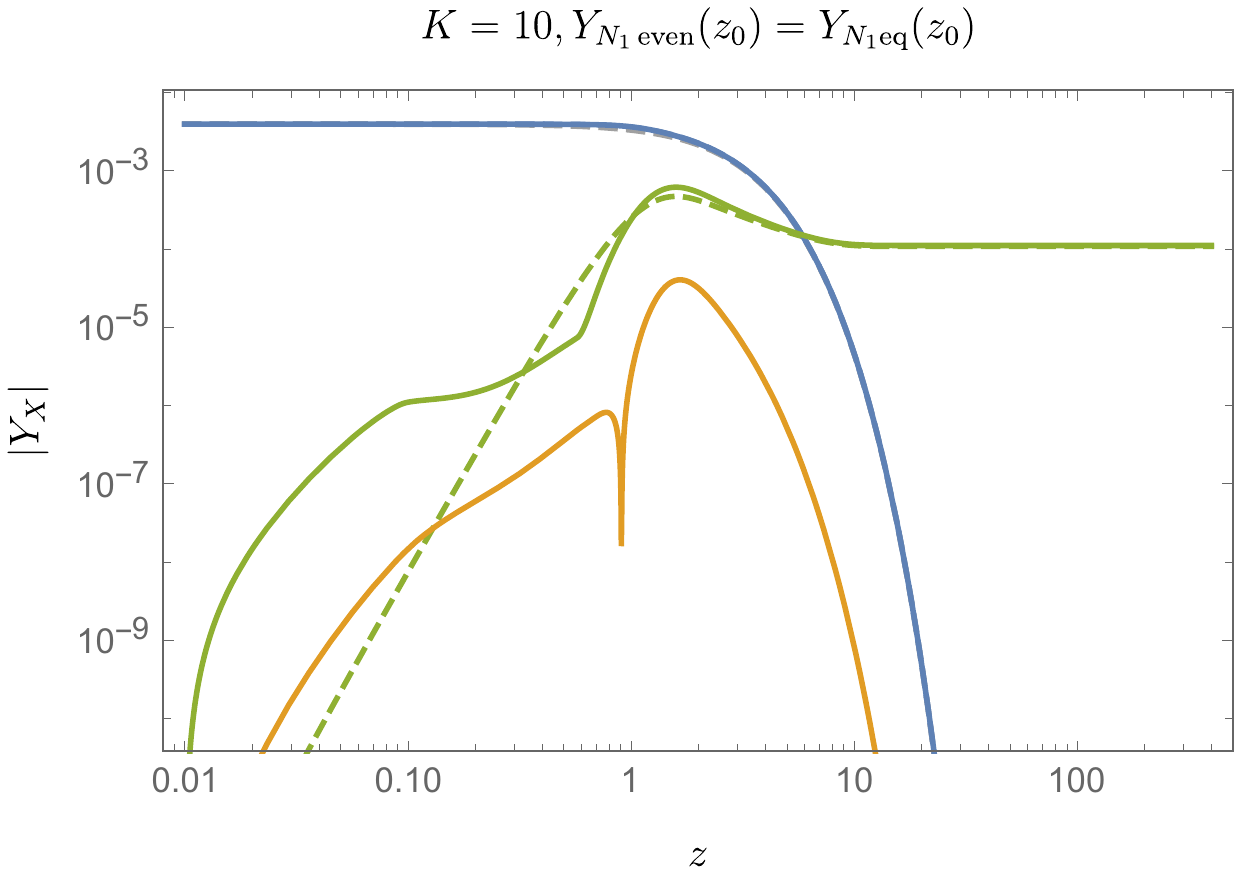}
\caption{\footnotesize\label{fig:Ys}%
Time evolution of the yields obtained by solving the fluid equations, for $K=0.001$ (top), 1 (middle) and 10 (bottom), as well as vanishing (left) and equilibrium  (right) initial conditions. 
Coloured solid lines correspond to the full relativistic equations and coloured dashed lines to their nonrelativistic approximations. 
In blue: $|Y_{N_1\rm even}(z)|$. In green: $|1/\epsilon_0Y_{B-L}(z)|$. In orange: $|1/\epsilon_0Y_{N_1\rm odd}(z)|$. The dashed gray line gives the equilibrium yield for $Y_{N_1}$. 
For the plots on the left, the dashed green and dashed blue lines overlap for early times, as explained in the text.}
\end{figure}

In Fig.~\ref{fig:eff_scan} we show the final $B-L$ asymmetry obtained for various initial conditions and values of $K$ between $10^{-4}$ and $10$. 
The fluid equations are integrated up to $z \approx 400$. 
For this value of $z$, and with the lower bound $K \approx 10^{-4}$, the above estimate gives $\abs{\delta(z = 400)} \lesssim 3 \cdot 10^{-4}$ for nonvanishing initial conditions and $\abs{\delta(z = 400)} \lesssim 1.3 \cdot 10^{-2}$ for vanishing initial conditions.
It is important to reiterate that this estimate does not account for theoretical uncertainties, as it captures only numerical deviations.
For both the nonrelativistic approximation and the fully relativistic result, one finds that there is a clear distinction in the behaviour of $\kappa_f$ for the weak washout regime with $K \lesssim 1$ and the strong washout regime with $K \gtrsim 1$. 
For a nonvanishing initial abundance of sterile neutrinos and $K\lesssim0.01$, the behaviour of the efficiency factor is well described by Eq.~\eqref{eq:efficiency_Y0}.
Furthermore, we confirm the estimate of Eq.~\eqref{eq:efficiency_weak_washout} and the sign flip of the asymmetry with respect to the nonrelativistic approximation for a vanishing initial abundance.
From the plot, the sign flip can be seen to occur at around $K \sim 0.2$.

In the strong washout regime, the fully relativistic and nonrelativistic predictions for the final $B-L$ asymmetry are in full agreement, as expected from the previous arguments.
The behaviour of the efficiency factor is well described by the approximate expression of Eqs.~\eqref{eq:efficiency_strong_washout} in conjunction with the value for $\overline{z}$ obtained by solving Eq.~\eqref{eq:saddle}.
For intermediate values of $K \sim 10^{-1}$, the relativistic fluid equations give a mild enhancement of less than 10\% for the final asymmetry. 

The explicit time evolution of the yields is illustrated in Fig.~\ref{fig:Ys} for several sample solutions with $K=0.001,1,10$ and for both vanishing and equilibrium initial conditions for the sterile neutrinos. 
The yields $Y_{B-L}, Y_{N_1\rm odd}$ are rescaled by $1/\epsilon_0$, in order to display the results in a way that is independent of the $CP$-violating source. 
Focusing first on the plots on the left, which correspond to vanishing initial conditions, it is clear that in the nonrelativistic limit $1/\epsilon_0Y_{B-L}(z)$ closely follows $-Y_{N_1\rm even}(z)$. 
This reflects the fact that, when washout is ignored, Eqs.~\eqref{eq:n_even_boltzmann_nr} and~\eqref{eq:y_bl_boltzmann_nr} imply that  $ \text{d}/\text{d}z (1/\epsilon_0 Y_{\Delta_\smallpara}+Y_{N_1\rm even})=0$.
This relation does not hold in the relativistic case due to the $z$ dependence of $\epsilon_{\rm eff}$. 
In particular, the enhancement of the source term at small $z$ always gives a larger relativistic asymmetry at early times. 
For weak washout, the relativistic asymmetry continues to be larger at late times, and for small $K$ (upper left plot) the early-time enhancement is enough to avoid a sign flip for the asymmetry after $Y_{N_1\rm even}$ reaches the equilibrium value and the sign of the source of the asymmetry changes. 
For larger $K$ (middle and lower left plots), the early-time enhancement is washed away and the sign flip happens as in the nonrelativistic case.
In the case of equilibrium initial conditions (plots on the right), the early-time enhancement in principle survives for small $K$, but becomes less relevant as most of the asymmetry is generated from late-time decays, when the source matches its nonrelativistic counterpart. 
This is illustrated by the  upper right plot. At high $K$, the early-time enhancement is washed out (see lower right plot), while for intermediate values near $K=1$ (middle right plot), 
at the sweet spot of parameter space in which $N_1$ decays not long after the source enhancement and when  the washout is not yet large, one finds the $<10\%$ enhancements of the efficiency seen in Fig.~\ref{fig:eff_scan}.

\subsection{Setup with partially equilibrated spectators}

\begin{figure}
\includegraphics[width = 0.49\textwidth]{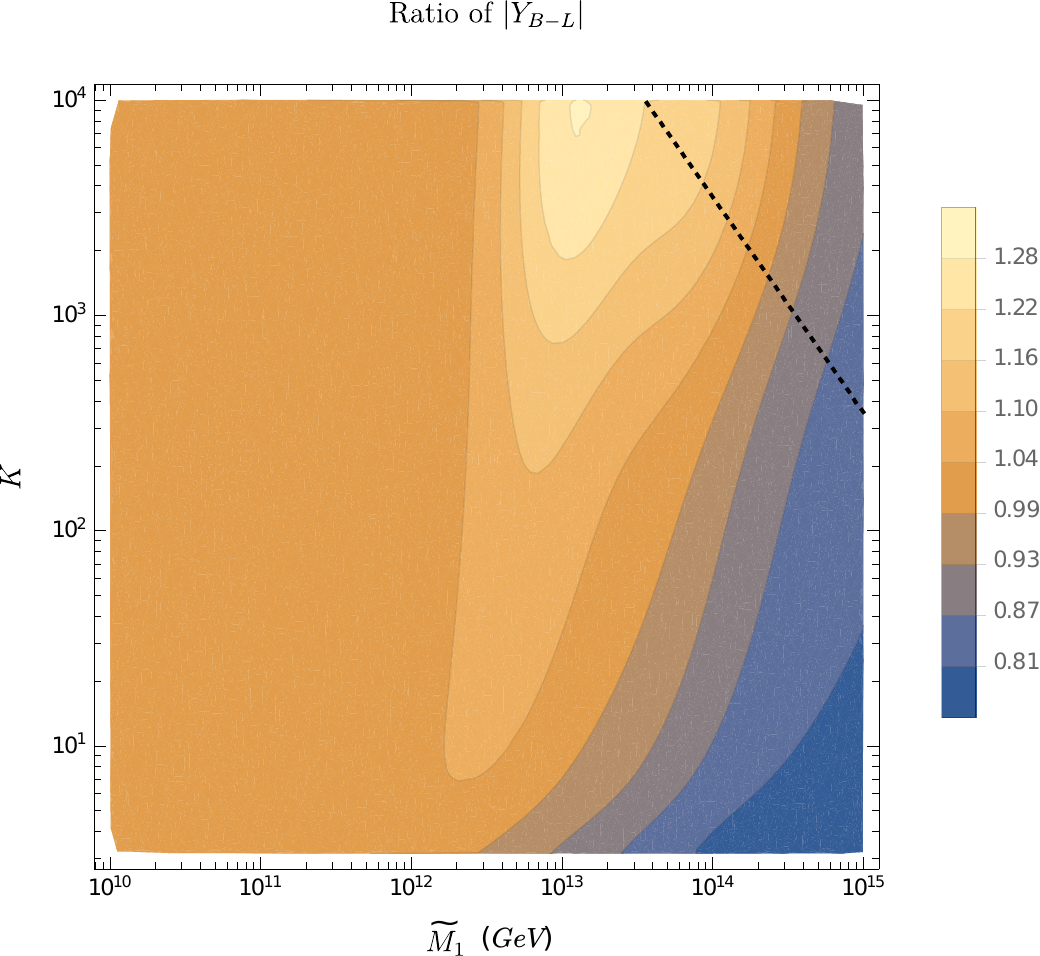}
\hskip1pt
\includegraphics[width = 0.49\textwidth,height=.45\textwidth]{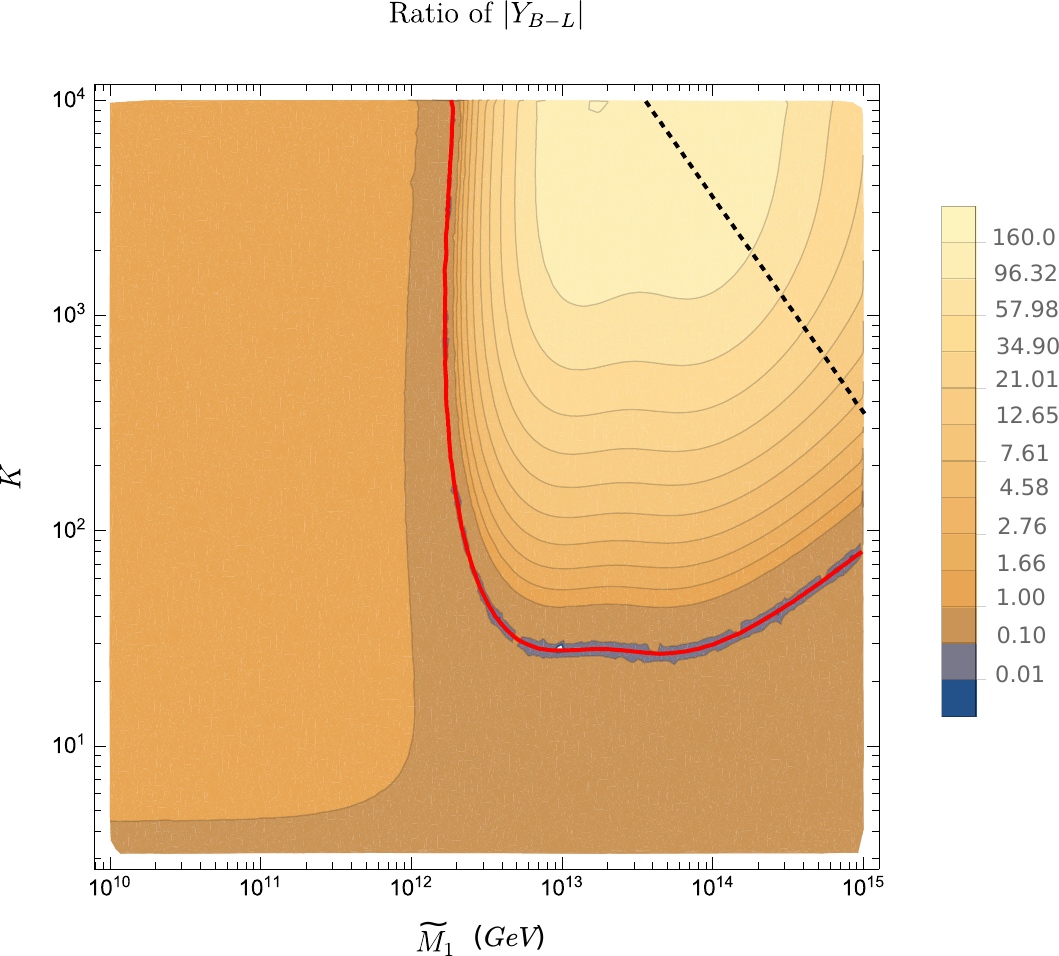}
\vspace{-10pt}
\caption{%
\footnotesize\label{fig:eff_scan_wi_spect}Ratios of final $B-L$ asymmetry obtained with and without including partially equilibrated bottom-Yukawa and weak sphalerons interactions. 
On the left: initially equilibrated sterile neutrinos; on the right: vanishing initial abundance of sterile neutrinos. %
The bright red line highlights the sign change of the final $B-L$ asymmetry and the dotted black lines indicates the region in which $\abs{F_{1}}^2 > 4 \pi $.
}
\end{figure}

We now turn towards the realistic setup for sterile neutrino masses of around $\unit[10^{13}]{GeV}$. 
To estimate the impact of spectators, we compare the final $B-L$ asymmetry obtained by solving the full set of equations~\eqref{eq:Boltzmann_n_even}--\eqref{eq:Boltzmann_weak_sphal} with that obtained by enforcing equilibration of both bottom-Yukawa and weak sphaleron interactions.
The result of a numerical parameter scan of the ratio of the two $B-L$ asymmetries in dependence of $\tilde M_1$ and $K$ is shown in Fig.~\ref{fig:eff_scan_wi_spect} for both $Y_{N_1 \text{even}} (0) = Y_{N_1 \text{eq}} (0)$ and $Y_{N_1 \text{even}} (0) = 0$ initial conditions.   
We restrict the scan to the strong washout regime, $K>3$, since the partially equilibrated spectators are expected to be most important in that case.
Further, we indicate in the plots where the Yukawa coupling $F_1$ becomes nonperturbatively large, what happens for large $K$ and large $M_1$.
At the high temperature scales under consideration, it can also be interesting to discuss bounds from reheating of the Universe and the
metastability of the scalar potential. While concrete constraints~\cite{Croon:2019dfw} can be derived e.g. for the Standard Model, these
will not be robust when further degrees of freedom are added, such that do not overlay our results with considerations of that kind in the present work.
We further note that since there is no dependence on the initial conditions for the sterile neutrinos of the final $B-L$ asymmetry (see Fig.~\ref{fig:eff_scan}) when neglecting partial equilibration of spectators, 
both plots in Fig.~\ref{fig:eff_scan_wi_spect} are in fact normalized to the same baseline asymmetry. \\

The first thing to notice is that there is no significant difference between the full treatment and the approximation using fully equilibrated spectators for $\tilde M_1 \ll \unit[10^{12}]{GeV}$. 
This is to be expected, since in this regime both bottom-Yukawa and electroweak sphaleron interactions reach chemical equilibrium well before washout processes decouple and $B-L$ freezes in. 
As a result, any asymmetry that may have been transferred to the spectator fields at early times is washed out. 

On the other hand, it is clear from Fig.~\ref{fig:eff_scan_wi_spect} that the final asymmetry computed using the full treatment with partially equilibrated spectators is strongly dependent on the initial conditions for masses $\tilde M_1 \gtrsim \unit[10^{12}]{GeV}$.
For initially equilibrated sterile neutrinos with $Y_{N_1\text{even}} (0) = Y_{N_1 \text{eq}} (0)$ the inclusion of partially equilibrated spectator interactions yields only order one corrections to the final $B-L$ asymmetry, 
which is consistent with the study of partially equilibrated spectators that was carried out in Ref.~\cite{1404.2915} within the nonrelativistic approximation. 
However, for $Y_{N_1 \text{even}} (0) = 0$ there are qualitative differences between the full and simplified treatments for a sufficiently strong washout.
In the full treatment, there is a sign flip in the final $B-L$ asymmetry for large $K$, starting at $K \gtrsim 30$ for $\tilde M_1$ near $\sim\unit[10^{13}]{GeV}$. 
The values of  parameters for  which the change of sign occurs are indicated by a red contour line in the right plot of Fig.~\ref{fig:eff_scan_wi_spect}. 

\begin{figure}
\includegraphics[width = 0.5\textwidth]{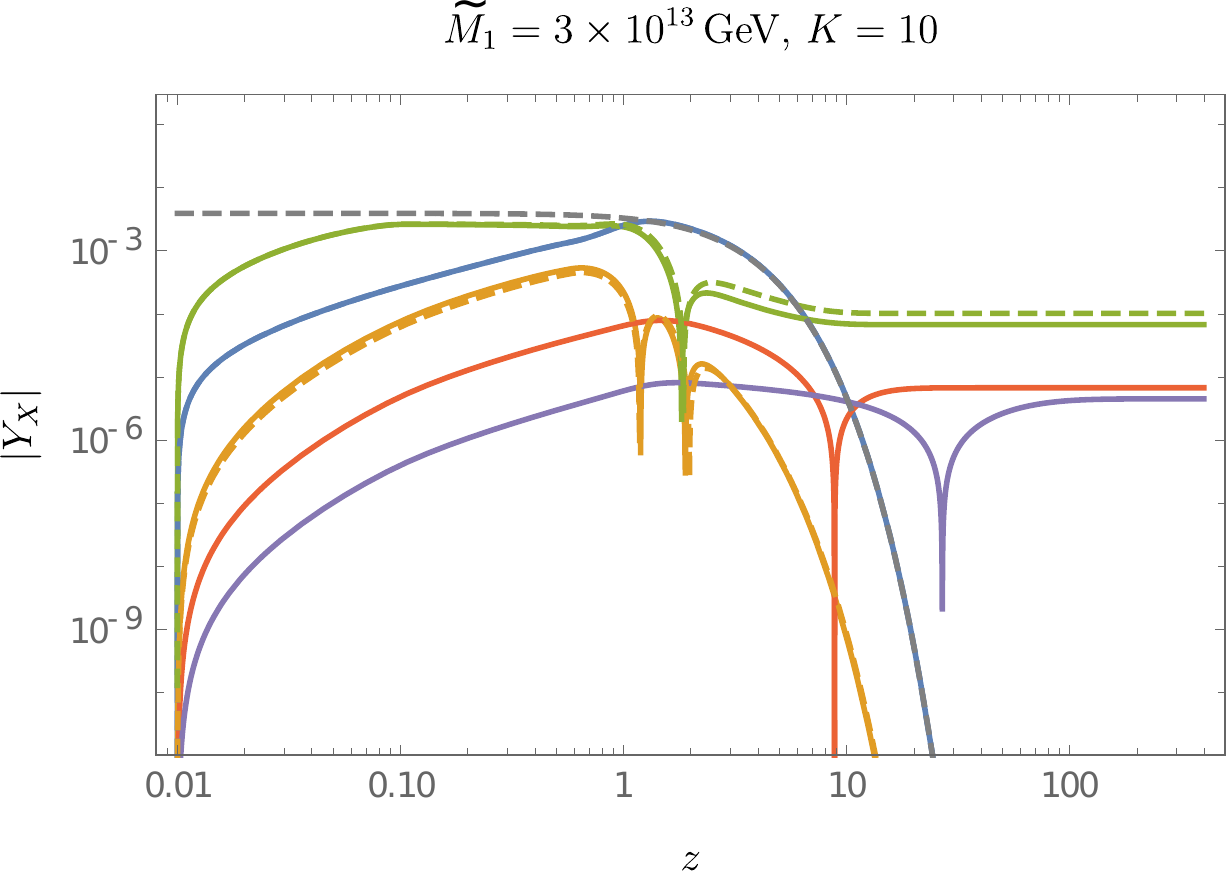}
\includegraphics[width = 0.5\textwidth]{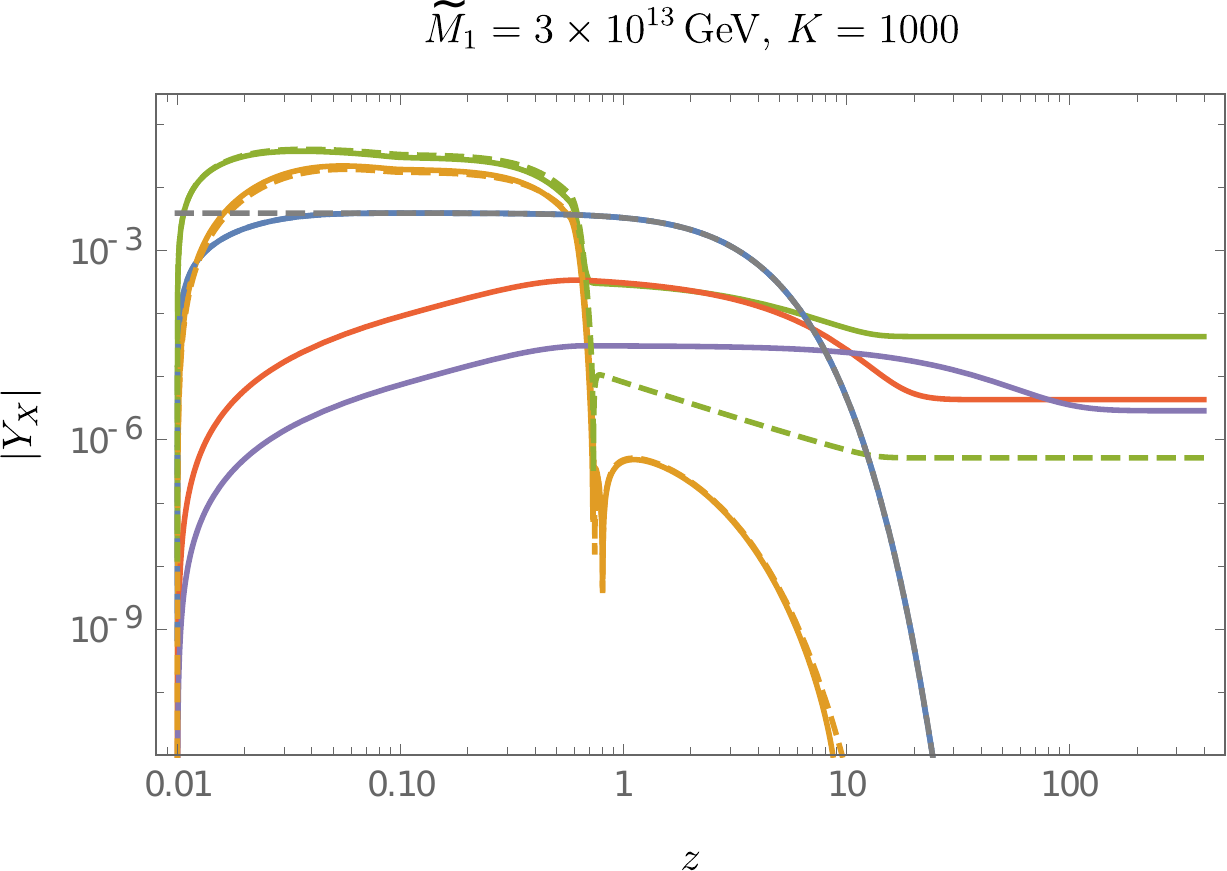}
\caption{\footnotesize\label{fig:Ys_spectators}%
Time evolution of the yields obtained by solving the fluid equations with partially (solid lines) or fully equilibrated (dashed lines) spectators, for $\tilde M_1=3\times 10^{13}$ GeV and $K=10$ (left) and 1000 (right). 
The particular yields are $|Y_{N_1\rm even}(z)|$ (blue), $|1/\epsilon_0Y_{B-L}(z)|$ (green), $|1/\epsilon_0Y_{N_1\rm odd}(z)|$ (orange), $|1/\epsilon_0Y_{\Delta_{\rm down}}(z)|$ (red) and $|1/\epsilon_0Y_{\Delta_{l_\perp}}(z)|$ (purple). 
The dashed gray line shows the equilibrium yield for $Y_{N_1}$. }
\end{figure}

Furthermore, inside the flipped regime, the final $B-L$ asymmetry is enhanced by up to two orders of magnitude. 
The mechanism responsible for the enhancement is illustrated for $\tilde M_1=3\times 10^{13}$ GeV and $K=10,1000$ in Fig.~\ref{fig:Ys_spectators}, 
which shows the yields with appropriate rescalings that make them independent of the magnitude of the $CP$-violating parameter $\epsilon_0$. 
For very large $K$, the large,spectator-induced values of the yield $Y_{\Delta\rm down}$ can alter the behaviour of $Y_{B-L}$, beyond the usual interplay of source and washout terms. 
As illustrated in the plot on the right of Fig.~\ref{fig:eff_scan_wi_spect}, once the sterile neutrino density has achieved equilibrium, the source term for the asymmetry changes sign and $Y_{B-L}$ starts decreasing in magnitude in the usual way, 
until it becomes close to  $Y_{\Delta\rm down}$; if the latter is large enough, $Y_{B-L}$ starts to track it, avoiding the sign flip due to the change in sign of the source, and escaping the decrease in magnitude associated with the usual effect of washout.

\section{Discussion and future prospects}
\label{sec:conclusions}

We have applied recent progress on the theory of relativistic effects for sterile neutrinos in a hot Standard Model plasma to leptogenesis in the weak washout regime and in the presence of partially equilibrated spectators. 
As expected from the fact that the generation of the asymmetry requires a departure from thermal equilibrium, 
the effects are most pronounced for nonthermal (and in particular vanishing) initial conditions for the sterile neutrinos, which in turn are expected to be generic in the framework of inflationary cosmology. 

With the mass spectrum and Yukawa couplings of the sterile neutrinos and the principle behind them unknown, 
weak washout is a perfectly viable option, in particular in the presence of more than two sterile neutrinos, which allows for weaker couplings of the lightest one. 
Thus, the developments that we have presented here on this topic are of phenomenological relevance. 
We have found that, when neglecting spectator interactions and for vanishing initial conditions in the weak washout regime, 
that is, for values of the washout parameter $K<0.1$, relativistic corrections give rise to a sign-flip and an enhancement of the absolute value of the asymmetry. 
Parametrically, this is characterized by the $\propto K$ scaling of the final asymmetry, compared to the $K^2$ scaling obtained in computations that do not account for all relativistic effects. 
This effect is due to an enhancement of the source in the relativistic regime. In keeping with previous results in the literature, 
we find that relativistic corrections tend to enhance all rates --as has been seen for light sterile neutrinos \cite{1012.3784,1202.1288,1302.0743,1303.5498}, 
or electrons \cite{Bodeker:2019ajh}-- but we found that the enhancement is more pronounced for the $CP$-violating source than for the washout rate.

Furthermore, we have studied the impact of partially equilibrated spectator fields, focusing on the particular example of the fields involved in bottom Yukawa  and weak sphaleron interactions. 
This choice was made for simplicity, as one may ignore lepton flavour effects in the relevant temperature range. 
However, we emphasize that the methods presented here are also applicable to spectator processes that equilibrate at lower temperatures. 
As in the case without partially equilibrated spectators, relativistic effects have a higher impact for vanishing initial conditions. 
In the very strong washout regime, for $K>30$, one obtains a sign-flip and an enhancement of the asymmetry of up to two orders of magnitude in the relevant sterile neutrino mass range of $\tilde{M}_1\gtrsim 10^{12}$ GeV, 
where the bottom Yukawa and weak sphaleron interaction remain partially equilibrated until after the freeze out of the lepton asymmetry.

As for the theoretical uncertainties, in the calculation of the rates we have limited ourselves  to leading-log accuracy. 
While leading order accurate calculations for the total, helicity-averaged relaxation rates of the sterile neutrinos in the intermediate regime between the relativistic and nonrelativistic limits are available~\cite{1302.0743,Ghisoiu:2014ena}, 
they have not yet been used in phenomenological calculations. 
Moreover, an important outcome of our work is that, rather than using the helicity-averaged rate, one needs to distinguish between LNV and LNC rates, for which the leading-order results had not been computed yet. 
The calculation of the $CP$-violating rates beyond leading-log accuracy in the intermediate regime appears challenging, while in the nonrelativistic regime they are already known to full NLO level \cite{Bodeker:2017deo}.
Also, it should be kept in mind that the leading order fluid expansion limit our accuracy to order one because of the unknown shape of the distribution functions of the sterile neutrinos.
It may therefore be interesting to revisit earlier studies~\cite{Basboll:2006yx,HahnWoernle:2009qn} that were looking into this issue for leptogenesis from heavy sterile neutrinos in view of the improved understanding of the relevant rates.
Another source of theoretical uncertainties lies in the fact that in the derivation of the relativistic fluid equations, we have neglected so-called ``double small'' terms, 
which are proportional to both the Standard Model yields and the deviation of the sterile neutrinos distribution functions from equilibrium, and therefore considered doubly suppressed in many treatments of leptogenesis. 
For initially thermalized sterile neutrinos, this is well justified. 
However, for a small or vanishing initial abundance of sterile neutrinos, implying a large departure from equilibrium, there may be sizable corrections in the final lepton asymmetry, 
especially if a significant portion of the asymmetry is produced at early times and retained until freeze out.

Leptogenesis in the type I seesaw scenario has been shown to work through large mass ranges of the sterile neutrinos and for different patterns of their couplings, that consequently imply very different dynamics of the mechanism. 
It is an interesting speculation that the viable regions in parameter space may be connected. 
The present work adds to this rich picture in the weak washout regime and for situations with partially equilibrated spectator processes.

\section*{Acknowledgements}
 CT and BG acknowledge support by the Collaborative Research Centre SFB1258 of the Deutsche Forschungsgemeinschaft (DFG). 
BG is grateful to the Physics Department of the University of Wisconsin-Madison for generous hospitality during the completion of this work.


\appendix

\section{Summary of the CTP formalism}
\label{sec:ctp_basics}

The basic objects of concern in nonequilibrium quantum field theory are time-dependent expectation values of operators for an arbitrary state (e.g. a thermal ensemble but notably also pure or mixed states deviating from equilibrium) represented by a given density matrix.  
The key idea in the application of the CTP formalism to nonequilibrium theory is to express these expectation values as a functional integration along a closed time-path typically chosen to go from an initial time $t = t_i$ forward to some $t = \tau>t_i$ and then back to $t_i$, 
with the density matrix encoded through external sources that are added to the Lagrangian~\cite{Calzetta:1986cq}. 
In complete analogy with the conventional path-integral formalism, the CTP path-integral can then be used to compute time-path ordered $n$-point functions. 
The only difference is that due to the closed time-path, there are now two distinct branches: a ``$+$'' branch, going forward in time with the usual time ordering, and a ``$-$'' branch with the opposite orientation and the reverse time-ordering. 
Since each local operator may be evaluated along either branch of the time path, there are four types of two-point functions, labeled as $++$ (or alternatively ``$T$''), $- -$  (or  ``$\bar T$''), $+-$ (or ``$<$''), and $-+$ (or ``$>$''). 
The $>,<$ two-point functions with the fields evaluated along different branches of the time path are called ``Wightman functions''. 
They correspond to correlations without time ordering and encode (among other things) the information necessary to compute the particle number densities.
For a given complex scalar field $\phi$, the four possible CTP propagators are given explicitly as\footnote{Our notation borrows heavily from \cite{1007.4783,hep-ph/0312110,1002.1326}, to which we refer for more details on the formalism.}
\begin{subequations}
\begin{align}
\label{eq:t_ordered_scalar_def}
\i\Delta^{++}_{X} (u,v) =&\,\langle T\,\phi_{X}(u)\,\phi^\dagger_{X}(v)\rangle \ ,
&
\i\Delta^{--}_{X} (u,v)=&\,\langle\overline{T}\,\phi_{X}(u)\, \phi^\dagger_{X}(v)\rangle \ ,
\\
\label{eq:Wightman_scalar_def}
\i\Delta^<_{X} (u,v)=&\,\langle\bar\phi^\dagger_{X}(v)\,\phi_{X,}(u)\rangle \ ,
&
\i\Delta^>_{X} (u,v)=&\,\langle\phi_{X}(u)\,\phi^\dagger_{X}(v)\rangle \ ,
\end{align}
\end{subequations}
where the $X$ signifies particle species and $T$ and $\overline{T}$ denote time ordering and anti-time ordering, respectively. 
It is useful to define the advanced and retarded propagators
\begin{subequations}
\begin{align}
\label{eq:scalar_ad_ret_def}
\i \Delta^{a}_X &\equiv \i \Delta_X^{T} - \i \Delta_X^{>} =  \i \Delta_X^{\overline T} -  \i \Delta_X^{<} \ ,
&
\i \Delta^{r}_X &\equiv \i \Delta_X^{T} - \i \Delta_X^{<} =  \i \Delta_X^{\overline T} -  \i \Delta_X^{>} \ ,
\intertext{as well as the hermitian and anti-hermitian combinations}
\label{eq:scalar_anti_herm_def}
\Delta^{\mathcal H}_X &\equiv \frac{1}{2}\big( \Delta^{a}_X + \Delta^{r}_X \big) \ ,
&
\Delta^{\mathcal A}_X &\equiv \frac{1}{2 \i}\big( \Delta^{a}_X - \Delta^{r}_X \big) \ .
\end{align}
\end{subequations}
The object $ \Delta^{\mathcal A}_X $ can be identified with the spectral function that is also familiar from quantum field theory at zero temperature. 
In the case of fermions, the various CTP two-point functions are defined completely analogously. 
For a given fermionic species $X$, one has
\begin{subequations}
\begin{align}
\label{eq:t_ordered_def}
\i S^{++}_{X,\alpha\beta} (u,v) =&\,\langle T\,\psi_{X,\alpha}(u)\,\bar\psi_{X,\beta}(v)\rangle \ ,
&
\i S^{--}_{X,\alpha\beta} (u,v)=&\,\langle\overline{T}\,\psi_{X,\alpha}(u)\,\bar\psi_{X,\beta}(v)\rangle \ ,
\\
\label{eq:Wightman_def}
\i S^<_{X,\alpha\beta} (u,v)=&\,-\langle\bar\psi_{X,\beta}(v)\,\psi_{X,\alpha}(u)\rangle \ ,
&
\i S^>_{X,\alpha\beta} (u,v)=&\,\langle\psi_{X,\alpha}(u)\,\bar\psi_{X,\beta}(v)\rangle \ .
\end{align}
\end{subequations}
As before, we define 
\begin{subequations}
\begin{align}
\label{eq:ad_ret_def}
\i S^{a}_X &\equiv \i S_X^{T} - \i S_X^{>} =  \i S_X^{\overline T} -  \i S_X^{<} \ ,
&
\i S^{r}_X &\equiv \i S_X^{T} - \i S_X^{<} =  \i S_X^{\overline T} -  \i S_X^{>} \ ,
\intertext{and}
\label{eq:anti_herm_def}
S^{\mathcal H}_X &\equiv \frac{1}{2}\big( S^{a}_X + S^{r}_X \big) \ ,
&
S^{\mathcal A}_X &\equiv \frac{1}{2 \i}\big( S^{a}_X - S^{r}_X \big) = \frac{\i}{2}\big( S^{>}_X - S^{<}_X \big) \ ,
\end{align}
\end{subequations}
where $S^{\mathcal A}_X$ is again the spectral function, but now for fermionic fields. \\

In what follows, we focus on the bosonic and fermionic Wightman functions~\eqref{eq:Wightman_scalar_def}, \eqref{eq:Wightman_def}, 
since these are what we need to compute the particle number densities $n_{N_1 h} (t)$ and $n^\pm_l(t)$.
Their connection to the number densities is seen most easily when working in Wigner space. 
For any two-point function $G(u,v)$, the Wigner transform is defined as the Fourier transform with respect to the relative coordinate $r=u-v$, 
and thus it depends on the Fourier momentum $k$ as well as the center of mass coordinate $x=1/2(u+v)$:
\begin{equation}
G (x,k) \equiv \int \text{d}^4 r \ e^{\i k r} G \Big(x + \frac12 r, x - \frac12 r \Big) \,.
\end{equation} 
In the special case of kinetic equilibrium at a temperature $T\equiv 1/\beta$, the Wigner transformed Wightman functions become independent of the $x$-coordinate due to spacetime translation invariance. 
They take the form
\begin{subequations}
\label{eq:Wightman_eq}
\begin{align}
\i\Delta_X^<(k)=&\,2 \Delta_X^{\cal A}(k) f_X(k) \ ,
&
\i\Delta_X^>(k)=&\,2 \Delta_X^{\cal A}(k)(1+f_X(k))\ ,
\\
\i S_X^<(k)=&\,-2 S_X^{\cal A}(k) f_X(k) \ ,
&
\i S_X^>(k)=&\,2 S_X^{\cal A}(k)(1-f_X(k)) \ ,
\end{align}
\end{subequations}
where $f_X(k)$ are the equilibrium distribution functions in the presence of chemical potentials $\mu_X$,
\begin{align}
\label{eq:dist_eq}
f_X(k)=&\,\frac{1}{e^{\beta (k^0-\mu_X)}+1}\quad \rm{(fermions)} \ ,
&
f_X(k)=&\,\frac{1}{e^{\beta (k^0-\mu_X)}-1}\quad \rm{(bosons)} \ .
\end{align}
We note that the above results are fully general.
In particular, they also hold when radiative effects are taken into account by including wave-function type loop corrections for the spectral functions $\Delta_X^{\cal A}(k),S_X^{\cal A}(k)$.  
At tree level, the spectral functions are given as
\begin{subequations}
\label{eq:spectral_eq_tree}
\begin{align}
\Delta_X^{\cal A,\rm tree}(k)=&\,\pi \delta(k^2-m^2_X)\,{\rm sign}(k^0) 
\ , \\
S_X^{\cal A,\rm tree}(k)=&\,\pi \delta(k^2-m^2_X)\,{\rm sign}(k^0) \p_X (\slashed{k}+m_X)
\ ,
\end{align}
\end{subequations}
where $\p_X=\p_{L/R}$ for left/right handed fields (in case of which $m_X=0$) and $\p_X=\mathbb{I}$ otherwise. 
Due to the explicit delta distributions, these tree-level results correspond to particle excitations of zero width. 
Radiative effects or out-of-equilibrium dynamics may give rise to a $\bf{k}$-dependence of the distribution functions, 
finite widths for the particle modes \cite{Garbrecht:2011xw}, and additional propagating modes such as holes.

Using the Wigner-transformed Wightman functions, the fermionic particle number densities $n_{N_1 h} (t)$ and $n^\pm_l(t)$
\footnote{In principle, the particle number densities are fully spacetime dependent, but for a spatially homogeneous and isotropic universe the spatial dependence is trivial, so we suppress it in the notation.} 
for the sterile neutrinos and Standard Model leptons \cite{Garbrecht:2008cb,1007.4783} can be obtained as
\begin{subequations}
\begin{align}
\label{eq:lep_density_def}
n(t)^\pm_{l_\smallpara} 
&= - \int\displaylimits_{0}^{\pm \infty} \frac{\text{d} k^0}{2\pi} \int \hspace{-3pt} \frac{\text{d}^3 \mathbf{k}}{(2\pi)^3} \tr \left[ \gamma^0 \i S_{l_\smallpara}^{<,>} (t,k) \right]
\intertext{and}
\label{eq:neut_density_def}
n(t)_{N_1 h} 
&= - \int\displaylimits_{0}^{\pm \infty} \frac{\text{d} k^0}{2\pi} \int \hspace{-3pt} \frac{\text{d}^3 \mathbf{k}}{(2\pi)^3} \tr \left[ \gamma^0 \p_h \i S_{N_1}^{<,>} (t,k)\right]
\\
\nonumber 
&=- \frac{1}{2} \int\displaylimits_{- \infty}^{+ \infty} \frac{\text{d} k^0}{2\pi} \int \hspace{-3pt} \frac{\text{d}^3 \mathbf{k}}{(2\pi)^3} \tr \Big[ \gamma^0 \p_h \big(\theta (k_0) \i S_{N_1}^{<}(t,k) - \theta (\text{-} k_0) \i S_{N_1}^{>}(t,k) \big)  \Big] 
\ .
\end{align}
\end{subequations}
In these equations, $\gamma^0$ is a Dirac-gamma matrix in one of the standard 
Weyl, Dirac or Majorana representations and $\p_h$ projects onto an helicity eigenstate with helicity $h$,
\begin{equation}
\label{eq:p_h}
\p_h \equiv \frac{1}{2} \big(\mathds{1} + h \hat{k}^i \gamma_0 \gamma_i \gamma_5 \big) \ .
\end{equation}
In Eq.~\eqref{eq:lep_density_def} we have omitted indices from the fundamental representation of $SU(2)$ for simplicity; note that both components of a doublet must yield identical asymmetries due to symmetric initial conditions and gauge symmetry. 
As before, the superscript $\pm$ on $n(t)^\pm_{l_\smallpara}$ indicates the particle and antiparticle number case, respectively. 
To obtain the second expression in Eq.~\eqref{eq:neut_density_def}, we have used the Majorana condition for the sterile neutrino spinors, which implies that particle and antiparticle number densities must be identical. 

Using the definitions \eqref{eq:lep_density_def} and \eqref{eq:neut_density_def}, one can obtain the time-evolution of the particle number densities $n_{N_1 h} (t)$ and $n^\pm_l(t)$ from that of the two-point functions, 
which satisfy a well known set of formally exact Schwinger-Dyson equations~\cite{Cornwall:1974vz,Calzetta:1986cq,Berges:2004yj}.
In Wigner space, these equations take the form~\cite{hep-ph/0312110}
\begin{subequations}
\begin{align}
\label{eq:schwinger_dyson_eq_1}
e^{-\i \diamond} \{ \slashed{k} - m_X - \Sigma_X^{a,r} \}\{ \i S_X^{a,r} \}&= \i \p_X \ ,
\\
\label{eq:schwinger_dyson_eq_2}
e^{-\i \diamond} \{ \slashed{k} - m_X - \Sigma_X^r \}\{ \i S_X^{<,>} \}
&= e^{-\i \diamond} \{ \Sigma_X^{<,>} \}\{ \i S_X^{a} \} \ ,
\end{align}
\end{subequations}
where $\p_{l_\smallpara} \equiv \p_R$ and $\p_{N_1} \equiv \mathds{1}$. 
The diamond operator $\diamond$ is defined via 
\begin{equation}
\diamond\{A (x,k) \}\{B (x,k)\} \equiv \frac12 \big[ (\partial_x^\mu A(x,k)) (\partial_{k,\mu} B(x,k)) - (\partial_k^\mu A(x,k)) (\partial_{x,\mu} B(x,k)) \big],
\end{equation}
while  $\slashed{\Sigma}_X^{\pm\pm}$ denotes the fermionic self-energies, which encode the loop corrections to the tree-level propagators. 
They are formally defined in terms of functional derivatives of the interacting part of the two-particle irreducible (2PI) effective action with respect to the CTP two-point functions
\begin{equation}
\label{eq:self_energy}
\Sigma_X^{ab} (u,v) \equiv - \i a b\, \frac{\delta \Gamma^{\text{2PI},\,\geq 2 \text{loop}}}{\delta S^{ba}_X (v,u)} \ .
\end{equation}
In analogy with Eq.~\eqref{eq:anti_herm_def}, one can define Hermitian and spectral combinations for the self-energies, $\slashed{\Sigma}_X^{\mathcal H}$ and $\slashed{\Sigma}_X^{\mathcal A}$, 
which contain information on quantum corrections to the dispersion relation and the width of the corresponding fermion. The self-energy
$\slashed{\Sigma}_X^{\mathcal A}$ is of particular interest, since it determines the equilibration rate of the associated particle, %
in analogy with the familiar optical theorem from zero-temperature quantum field theory, relating the imaginary part of a two-point function to the total decay rate. 

To obtain the fluid equations for leptogenesis, we perform a simultaneous expansion of the Schwinger-Dyson equations~\eqref{eq:schwinger_dyson_eq_1}, \eqref{eq:schwinger_dyson_eq_2} in the gradients $\partial_x$ and the coupling constants of the underlying theory.
To understand why this is reasonable, consider the typical scales associated with the derivatives $\partial_k$ and $\partial_x$ appearing in the diamond operator in Eqs.~\eqref{eq:schwinger_dyson_eq_1}, \eqref{eq:schwinger_dyson_eq_2}. 
The derivatives in the momentum $k$ are given by the typical microscopic energy scale, so that close to kinetic equilibrium $ \partial_k \sim \nicefrac{1}{T} $. 
The same is not necessarily true for the change in the average macroscopic coordinate $x$, in fact one often has $\partial_x \ll T$.
In our case, there are two sources of nonvanishing macroscopic gradients $\partial_x$: first, the expansion of the universe, and second, interactions such as scatterings, decays, and oscillations. 
For the expansion of the universe, the associated time scale (in conformal time) is $\nicefrac{\partial_t a(t) }{a(t)} \approx a(t) H$. 
For the $B-L$ violating interactions induced by the sterile neutrinos, the time scale is $\nicefrac{\partial_t \mu_X}{\mu_X} \sim |F|^2 T$. 
In the weak washout regime, the Hubble expansion rate as a source for gradient corrections always dominates over the rate of $B-L$ violating interactions before freeze-out,
whereas for strong washout, the expansion dominates at early times and the $B-L$ violating processes dominate later, prior to freeze-out.
For the gradients from the expansion of the Universe, one finds that $\partial_x \cdot \partial_k \sim \nicefrac{a(t) H}{T}=\nicefrac{H}{T_{\rm phys}} \ll 1$ and for those from $B-L$ violation $\partial_x \cdot \partial_k \sim F^2 \ll 1$.
Notice that for the second case the expansion in the gradients is equivalent to a perturbative expansion in the sterile neutrino Yukawa couplings. 
Since the self energies are of second order in the Yukawa couplings, this implies that terms such as $\partial_t \Sigma^{\cal H/A}$ , $\partial_t \Sigma^{a/r}$ are of subleading order in the simultaneous expansion. 
Taking this into account and expanding to zeroth order in the gradients and to leading order in the Yukawa couplings of the sterile neutrinos, the Schwinger-Dyson equations become~\cite{1002.1326,1007.4783}
\begin{subequations}
\label{eq:SDEs:Wigner}
\begin{align}
\label{eq:schwinger_dyson_eq_1_after_gradient}
\big( \slashed{k} - m_X -  \Sigma_X^{a,r} \big) \i S_X^{a,r} &= \i \p_X \ ,
\\
\label{eq:schwinger_dyson_eq_2_after_gradient}
\big( \slashed{k} - m_X - \Sigma_X^r \big) \i S_X^{<,>} +\frac{ \i }{2} \slashed{\partial}_x( \i S_X^{<,>})
&=  \Sigma_X^{<,>} \, \i S_X^{a} \ .
\end{align}
\end{subequations}

\printbibliography

@article{Sakharov:1967dj,
      author         = "Sakharov, A. D.",
      title          = "{Violation of CP Invariance, c Asymmetry, and Baryon
                        Asymmetry of the Universe}",
      journal        = "Pisma Zh. Eksp. Teor. Fiz.",
      volume         = "5",
      year           = "1967",
      pages          = "32-35",
      doi            = "10.1070/PU1991v034n05ABEH002497",
      note           = "[Usp. Fiz. Nauk161,61(1991)]",
      SLACcitation   = "%%CITATION = ZFPRA,5,32;%%"
}

@article{hep-ph/0202239,
      author         = "Davidson, Sacha and Ibarra, Alejandro",
      title          = "{A Lower bound on the right-handed neutrino mass from
                        leptogenesis}",
      journal        = "Phys. Lett.",
      volume         = "B535",
      year           = "2002",
      pages          = "25-32",
      doi            = "10.1016/S0370-2693(02)01735-5",
      eprint         = "hep-ph/0202239",
      archivePrefix  = "arXiv",
      primaryClass   = "hep-ph",
      reportNumber   = "OUTP-02-10P, IPPP-02-16, DCPT-02-32",
      SLACcitation   = "%%CITATION = HEP-PH/0202239;%%"
}

@article{hep-ph/0401240,
      author         = "Buchmuller, W. and Di Bari, P. and Plumacher, M.",
      title          = "{Leptogenesis for pedestrians}",
      journal        = "Annals Phys.",
      volume         = "315",
      year           = "2005",
      pages          = "305-351",
      doi            = "10.1016/j.aop.2004.02.003",
      eprint         = "hep-ph/0401240",
      archivePrefix  = "arXiv",
      primaryClass   = "hep-ph",
      reportNumber   = "DESY-03-100, UAB-FT-551, CERN-TH-2003-199",
      SLACcitation   = "%%CITATION = HEP-PH/0401240;%%"
}

@article{Garbrecht:2008cb,
      author         = "Garbrecht, Bjorn and Konstandin, Thomas",
      title          = "{Separation of Equilibration Time-Scales in the Gradient
                        Expansion}",
      journal        = "Phys. Rev.",
      volume         = "D79",
      year           = "2009",
      pages          = "085003",
      doi            = "10.1103/PhysRevD.79.085003",
      eprint         = "0810.4016",
      archivePrefix  = "arXiv",
      primaryClass   = "hep-ph",
      reportNumber   = "UAB-FT-657, NPAC-08-22",
      SLACcitation   = "%%CITATION = ARXIV:0810.4016;%%"
}

@article{1404.2915,
      author         = "Garbrecht, Björn and Schwaller, Pedro",
      title          = "{Spectator Effects during Leptogenesis in the Strong
                        Washout Regime}",
      journal        = "JCAP",
      volume         = "1410",
      year           = "2014",
      number         = "10",
      pages          = "012",
      doi            = "10.1088/1475-7516/2014/10/012",
      eprint         = "1404.2915",
      archivePrefix  = "arXiv",
      primaryClass   = "hep-ph",
      reportNumber   = "TUM-HEP-938-14, CERN-PH-TH-2014-059",
      SLACcitation   = "%%CITATION = ARXIV:1404.2915;%%"
}

@article{1606.06690,
      author         = "Drewes, Marco and Garbrecht, Bjorn and Gueter, Dario and
                        Klaric, Juraj",
      title          = "{Leptogenesis from Oscillations of Heavy Neutrinos with
                        Large Mixing Angles}",
      journal        = "JHEP",
      volume         = "12",
      year           = "2016",
      pages          = "150",
      doi            = "10.1007/JHEP12(2016)150",
      eprint         = "1606.06690",
      archivePrefix  = "arXiv",
      primaryClass   = "hep-ph",
      reportNumber   = "TUM-HEP-1050-16",
      SLACcitation   = "%%CITATION = ARXIV:1606.06690;%%"
}

@article{1007.4783,
      author         = "Beneke, Martin and Garbrecht, Bjorn and Fidler, Christian
                        and Herranen, Matti and Schwaller, Pedro",
      title          = "{Flavoured Leptogenesis in the CTP Formalism}",
      journal        = "Nucl. Phys.",
      volume         = "B843",
      year           = "2011",
      pages          = "177-212",
      doi            = "10.1016/j.nuclphysb.2010.10.001",
      eprint         = "1007.4783",
      archivePrefix  = "arXiv",
      primaryClass   = "hep-ph",
      reportNumber   = "TTK-10-44, ZU-TH-09-10",
      SLACcitation   = "%%CITATION = ARXIV:1007.4783;%%"
}

@article{hep-ph/0312110,
      author         = "Prokopec, Tomislav and Schmidt, Michael G. and Weinstock,
                        Steffen",
      title          = "{Transport equations for chiral fermions to order h bar
                        and electroweak baryogenesis. Part 1}",
      journal        = "Annals Phys.",
      volume         = "314",
      year           = "2004",
      pages          = "208-265",
      doi            = "10.1016/j.aop.2004.06.002",
      eprint         = "hep-ph/0312110",
      archivePrefix  = "arXiv",
      primaryClass   = "hep-ph",
      reportNumber   = "BNL-72343-2004-JA, HD-THEP-03-62",
      SLACcitation   = "%%CITATION = HEP-PH/0312110;%%"
}

@article{Berges:2004yj,
      author         = "Berges, Juergen",
      title          = "{Introduction to nonequilibrium quantum field theory}",
      booktitle      = "{Proceedings, 9th Hadron Physics and 7th Relativistic
                        Aspects of Nuclear Physics (HADRON-RANP 2004): A Joint
                        Meeting on QCD and QGP: Rio de Janeiro, Brazil, March
                        28-April 3, 2004}",
      journal        = "AIP Conf. Proc.",
      volume         = "739",
      year           = "2004",
      number         = "1",
      pages          = "3-62",
      doi            = "10.1063/1.1843591",
      eprint         = "hep-ph/0409233",
      archivePrefix  = "arXiv",
      primaryClass   = "hep-ph",
      SLACcitation   = "%%CITATION = HEP-PH/0409233;%%"
}

@article{1112.5954,
      author         = "Garbrecht, Bjorn and Herranen, Matti",
      title          = "{Effective Theory of Resonant Leptogenesis in the
                        Closed-Time-Path Approach}",
      journal        = "Nucl. Phys.",
      volume         = "B861",
      year           = "2012",
      pages          = "17-52",
      doi            = "10.1016/j.nuclphysb.2012.03.009",
      eprint         = "1112.5954",
      archivePrefix  = "arXiv",
      primaryClass   = "hep-ph",
      SLACcitation   = "%%CITATION = ARXIV:1112.5954;%%"
}

@article{hep-ph/9710460,
      author         = "Buchmuller, W. and Plumacher, M.",
      title          = "{CP asymmetry in Majorana neutrino decays}",
      journal        = "Phys. Lett.",
      volume         = "B431",
      year           = "1998",
      pages          = "354-362",
      doi            = "10.1016/S0370-2693(97)01548-7",
      eprint         = "hep-ph/9710460",
      archivePrefix  = "arXiv",
      primaryClass   = "hep-ph",
      reportNumber   = "DESY-97-190",
      SLACcitation   = "%%CITATION = HEP-PH/9710460;%%"
}

@article{1002.1326,
      author         = "Beneke, Martin and Garbrecht, Bjorn and Herranen, Matti
                        and Schwaller, Pedro",
      title          = "{Finite Number Density Corrections to Leptogenesis}",
      journal        = "Nucl. Phys.",
      volume         = "B838",
      year           = "2010",
      pages          = "1-27",
      doi            = "10.1016/j.nuclphysb.2010.05.003",
      eprint         = "1002.1326",
      archivePrefix  = "arXiv",
      primaryClass   = "hep-ph",
      reportNumber   = "TTK-10-16, ZU-TH-02-10",
      SLACcitation   = "%%CITATION = ARXIV:1002.1326;%%"
}

@article{1303.5498,
      author         = "Garbrecht, Björn and Glowna, Frank and Schwaller, Pedro",
      title          = "{Scattering Rates For Leptogenesis: Damping of Lepton
                        Flavour Coherence and Production of Singlet Neutrinos}",
      journal        = "Nucl. Phys.",
      volume         = "B877",
      year           = "2013",
      pages          = "1-35",
      doi            = "10.1016/j.nuclphysb.2013.08.020",
      eprint         = "1303.5498",
      archivePrefix  = "arXiv",
      primaryClass   = "hep-ph",
      reportNumber   = "TUM-HEP-880-13, TTK-13-07, ANL-HEP-PR-13-19",
      SLACcitation   = "%%CITATION = ARXIV:1303.5498;%%"
}

@article{1202.1288,
      author         = "Besak, Denis and Bodeker, Dietrich",
      title          = "{Thermal production of ultrarelativistic right-handed
                        neutrinos: Complete leading-order results}",
      journal        = "JCAP",
      volume         = "1203",
      year           = "2012",
      pages          = "029",
      doi            = "10.1088/1475-7516/2012/03/029",
      eprint         = "1202.1288",
      archivePrefix  = "arXiv",
      primaryClass   = "hep-ph",
      reportNumber   = "BI-TP-2012-05",
      SLACcitation   = "%%CITATION = ARXIV:1202.1288;%%"
}

@article{1302.0743,
      author         = "Garbrecht, Björn and Glowna, Frank and Herranen, Matti",
      title          = "{Right-Handed Neutrino Production at Finite Temperature:
                        Radiative Corrections, Soft and Collinear Divergences}",
      journal        = "JHEP",
      volume         = "04",
      year           = "2013",
      pages          = "099",
      doi            = "10.1007/JHEP04(2013)099",
      eprint         = "1302.0743",
      archivePrefix  = "arXiv",
      primaryClass   = "hep-ph",
      reportNumber   = "TUM-HEP-875-13, TTK-13-02",
      SLACcitation   = "%%CITATION = ARXIV:1302.0743;%%"
}

@phdthesis{Glowna:2015aos,
      author         = "Glowna, Frank",
      title          = "{Right-handed Neutrino Production at Finite
                        Temperatures}",
      school         = "Munich, Tech. U.",
      year           = "2015",
      url            = "http://mediatum.ub.tum.de?id=1253887",
      SLACcitation   = "%%CITATION = INSPIRE-1604369;%%"
}

@article{Weldon:1982bn,
      author         = "Weldon, H. Arthur",
      title          = "{Effective Fermion Masses of Order gT in High Temperature
                        Gauge Theories with Exact Chiral Invariance}",
      journal        = "Phys. Rev.",
      volume         = "D26",
      year           = "1982",
      pages          = "2789",
      doi            = "10.1103/PhysRevD.26.2789",
      reportNumber   = "PRINT-82-0423 (PENN)",
      SLACcitation   = "%%CITATION = PHRVA,D26,2789;%%"
}

@book{Bellac:2011kqa,
      author         = "Bellac, Michel Le",
      title          = "{Thermal Field Theory}",
      publisher      = "Cambridge University Press",
      year           = "2011",
      url            = "http://www.cambridge.org/mw/academic/subjects/physics/theoretical-physics-and-mathematical-physics/thermal-field-theory?format=AR",
      ISBN           = "9780511885068, 9780521654777",
      SLACcitation   = "%%CITATION = INSPIRE-1384874;%%"
}

@article{1012.3784,
      author         = "Anisimov, Alexey and Besak, Denis and Bodeker, Dietrich",
      title          = "{Thermal production of relativistic Majorana neutrinos:
                        Strong enhancement by multiple soft scattering}",
      journal        = "JCAP",
      volume         = "1103",
      year           = "2011",
      pages          = "042",
      doi            = "10.1088/1475-7516/2011/03/042",
      eprint         = "1012.3784",
      archivePrefix  = "arXiv",
      primaryClass   = "hep-ph",
      reportNumber   = "BI-TP-2010-48",
      SLACcitation   = "%%CITATION = ARXIV:1012.3784;%%"
}

@inproceedings{Moore:2000ara,
      author         = "Moore, Guy D.",
      title          = "{Do we understand the sphaleron rate?}",
      booktitle      = "{Strong and electroweak matter. Proceedings, Meeting,
                        SEWM 2000, Marseille, France, June 13-17, 2000}",
      year           = "2000",
      pages          = "82-94",
      doi            = "10.1142/9789812799913_0007",
      eprint         = "hep-ph/0009161",
      archivePrefix  = "arXiv",
      primaryClass   = "hep-ph",
      SLACcitation   = "%%CITATION = HEP-PH/0009161;%%"
}

@article{Fukugita:1986hr,
      author         = "Fukugita, M. and Yanagida, T.",
      title          = "{Baryogenesis Without Grand Unification}",
      journal        = "Phys. Lett.",
      volume         = "B174",
      year           = "1986",
      pages          = "45-47",
      doi            = "10.1016/0370-2693(86)91126-3",
      reportNumber   = "RIFP-641",
      SLACcitation   = "%%CITATION = PHLTA,B174,45;%%"
}

@article{Aaboud:2018wps,
      author         = "Aaboud, Morad and others",
      title          = "{Measurement of the Higgs boson mass in the $H\rightarrow
                        ZZ^* \rightarrow 4\ell$ and $H \rightarrow \gamma\gamma$
                        channels with $\sqrt{s}=13$ TeV $pp$ collisions using the
                        ATLAS detector}",
      collaboration  = "ATLAS",
      journal        = "Phys. Lett.",
      volume         = "B784",
      year           = "2018",
      pages          = "345-366",
      doi            = "10.1016/j.physletb.2018.07.050",
      eprint         = "1806.00242",
      archivePrefix  = "arXiv",
      primaryClass   = "hep-ex",
      reportNumber   = "CERN-EP-2018-085",
      SLACcitation   = "%%CITATION = ARXIV:1806.00242;%%"
}

@article{Aaboud:2018zbu,
      author         = "Aaboud, Morad and others",
      title          = "{Measurement of the top quark mass in the $t\bar{t}\to$
                        lepton+jets channel from $\sqrt{s}=8$ TeV ATLAS data and
                        combination with previous results}",
      collaboration  = "ATLAS",
      journal        = "Submitted to: Eur. Phys. J.",
      year           = "2018",
      eprint         = "1810.01772",
      archivePrefix  = "arXiv",
      primaryClass   = "hep-ex",
      reportNumber   = "CERN-EP-2018-238",
      SLACcitation   = "%%CITATION = ARXIV:1810.01772;%%"
}

@article{Tanabashi:2018oca,
      author         = "Tanabashi, M. and others",
      title          = "{Review of Particle Physics}",
      collaboration  = "Particle Data Group",
      journal        = "Phys. Rev.",
      volume         = "D98",
      year           = "2018",
      number         = "3",
      pages          = "030001",
      doi            = "10.1103/PhysRevD.98.030001",
      SLACcitation   = "%%CITATION = PHRVA,D98,030001;%%"
}

@article{Hempfling:1994ar,
      author         = "Hempfling, Ralf and Kniehl, Bernd A.",
      title          = "{On the relation between the fermion pole mass and MS
                        Yukawa coupling in the standard model}",
      journal        = "Phys. Rev.",
      volume         = "D51",
      year           = "1995",
      pages          = "1386-1394",
      doi            = "10.1103/PhysRevD.51.1386",
      eprint         = "hep-ph/9408313",
      archivePrefix  = "arXiv",
      primaryClass   = "hep-ph",
      reportNumber   = "KEK-TH-408, KEK-PREPRINT-94-74",
      SLACcitation   = "%%CITATION = HEP-PH/9408313;%%"
}

@article{Chetyrkin:1999qi,
      author         = "Chetyrkin, K. G. and Steinhauser, M.",
      title          = "{The Relation between the MS-bar and the on-shell quark
                        mass at order alpha(s)**3}",
      journal        = "Nucl. Phys.",
      volume         = "B573",
      year           = "2000",
      pages          = "617-651",
      doi            = "10.1016/S0550-3213(99)00784-1",
      eprint         = "hep-ph/9911434",
      archivePrefix  = "arXiv",
      primaryClass   = "hep-ph",
      reportNumber   = "DESY-99-174, TTP-99-47",
      SLACcitation   = "%%CITATION = HEP-PH/9911434;%%"
}

@article{Melnikov:2000qh,
      author         = "Melnikov, Kirill and Ritbergen, Timo van",
      title          = "{The Three loop relation between the MS-bar and the pole
                        quark masses}",
      journal        = "Phys. Lett.",
      volume         = "B482",
      year           = "2000",
      pages          = "99-108",
      doi            = "10.1016/S0370-2693(00)00507-4",
      eprint         = "hep-ph/9912391",
      archivePrefix  = "arXiv",
      primaryClass   = "hep-ph",
      reportNumber   = "SLAC-PUB-8321, TTP-99-51",
      SLACcitation   = "%%CITATION = HEP-PH/9912391;%%"
}

@article{Luo:2002ey,
      author         = "Luo, Ming-xing and Xiao, Yong",
      title          = "{Two loop renormalization group equations in the standard
                        model}",
      journal        = "Phys. Rev. Lett.",
      volume         = "90",
      year           = "2003",
      pages          = "011601",
      doi            = "10.1103/PhysRevLett.90.011601",
      eprint         = "hep-ph/0207271",
      archivePrefix  = "arXiv",
      primaryClass   = "hep-ph",
      SLACcitation   = "%%CITATION = HEP-PH/0207271;%%"
}

@article{Degrassi:2012ry,
      author         = "Degrassi, Giuseppe and Di Vita, Stefano and Elias-Miro,
                        Joan and Espinosa, Jose R. and Giudice, Gian F. and
                        Isidori, Gino and Strumia, Alessandro",
      title          = "{Higgs mass and vacuum stability in the Standard Model at
                        NNLO}",
      journal        = "JHEP",
      volume         = "08",
      year           = "2012",
      pages          = "098",
      doi            = "10.1007/JHEP08(2012)098",
      eprint         = "1205.6497",
      archivePrefix  = "arXiv",
      primaryClass   = "hep-ph",
      reportNumber   = "CERN-PH-TH-2012-134, RM3-TH-12-9",
      SLACcitation   = "%%CITATION = ARXIV:1205.6497;%%"
}

@article{Keldysh:1964ud,
      author         = "Keldysh, L. V.",
      title          = "{Diagram technique for nonequilibrium processes}",
      journal        = "Zh. Eksp. Teor. Fiz.",
      volume         = "47",
      year           = "1964",
      pages          = "1515-1527",
      note           = "[Sov. Phys. JETP20,1018(1965)]",
      SLACcitation   = "%%CITATION = ZETFA,47,1515;%%"
}

@article{Schwinger:1960qe,
      author         = "Schwinger, Julian S.",
      title          = "{Brownian motion of a quantum oscillator}",
      journal        = "J. Math. Phys.",
      volume         = "2",
      year           = "1961",
      pages          = "407-432",
      doi            = "10.1063/1.1703727",
      SLACcitation   = "%%CITATION = JMAPA,2,407;%%"
}

@article{Calzetta:1986cq,
      author         = "Calzetta, E. and Hu, B. L.",
      title          = "{Nonequilibrium Quantum Fields: Closed Time Path
                        Effective Action, Wigner Function and Boltzmann Equation}",
      journal        = "Phys. Rev.",
      volume         = "D37",
      year           = "1988",
      pages          = "2878",
      doi            = "10.1103/PhysRevD.37.2878",
      reportNumber   = "MDDP-PP-87-104",
      SLACcitation   = "%%CITATION = PHRVA,D37,2878;%%"
}

@article{Braaten:1989mz,
      author         = "Braaten, Eric and Pisarski, Robert D.",
      title          = "{Soft Amplitudes in Hot Gauge Theories: A General
                        Analysis}",
      journal        = "Nucl. Phys.",
      volume         = "B337",
      year           = "1990",
      pages          = "569-634",
      doi            = "10.1016/0550-3213(90)90508-B",
      reportNumber   = "BNL-43293, FERMILAB-PUB-89-152-T, NUHEP-TH-89-7",
      SLACcitation   = "%%CITATION = NUPHA,B337,569;%%"
}

@article{Hambye:2017elz,
      author         = "Hambye, Thomas and Teresi, Daniele",
      title          = "{Baryogenesis from L-violating Higgs-doublet decay in the
                        density-matrix formalism}",
      journal        = "Phys. Rev.",
      volume         = "D96",
      year           = "2017",
      number         = "1",
      pages          = "015031",
      doi            = "10.1103/PhysRevD.96.015031",
      eprint         = "1705.00016",
      archivePrefix  = "arXiv",
      primaryClass   = "hep-ph",
      reportNumber   = "ULB-TH-17-07",
      SLACcitation   = "%%CITATION = ARXIV:1705.00016;%%"
}

@article{Hambye:2016sby,
      author         = "Hambye, Thomas and Teresi, Daniele",
      title          = "{Higgs doublet decay as the origin of the baryon
                        asymmetry}",
      journal        = "Phys. Rev. Lett.",
      volume         = "117",
      year           = "2016",
      number         = "9",
      pages          = "091801",
      doi            = "10.1103/PhysRevLett.117.091801",
      eprint         = "1606.00017",
      archivePrefix  = "arXiv",
      primaryClass   = "hep-ph",
      reportNumber   = "ULB-TH-16-08",
      SLACcitation   = "%%CITATION = ARXIV:1606.00017;%%"
}

@article{Ghiglieri:2017gjz,
      author         = "Ghiglieri, J. and Laine, M.",
      title          = "{GeV-scale hot sterile neutrino oscillations: a
                        derivation of evolution equations}",
      journal        = "JHEP",
      volume         = "05",
      year           = "2017",
      pages          = "132",
      doi            = "10.1007/JHEP05(2017)132",
      eprint         = "1703.06087",
      archivePrefix  = "arXiv",
      primaryClass   = "hep-ph",
      SLACcitation   = "%%CITATION = ARXIV:1703.06087;%%"
}

@article{Eijima:2017anv,
      author         = "Eijima, Shintaro and Shaposhnikov, Mikhail",
      title          = "{Fermion number violating effects in low scale
                        leptogenesis}",
      journal        = "Phys. Lett.",
      volume         = "B771",
      year           = "2017",
      pages          = "288-296",
      doi            = "10.1016/j.physletb.2017.05.068",
      eprint         = "1703.06085",
      archivePrefix  = "arXiv",
      primaryClass   = "hep-ph",
      SLACcitation   = "%%CITATION = ARXIV:1703.06085;%%"
}

@article{Antusch:2017pkq,
      author         = "Antusch, Stefan and Cazzato, Eros and Drewes, Marco and
                        Fischer, Oliver and Garbrecht, Bjorn and Gueter, Dario and
                        Klaric, Juraj",
      title          = "{Probing Leptogenesis at Future Colliders}",
      journal        = "JHEP",
      volume         = "09",
      year           = "2018",
      pages          = "124",
      doi            = "10.1007/JHEP09(2018)124",
      eprint         = "1710.03744",
      archivePrefix  = "arXiv",
      primaryClass   = "hep-ph",
      reportNumber   = "TUM-1160/18, CP3-17-48",
      SLACcitation   = "%%CITATION = ARXIV:1710.03744;%%"
}

@article{Akhmedov:1998qx,
      author         = "Akhmedov, Evgeny K. and Rubakov, V. A. and Smirnov, A.
                        {\relax Yu}.",
      title          = "{Baryogenesis via neutrino oscillations}",
      journal        = "Phys. Rev. Lett.",
      volume         = "81",
      year           = "1998",
      pages          = "1359-1362",
      doi            = "10.1103/PhysRevLett.81.1359",
      eprint         = "hep-ph/9803255",
      archivePrefix  = "arXiv",
      primaryClass   = "hep-ph",
      reportNumber   = "IC-98-22, INR-98-14-T",
      SLACcitation   = "%%CITATION = HEP-PH/9803255;%%"
}

@article{Asaka:2005pn,
      author         = "Asaka, Takehiko and Shaposhnikov, Mikhail",
      title          = "{The nuMSM, dark matter and baryon asymmetry of the
                        universe}",
      journal        = "Phys. Lett.",
      volume         = "B620",
      year           = "2005",
      pages          = "17-26",
      doi            = "10.1016/j.physletb.2005.06.020",
      eprint         = "hep-ph/0505013",
      archivePrefix  = "arXiv",
      primaryClass   = "hep-ph",
      SLACcitation   = "%%CITATION = HEP-PH/0505013;%%"
}

@article{Barbieri:1999ma,
      author         = "Barbieri, Riccardo and Creminelli, Paolo and Strumia,
                        Alessandro and Tetradis, Nikolaos",
      title          = "{Baryogenesis through leptogenesis}",
      journal        = "Nucl. Phys.",
      volume         = "B575",
      year           = "2000",
      pages          = "61-77",
      doi            = "10.1016/S0550-3213(00)00011-0",
      eprint         = "hep-ph/9911315",
      archivePrefix  = "arXiv",
      primaryClass   = "hep-ph",
      reportNumber   = "SNS-PH-99-15, IFUP-TH-99-55",
      SLACcitation   = "%%CITATION = HEP-PH/9911315;%%"
}

@article{Basboll:2006yx,
      author         = "Basboll, Anders and Hannestad, Steen",
      title          = "{Decay of heavy Majorana neutrinos using the full
                        Boltzmann equation including its implications for
                        leptogenesis}",
      journal        = "JCAP",
      volume         = "0701",
      year           = "2007",
      pages          = "003",
      doi            = "10.1088/1475-7516/2007/01/003",
      eprint         = "hep-ph/0609025",
      archivePrefix  = "arXiv",
      primaryClass   = "hep-ph",
      SLACcitation   = "%%CITATION = HEP-PH/0609025;%%"
}

@article{HahnWoernle:2009qn,
      author         = "Hahn-Woernle, F. and Plumacher, M. and Wong, Y. Y. Y.",
      title          = "{Full Boltzmann equations for leptogenesis including
                        scattering}",
      journal        = "JCAP",
      volume         = "0908",
      year           = "2009",
      pages          = "028",
      doi            = "10.1088/1475-7516/2009/08/028",
      eprint         = "0907.0205",
      archivePrefix  = "arXiv",
      primaryClass   = "hep-ph",
      reportNumber   = "CERN-PH-TH-2009-107, MPP-2009-85, PITHA-09-15",
      SLACcitation   = "%%CITATION = ARXIV:0907.0205;%%"
}

@article{Asaka:2011wq,
      author         = "Asaka, Takehiko and Eijima, Shintaro and Ishida, Hiroyuki",
      title          = "{Kinetic Equations for Baryogenesis via Sterile Neutrino
                        Oscillation}",
      journal        = "JCAP",
      volume         = "1202",
      year           = "2012",
      pages          = "021",
      doi            = "10.1088/1475-7516/2012/02/021",
      eprint         = "1112.5565",
      archivePrefix  = "arXiv",
      primaryClass   = "hep-ph",
      SLACcitation   = "%%CITATION = ARXIV:1112.5565;%%"
}

@article{Ghiglieri:2017csp,
      author         = "Ghiglieri, J. and Laine, M.",
      title          = "{GeV-scale hot sterile neutrino oscillations: a numerical
                        solution}",
      journal        = "JHEP",
      volume         = "02",
      year           = "2018",
      pages          = "078",
      doi            = "10.1007/JHEP02(2018)078",
      eprint         = "1711.08469",
      archivePrefix  = "arXiv",
      primaryClass   = "hep-ph",
      reportNumber   = "CERN-TH-2018-022",
      SLACcitation   = "%%CITATION = ARXIV:1711.08469;%%"
}

@article{Endoh:2003mz,
      author         = "Endoh, Tomohiro and Morozumi, Takuya and Xiong, Zhao-hua",
      title          = "{Primordial lepton family asymmetries in seesaw model}",
      journal        = "Prog. Theor. Phys.",
      volume         = "111",
      year           = "2004",
      pages          = "123-149",
      doi            = "10.1143/PTP.111.123",
      eprint         = "hep-ph/0308276",
      archivePrefix  = "arXiv",
      primaryClass   = "hep-ph",
      SLACcitation   = "%%CITATION = HEP-PH/0308276;%%"
}

@article{Pilaftsis:2003gt,
      author         = "Pilaftsis, Apostolos and Underwood, Thomas E. J.",
      title          = "{Resonant leptogenesis}",
      journal        = "Nucl. Phys.",
      volume         = "B692",
      year           = "2004",
      pages          = "303-345",
      doi            = "10.1016/j.nuclphysb.2004.05.029",
      eprint         = "hep-ph/0309342",
      archivePrefix  = "arXiv",
      primaryClass   = "hep-ph",
      reportNumber   = "MC-TH-2003-09",
      SLACcitation   = "%%CITATION = HEP-PH/0309342;%%"
}

@article{Pilaftsis:2005rv,
      author         = "Pilaftsis, Apostolos and Underwood, Thomas E. J.",
      title          = "{Electroweak-scale resonant leptogenesis}",
      journal        = "Phys. Rev.",
      volume         = "D72",
      year           = "2005",
      pages          = "113001",
      doi            = "10.1103/PhysRevD.72.113001",
      eprint         = "hep-ph/0506107",
      archivePrefix  = "arXiv",
      primaryClass   = "hep-ph",
      SLACcitation   = "%%CITATION = HEP-PH/0506107;%%"
}

@article{Abada:2006fw,
      author         = "Abada, Asmaa and Davidson, Sacha and Josse-Michaux,
                        Francois-Xavier and Losada, Marta and Riotto, Antonio",
      title          = "{Flavor issues in leptogenesis}",
      journal        = "JCAP",
      volume         = "0604",
      year           = "2006",
      pages          = "004",
      doi            = "10.1088/1475-7516/2006/04/004",
      eprint         = "hep-ph/0601083",
      archivePrefix  = "arXiv",
      primaryClass   = "hep-ph",
      reportNumber   = "CERN-PH-TH-2006-001, DNI-UAN-06-03, LPT-ORSAY-06-03,
                        LYCEN-2006-03",
      SLACcitation   = "%%CITATION = HEP-PH/0601083;%%"
}

@article{Nardi:2006fx,
      author         = "Nardi, Enrico and Nir, Yosef and Roulet, Esteban and
                        Racker, Juan",
      title          = "{The Importance of flavor in leptogenesis}",
      journal        = "JHEP",
      volume         = "01",
      year           = "2006",
      pages          = "164",
      doi            = "10.1088/1126-6708/2006/01/164",
      eprint         = "hep-ph/0601084",
      archivePrefix  = "arXiv",
      primaryClass   = "hep-ph",
      SLACcitation   = "%%CITATION = HEP-PH/0601084;%%"
}

@article{Covi:1996wh,
      author         = "Covi, Laura and Roulet, Esteban and Vissani, Francesco",
      title          = "{CP violating decays in leptogenesis scenarios}",
      journal        = "Phys. Lett.",
      volume         = "B384",
      year           = "1996",
      pages          = "169-174",
      doi            = "10.1016/0370-2693(96)00817-9",
      eprint         = "hep-ph/9605319",
      archivePrefix  = "arXiv",
      primaryClass   = "hep-ph",
      reportNumber   = "SISSA-66-96-EP, IC-96-73",
      SLACcitation   = "%%CITATION = HEP-PH/9605319;%%"
}

@article{Salvio:2011sf,
      author         = "Salvio, Alberto and Lodone, Paolo and Strumia,
                        Alessandro",
      title          = "{Towards leptogenesis at NLO: the right-handed neutrino
                        interaction rate}",
      journal        = "JHEP",
      volume         = "08",
      year           = "2011",
      pages          = "116",
      doi            = "10.1007/JHEP08(2011)116",
      eprint         = "1106.2814",
      archivePrefix  = "arXiv",
      primaryClass   = "hep-ph",
      reportNumber   = "IFUP-TH-2008-19",
      SLACcitation   = "%%CITATION = ARXIV:1106.2814;%%"
}

@article{Laine:2011pq,
      author         = "Laine, M. and Schroder, Y.",
      title          = "{Thermal right-handed neutrino production rate in the
                        non-relativistic regime}",
      journal        = "JHEP",
      volume         = "02",
      year           = "2012",
      pages          = "068",
      doi            = "10.1007/JHEP02(2012)068",
      eprint         = "1112.1205",
      archivePrefix  = "arXiv",
      primaryClass   = "hep-ph",
      SLACcitation   = "%%CITATION = ARXIV:1112.1205;%%"
}

@article{Giudice:2003jh,
      author         = "Giudice, G. F. and Notari, A. and Raidal, M. and Riotto,
                        A. and Strumia, A.",
      title          = "{Towards a complete theory of thermal leptogenesis in the
                        SM and MSSM}",
      journal        = "Nucl. Phys.",
      volume         = "B685",
      year           = "2004",
      pages          = "89-149",
      doi            = "10.1016/j.nuclphysb.2004.02.019",
      eprint         = "hep-ph/0310123",
      archivePrefix  = "arXiv",
      primaryClass   = "hep-ph",
      reportNumber   = "IFUP-TH-2003-37, CERN-TH-2003-240",
      SLACcitation   = "%%CITATION = HEP-PH/0310123;%%"
}

@article{Kiessig:2011fw,
      author         = "Kiessig, Clemens and Plumacher, Michael",
      title          = "{Hard-Thermal-Loop Corrections in Leptogenesis I:
                        CP-Asymmetries}",
      journal        = "JCAP",
      volume         = "1207",
      year           = "2012",
      pages          = "014",
      doi            = "10.1088/1475-7516/2012/07/014",
      eprint         = "1111.1231",
      archivePrefix  = "arXiv",
      primaryClass   = "hep-ph",
      reportNumber   = "MPP-2011-61",
      SLACcitation   = "%%CITATION = ARXIV:1111.1231;%%"
}

@article{Kiessig:2011ga,
      author         = "Kiessig, Clemens and Plumacher, Michael",
      title          = "{Hard-Thermal-Loop Corrections in Leptogenesis II:
                        Solving the Boltzmann Equations}",
      journal        = "JCAP",
      volume         = "1209",
      year           = "2012",
      pages          = "012",
      doi            = "10.1088/1475-7516/2012/09/012",
      eprint         = "1111.1235",
      archivePrefix  = "arXiv",
      primaryClass   = "hep-ph",
      reportNumber   = "MPP-2011-62",
      SLACcitation   = "%%CITATION = ARXIV:1111.1235;%%"
}

@article{Buchmuller:2000nd,
      author         = "Buchmuller, Wilfried and Fredenhagen, Stefan",
      title          = "{Quantum mechanics of baryogenesis}",
      journal        = "Phys. Lett.",
      volume         = "B483",
      year           = "2000",
      pages          = "217-224",
      doi            = "10.1016/S0370-2693(00)00573-6",
      eprint         = "hep-ph/0004145",
      archivePrefix  = "arXiv",
      primaryClass   = "hep-ph",
      reportNumber   = "DESY-00-056",
      SLACcitation   = "%%CITATION = HEP-PH/0004145;%%"
}

@article{DeSimone:2007gkc,
      author         = "De Simone, Andrea and Riotto, Antonio",
      title          = "{Quantum Boltzmann Equations and Leptogenesis}",
      journal        = "JCAP",
      volume         = "0708",
      year           = "2007",
      pages          = "002",
      doi            = "10.1088/1475-7516/2007/08/002",
      eprint         = "hep-ph/0703175",
      archivePrefix  = "arXiv",
      primaryClass   = "hep-ph",
      reportNumber   = "DFPD-07-TH-04, MIT-CTP-3817",
      SLACcitation   = "%%CITATION = HEP-PH/0703175;%%"
}

@article{Garny:2009rv,
      author         = "Garny, M. and Hohenegger, A. and Kartavtsev, A. and
                        Lindner, M.",
      title          = "{Systematic approach to leptogenesis in nonequilibrium
                        QFT: Vertex contribution to the CP-violating parameter}",
      journal        = "Phys. Rev.",
      volume         = "D80",
      year           = "2009",
      pages          = "125027",
      doi            = "10.1103/PhysRevD.80.125027",
      eprint         = "0909.1559",
      archivePrefix  = "arXiv",
      primaryClass   = "hep-ph",
      reportNumber   = "TUM-HEP-735-09",
      SLACcitation   = "%%CITATION = ARXIV:0909.1559;%%"
}

@article{Garny:2009qn,
      author         = "Garny, M. and Hohenegger, A. and Kartavtsev, A. and
                        Lindner, M.",
      title          = "{Systematic approach to leptogenesis in nonequilibrium
                        QFT: Self-energy contribution to the CP-violating
                        parameter}",
      journal        = "Phys. Rev.",
      volume         = "D81",
      year           = "2010",
      pages          = "085027",
      doi            = "10.1103/PhysRevD.81.085027",
      eprint         = "0911.4122",
      archivePrefix  = "arXiv",
      primaryClass   = "hep-ph",
      reportNumber   = "TUM-HEP-740-09",
      SLACcitation   = "%%CITATION = ARXIV:0911.4122;%%"
}

@article{Garny:2010nj,
      author         = "Garny, M. and Hohenegger, A. and Kartavtsev, A.",
      title          = "{Medium corrections to the CP-violating parameter in
                        leptogenesis}",
      journal        = "Phys. Rev.",
      volume         = "D81",
      year           = "2010",
      pages          = "085028",
      doi            = "10.1103/PhysRevD.81.085028",
      eprint         = "1002.0331",
      archivePrefix  = "arXiv",
      primaryClass   = "hep-ph",
      reportNumber   = "TUM-HEP-749-10",
      SLACcitation   = "%%CITATION = ARXIV:1002.0331;%%"
}

@article{Garny:2010nz,
      author         = "Garny, M. and Hohenegger, A. and Kartavtsev, A.",
      title          = "{Quantum corrections to leptogenesis from the gradient
                        expansion}",
      year           = "2010",
      eprint         = "1005.5385",
      archivePrefix  = "arXiv",
      primaryClass   = "hep-ph",
      SLACcitation   = "%%CITATION = ARXIV:1005.5385;%%"
}

@article{Anisimov:2010dk,
      author         = "Anisimov, A. and Buchmüller, W. and Drewes, M. and
                        Mendizabal, S.",
      title          = "{Quantum Leptogenesis I}",
      journal        = "Annals Phys.",
      volume         = "326",
      year           = "2011",
      pages          = "1998-2038",
      doi            = "10.1016/j.aop.2011.02.002, 10.1016/j.aop.2013.05.00",
      note           = "[Erratum: Annals Phys.338,376(2011)]",
      eprint         = "1012.5821",
      archivePrefix  = "arXiv",
      primaryClass   = "hep-ph",
      reportNumber   = "DESY-10-218",
      SLACcitation   = "%%CITATION = ARXIV:1012.5821;%%"
}

@article{Garbrecht:2011aw,
      author         = "Garbrecht, Bjorn and Herranen, Matti",
      title          = "{Effective Theory of Resonant Leptogenesis in the
                        Closed-Time-Path Approach}",
      journal        = "Nucl. Phys.",
      volume         = "B861",
      year           = "2012",
      pages          = "17-52",
      doi            = "10.1016/j.nuclphysb.2012.03.009",
      eprint         = "1112.5954",
      archivePrefix  = "arXiv",
      primaryClass   = "hep-ph",
      SLACcitation   = "%%CITATION = ARXIV:1112.5954;%%"
}

@article{DeSimone:2006nrs,
      author         = "De Simone, Andrea and Riotto, Antonio",
      title          = "{On the impact of flavour oscillations in leptogenesis}",
      journal        = "JCAP",
      volume         = "0702",
      year           = "2007",
      pages          = "005",
      doi            = "10.1088/1475-7516/2007/02/005",
      eprint         = "hep-ph/0611357",
      archivePrefix  = "arXiv",
      primaryClass   = "hep-ph",
      reportNumber   = "CERN-PH-TH-2006-242, MIT-CTP-3789",
      SLACcitation   = "%%CITATION = HEP-PH/0611357;%%"
}

@article{Ghiglieri:2018wbs,
      author         = "Ghiglieri, J. and Laine, M.",
      title          = "{Precision study of GeV-scale resonant leptogenesis}",
      journal        = "JHEP",
      volume         = "02",
      year           = "2019",
      pages          = "014",
      doi            = "10.1007/JHEP02(2019)014",
      eprint         = "1811.01971",
      archivePrefix  = "arXiv",
      primaryClass   = "hep-ph",
      reportNumber   = "CERN-TH-2018-232",
      SLACcitation   = "%%CITATION = ARXIV:1811.01971;%%"
}

@article{Eijima:2017cxr,
      author         = "Eijima, S. and Shaposhnikov, M. and Timiryasov, I.",
      title          = "{Freeze-out of baryon number in low-scale leptogenesis}",
      journal        = "JCAP",
      volume         = "1711",
      year           = "2017",
      number         = "11",
      pages          = "030",
      doi            = "10.1088/1475-7516/2017/11/030",
      eprint         = "1709.07834",
      archivePrefix  = "arXiv",
      primaryClass   = "hep-ph",
      SLACcitation   = "%%CITATION = ARXIV:1709.07834;%%"
}

@article{Eijima:2018qke,
      author         = "Eijima, S. and Shaposhnikov, M. and Timiryasov, I.",
      title          = "{Parameter space of baryogenesis in the $\nu$MSM}",
      year           = "2018",
      eprint         = "1808.10833",
      archivePrefix  = "arXiv",
      primaryClass   = "hep-ph",
      SLACcitation   = "%%CITATION = ARXIV:1808.10833;%%"
}

@article{Ghisoiu:2014ena,
      author         = "Ghisoiu, I. and Laine, M.",
      title          = "{Right-handed neutrino production rate at T > 160 GeV}",
      journal        = "JCAP",
      volume         = "1412",
      year           = "2014",
      number         = "12",
      pages          = "032",
      doi            = "10.1088/1475-7516/2014/12/032",
      eprint         = "1411.1765",
      archivePrefix  = "arXiv",
      primaryClass   = "hep-ph",
      SLACcitation   = "%%CITATION = ARXIV:1411.1765;%%"
}

@article{Cornwall:1974vz,
      author         = "Cornwall, John M. and Jackiw, R. and Tomboulis, E.",
      title          = "{Effective Action for Composite Operators}",
      journal        = "Phys. Rev.",
      volume         = "D10",
      year           = "1974",
      pages          = "2428-2445",
      doi            = "10.1103/PhysRevD.10.2428",
      reportNumber   = "MIT-CTP-419",
      SLACcitation   = "%%CITATION = PHRVA,D10,2428;%%"
}

@article{Biondini:2017rpb,
      author         = "Biondini, Simone and others",
      title          = "{Status of rates and rate equations for thermal
                        leptogenesis}",
      journal        = "Int. J. Mod. Phys.",
      volume         = "A33",
      year           = "2018",
      number         = "05n06",
      pages          = "1842004",
      doi            = "10.1142/S0217751X18420046",
      eprint         = "1711.02864",
      archivePrefix  = "arXiv",
      primaryClass   = "hep-ph",
      SLACcitation   = "%%CITATION = ARXIV:1711.02864;%%"
}

@article{Buchmuller:2001sr,
      author         = "Buchmuller, W. and Plumacher, M.",
      title          = "{Spectator processes and baryogenesis}",
      journal        = "Phys. Lett.",
      volume         = "B511",
      year           = "2001",
      pages          = "74-76",
      doi            = "10.1016/S0370-2693(01)00614-1",
      eprint         = "hep-ph/0104189",
      archivePrefix  = "arXiv",
      primaryClass   = "hep-ph",
      reportNumber   = "DESY-01-046, OUTP-01-19-P",
      SLACcitation   = "%%CITATION = HEP-PH/0104189;%%"
}

@article{Buchmuller:2005eh,
      author         = "Buchmuller, W. and Peccei, R. D. and Yanagida, T.",
      title          = "{Leptogenesis as the origin of matter}",
      journal        = "Ann. Rev. Nucl. Part. Sci.",
      volume         = "55",
      year           = "2005",
      pages          = "311-355",
      doi            = "10.1146/annurev.nucl.55.090704.151558",
      eprint         = "hep-ph/0502169",
      archivePrefix  = "arXiv",
      primaryClass   = "hep-ph",
      reportNumber   = "DESY-05-031",
      SLACcitation   = "%%CITATION = HEP-PH/0502169;%%"
}

@article{Davidson:2008bu,
      author         = "Davidson, Sacha and Nardi, Enrico and Nir, Yosef",
      title          = "{Leptogenesis}",
      journal        = "Phys. Rept.",
      volume         = "466",
      year           = "2008",
      pages          = "105-177",
      doi            = "10.1016/j.physrep.2008.06.002",
      eprint         = "0802.2962",
      archivePrefix  = "arXiv",
      primaryClass   = "hep-ph",
      SLACcitation   = "%%CITATION = ARXIV:0802.2962;%%"
}

@article{Garbrecht:2018mrp,
      author         = "Garbrecht, Björn",
      title          = "{Why is there more matter than antimatter? Calculational
                        methods for leptogenesis and electroweak baryogenesis}",
      year           = "2018",
      eprint         = "1812.02651",
      archivePrefix  = "arXiv",
      primaryClass   = "hep-ph",
      reportNumber   = "TUM-HEP-1177-18",
      SLACcitation   = "%%CITATION = ARXIV:1812.02651;%%"
}

@article{Biondini:2013xua,
      author         = "Biondini, Simone and Brambilla, Nora and Escobedo, Miguel
                        Angel and Vairo, Antonio",
      title          = "{An effective field theory for non-relativistic Majorana
                        neutrinos}",
      journal        = "JHEP",
      volume         = "12",
      year           = "2013",
      pages          = "028",
      doi            = "10.1007/JHEP12(2013)028",
      eprint         = "1307.7680",
      archivePrefix  = "arXiv",
      primaryClass   = "hep-ph",
      reportNumber   = "TUM-EFT-34-12",
      SLACcitation   = "%%CITATION = ARXIV:1307.7680;%%"
}

@article{Biondini:2015gyw,
      author         = "Biondini, Simone and Brambilla, Nora and Escobedo, Miguel
                        Angel and Vairo, Antonio",
      title          = "{CP asymmetry in heavy Majorana neutrino decays at finite
                        temperature: the nearly degenerate case}",
      journal        = "JHEP",
      volume         = "03",
      year           = "2016",
      pages          = "191",
      doi            = "10.1007/JHEP03(2016)191, 10.1007/JHEP08(2016)072",
      note           = "[Erratum: JHEP08,072(2016)]",
      eprint         = "1511.02803",
      archivePrefix  = "arXiv",
      primaryClass   = "hep-ph",
      SLACcitation   = "%%CITATION = ARXIV:1511.02803;%%"
}

@article{Biondini:2016arl,
      author         = "Biondini, Simone and Brambilla, Nora and Vairo, Antonio",
      title          = "{CP asymmetry in heavy Majorana neutrino decays at finite
                        temperature: the hierarchical case}",
      journal        = "JHEP",
      volume         = "09",
      year           = "2016",
      pages          = "126",
      doi            = "10.1007/JHEP09(2016)126",
      eprint         = "1608.01979",
      archivePrefix  = "arXiv",
      primaryClass   = "hep-ph",
      SLACcitation   = "%%CITATION = ARXIV:1608.01979;%%"
}

@article{Garny:2011hg,
      author         = "Garny, Mathias and Kartavtsev, Alexander and Hohenegger,
                        Andreas",
      title          = "{Leptogenesis from first principles in the resonant
                        regime}",
      journal        = "Annals Phys.",
      volume         = "328",
      year           = "2013",
      pages          = "26-63",
      doi            = "10.1016/j.aop.2012.10.007",
      eprint         = "1112.6428",
      archivePrefix  = "arXiv",
      primaryClass   = "hep-ph",
      reportNumber   = "DESY-11-264",
      SLACcitation   = "%%CITATION = ARXIV:1112.6428;%%"
}

@article{Drewes:2016gmt,
      author         = "Drewes, Marco and Garbrecht, Bjorn and Gueter, Dario and
                        Klaric, Juraj",
      title          = "{Leptogenesis from Oscillations of Heavy Neutrinos with
                        Large Mixing Angles}",
      journal        = "JHEP",
      volume         = "12",
      year           = "2016",
      pages          = "150",
      doi            = "10.1007/JHEP12(2016)150",
      eprint         = "1606.06690",
      archivePrefix  = "arXiv",
      primaryClass   = "hep-ph",
      reportNumber   = "TUM-HEP-1050-16",
      SLACcitation   = "%%CITATION = ARXIV:1606.06690;%%"
}

@article{Garbrecht:2011xw,
      author         = "Garbrecht, Bjorn and Garny, Mathias",
      title          = "{Finite Width in out-of-Equilibrium Propagators and
                        Kinetic Theory}",
      journal        = "Annals Phys.",
      volume         = "327",
      year           = "2012",
      pages          = "914-934",
      doi            = "10.1016/j.aop.2011.10.005",
      eprint         = "1108.3688",
      archivePrefix  = "arXiv",
      primaryClass   = "hep-ph",
      reportNumber   = "TTK-11-37, TUM-HEP-816-11",
      SLACcitation   = "%%CITATION = ARXIV:1108.3688;%%"
}

@article{Croon:2019dfw,
      author         = "Croon, Djuna and Fernandez, Nicolas and McKeen, David and
                        White, Graham",
      title          = "{Stability, reheating and leptogenesis}",
      year           = "2019",
      eprint         = "1903.08658",
      archivePrefix  = "arXiv",
      primaryClass   = "hep-ph",
      reportNumber   = "SCIPP 18/07",
      SLACcitation   = "%%CITATION = ARXIV:1903.08658;%%"
}

@article{Bodeker:2017deo,
      author         = "Bodeker, Dietrich and Sangel, Marc",
      title          = "{Lepton asymmetry rate from quantum field theory: NLO in
                        the hierarchical limit}",
      journal        = "JCAP",
      volume         = "1706",
      year           = "2017",
      number         = "06",
      pages          = "052",
      doi            = "10.1088/1475-7516/2017/06/052",
      eprint         = "1702.02155",
      archivePrefix  = "arXiv",
      primaryClass   = "hep-ph",
      SLACcitation   = "%%CITATION = ARXIV:1702.02155;%%"
}

@article{Bodeker:2019ajh,
      author         = "Bödeker, Dietrich and Schröder, Dennis",
      title          = "{Equilibration of right-handed electrons}",
      journal        = "JCAP",
      volume         = "1905",
      year           = "2019",
      number         = "05",
      pages          = "010",
      doi            = "10.1088/1475-7516/2019/05/010",
      eprint         = "1902.07220",
      archivePrefix  = "arXiv",
      primaryClass   = "hep-ph",
      SLACcitation   = "%%CITATION = ARXIV:1902.07220;%%"
}

\end{document}